\documentclass[useAMS,twocolumn,fleqn,usenatbib]{mn2e}
\usepackage{xr}
\usepackage{url}
\usepackage[dvips]{graphicx,epsfig,color}
\usepackage{amsmath,amssymb}
\usepackage[T1]{fontenc}
\usepackage{aecompl}
\usepackage{subfigure}
\usepackage[light,bottom,none]{draftcopy}
\usepackage{times}

\newcommand\sersic{S\'{e}rsic}

\def\kms{\ifmmode km\thinspace \mbox{s}^{-1}\else km\thinspace$\mbox{s}^{-1}$\fi}     
\def\deg{\ifmmode^\circ\else$^\circ$\fi}  
\def\arcs{\ifmmode {'' }\else $'' $\fi}  
\def\arcm{\ifmmode {' }\else $' $\fi}    

\def\msun{$M_\odot$}
\def\kinms{\texttt{KinMS}}

\def\mbh{\ifmmode{M_{\mbox{\tiny BH}}}\else $M_{\mbox{\tiny BH}}$\fi}
\def\vdiff{\ifmmode{V_{\mbox{\tiny diff}}}\else $V_{\mbox{\tiny diff}}$\fi}
\def\rsoi{\ifmmode{r_{\mbox{\tiny SOI}}}\else $r_{\mbox{\tiny SOI}}$\fi}
\def\reqv{\ifmmode{r_{\mbox{\tiny eqv}}}\else $r_{\mbox{\tiny eqv}}$\fi}
\def\vgal{\ifmmode{V(r)_{\mbox{\tiny gal}}}\else $V(r)_{\mbox{\tiny gal}}$\fi}
\def\vrot{\ifmmode{V(r)_{\mbox{\tiny rot}}}\else $V(r)_{\mbox{\tiny rot}}$\fi}
\def\vobs{\ifmmode{V(r)_{\mbox{\tiny obs}}}\else $V(r)_{\mbox{\tiny obs}}$\fi}

\def\h2{$\mbox{H}_2$}
\def\lapp{\ifmmode {_<\atop{^\sim}} \else {$_<\atop{^\sim}$}\fi} 
\def\gapp{\ifmmode {_>\atop{^\sim}} \else {$_>\atop{^\sim}$}\fi} 

\title[SMBH mass measurement using ALMA]
{Black hole mass measurement using molecular gas kinematics: What ALMA can do}
\author[Ilsang Yoon]{Ilsang Yoon$^{1,2}$\thanks{E-mail: iyoon@nrao.edu}\\
$^{1}$National Radio Astronomy Observatory, 520 Edgemont Road, Charlottesville, VA 22903, USA\\
$^{2}$Leiden Observatory, Leiden University, P.O. Box 9513, 2300 RA Leiden, The Netherlands}

\begin{document}

\date{Accepted 1988 December 15. Received 1988 December 14; in original form 1988 October 11}

\pagerange{\pageref{firstpage}--\pageref{lastpage}} \pubyear{2002}

\maketitle

\label{firstpage}

\begin{abstract}
We study the limits of spatial and velocity resolution of radio interferometry to infer 
the mass of supermassive black hole (SMBH) in galactic centre using the kinematics of circum nuclear molecular 
gas, by considering the shapes of galaxy surface brightness profile, signal-to-noise ratios (S/N) of position-velocity 
diagram (PVD) and systematic errors due to spatial and velocity structure of the molecular gas. 
We argue that for fixed 
galaxy stellar mass and SMBH mass, the spatial and velocity scale that needs to be resolved  
increases and decreases respectively with decreasing \sersic\ index of the galaxy surface brightness profile. 
We validate our arguments using simulated PVDs for varying beam size and velocity channel width.
Furthermore, we consider the systematic effects to the inference of the SMBH mass by 
simulating PVDs including spatial and velocity structure of the 
molecular gas, which demonstrates that their impacts are not significant for the PVD with good S/N unless 
the spatial and velocity scale associated with the systematic effects are comparable to or larger than
the angular resolution and velocity channel width of the PVD from pure circular motion.
Also, we caution that a bias in galaxy surface brightness profile owing to the poor resolution of 
galaxy photometric image can largely bias the SMBH mass by an order of magnitude. 
This study shows the promise and the limit of ALMA observation to measure the SMBH mass using molecular 
gas kinematics and provides a useful technical justification for ALMA proposal with science goal of measuring 
SMBH mass.
\end{abstract}

\begin{keywords}
 black hole physics -- methods: observational -- ISM: kinematics and dynamics -- galaxies: nuclei
\end{keywords}

\section{Introduction}\label{introduction}
Supermassive black holes (SMBH) in galactic centres influence the formation and evolution of the host 
galaxies \citep{richstone_etal_1998,kormendy_and_ho_2013} as suggested by well-known empirical relations between 
the SMBH mass and host galaxy properties \citep[e.g.,][]{magorrian_etal_1998,ferrarese_and_merritt_2000,gebhardt_etal_2000,tremaine_etal_2002,marconi_and_hunt_2003,haring_and_rix_2004,gultekin_etal_2009}.
Therefore understanding the formation and evolution of the SMBH is an integral part of the current galaxy formation 
theory \citep[][]{silk_and_mamon_2012}. In observation, a first step toward the understanding of the SMBH is a demographic 
study for wide range of black hole masses, however it is far from complete: the number of SMBHs with dynamical mass measurement
is currently 70-80 primarily due to the difficulty of achieving the required depth and resolution \citep[][]{shankar_etal_2016} and 
the detection of SMBHs with low \citep[e.g.,][for \mbh$<10^{6}$\msun]{brok_etal_2015,seth_etal_2010} and high \citep[e.g.,][for \mbh$>10^{10}$\msun]{mcconnell_etal_2011} 
dynamical mass has been rarely reported.

Measuring the dynamical mass of SMBH for wide range of black hole mass and host galaxy type is required to make a robust inference of the co-evolution between SMBH and 
host galaxy and also important to calibrate other empirical methods (e.g., X-ray luminosity) for indirect mass measurement in order to 
increase the sample size of SMBHs. However, dynamical measurement of the SMBH mass is challenging 
due to the requirement of high angular and velocity resolution \citep[][]{ferrarese_and_ford_2005}. Except for the rare population of 
galaxies with nuclear maser \citep[e.g.,][]{miyoshi_etal_1995}, stellar 
kinematics \citep[mostly in early-type galaxies; e.g.,][]{kormendy_1988,bower_etal_2001,mcconnell_etal_2011} and ionised gas 
kinematics \citep[in spiral and some early-type galaxies; e.g.,][]{ferrarese_etal_1996,sarzi_etal_2001,barth_etal_2001}  
have been used to measure the dynamical mass. As a result, most of the SMBH mass measurements have been obtained for a sample of 
early-type galaxies and spiral galaxies with prominent bulge because they have a well developed dynamically 
relaxed stellar system and powerful energy source ionising the gas.

For comprehensive understanding of the connection between the SMBH and its host galaxy, it is important to expand the measurement of 
SMBH mass to the regime of small bulgeless galaxies hosting lower mass SMBH. A recent study based on a large compilation of SMBHs shows that the empirical 
relation between SMBH mass and stellar velocity dispersion known as $M-\sigma$ relation \citep[e.g.,][]{tremaine_etal_2002} has a different normalization 
for early- and late-type galaxies \citep[][]{mcconnell_and_ma_2013}. In addition, there are strong evidences of the low luminosity AGN in bulgeless galaxies 
or dwarf galaxies hosting low mass SMBH \citep[e.g.,][]{filippenko_and_ho_2003,reines_etal_2011,mcalpine_etal_2011,araya_salvo_etal_2012} whose 
inferred masses are largely uncertain without accurate dynamical mass measurement. Furthermore, a large number of galaxies presumably hosting dust obscured 
AGN \citep[e.g.,][]{mateos_etal_2013,satyapal_etal_2014} has been revealed by red mid-IR colour \citep[][]{stern_etal_2012}. For those low mass or dust obscured galaxies, 
it may be even more difficult to apply the current methods using stellar and ionised gas kinematics if they do not have a well-developed 
bulge for reliable radial velocity dispersion measurement or have dust obscuring the ionised gas and also affecting the galaxy image analysis for mass
modeling. 

Recently the high angular and velocity resolution radio interferometry enables to make a use of the observation of molecular gas kinematics for measuring the dynamical 
mass of SMBH \citep[e.g.,][]{davis_etal_2013,onishi_etal_2015,barth_etal_2016}. Molecular gas as a kinematic tracer of the rotation velocity in galactic centre 
has advantages; in principle it is possible in any galaxy type with associated molecular gas if it exists, and the high angular resolutions routinely achievable 
by new (sub)-millimetre interferometry (e.g. the Atacama Large Millimetre/sub-millimetre Array; ALMA) mean it may be possible to probe larger volumes of 
the universe than ever before, enabling a complete mass limited census of SMBH \citep[][]{davis_2014}. 

With increasing importance and potential of the molecular gas kinematics for SMBH mass measurement, a figure of merit of the molecular gas kinematics has been 
discussed in \citet{davis_2014} which shows that the difference of rotation velocity with and without SMBH can be detected by the observation of molecular 
gas kinematics outside the black hole sphere of influence using the simple argument based on an assumed galaxy rotation velocity at desired angular 
resolution of the interferometry and that the required angular resolution  becomes larger with 
decreasing concentration (measured by logarithmic slope within 0.1\arcs\ radius) of the galaxy surface brightness profile.

Although some useful formulae and discussions using galaxy samples with velocity measurement from 
$\mbox{ATLAS}^{\mbox{\tiny 3D}}$ survey \citep[][]{krajnovic_etal_2013} have been provided in \cite{davis_2014}, their arguments are based on a single 
molecular gas parcel rotating around SMBH with assumed rotation velocity (i.e., 100 km/s) at the desired angular radius, without considering observational effect (e.g. beam smearing), 
systematic motions (e.g., non-circular motion) and spatial distribution of molecular gas. 
In addition, unlike $\mbox{ATLAS}^{\mbox{\tiny 3D}}$ survey, detail kinematic information is not available for most of galaxies, which is inconvenient for adopting simple 
relation in \cite{davis_2014} to determine the relevant physical scale for a given galaxy to resolve the gas kinematics for the SMBH mass measurement. 
It is useful to generalise and expand the work in \cite{davis_2014} to the wide range of observing and galaxy parameter 
space to provide a convenient and useful guideline for estimating the angular and velocity resolution of the proposed interferometry observation without relying on 
a prior knowledge of galaxy rotation velocity.

Indeed, molecular gas kinematics in galactic centre is complicated. It is governed by the SMBH and galaxy stellar mass distribution determined by galaxy surface 
brightness profile shape that varies with different galaxy types. Also signal-to-noise ratio (S/N) of the gas density 
profile and the profile shape, and other complicated velocity structures including inflow/outflow, random velocity motion and disc warp may contaminate the observed 
molecular gas kinematics and affect our inference of the SMBH mass. We consider these effects by simulating the position-velocity diagrams (hereafter PVD) 
for a range of observing parameters and properties of the galaxy and the circum nuclear molecular gas. 
In particular, the limits of the measurable SMBH mass given these systematic effects and the resolution of interferometry have not been discussed in detail 
using realistic simulations of the observed gas kinematics, which will be useful for the SMBH mass measurement using ALMA observations.
In addition, it will also serve as a useful guideline for the technical justification of ALMA proposal to observe molecular gas kinematics in galaxies for different 
galaxy surface brightness profiles and realistic observing conditions.

In this work, we study the parameter space of the angular and velocity resolution of radio interferometry for ranges of the galaxy types and the structure of circum nuclear
molecular gas in order to characterise how the information of molecular gas kinematics can be best utilised for the SMBH mass measurement. In Section \ref{sec:theory}, we provide a simple 
argument to derive relevant spatial and velocity resolution for given galaxy surface brightness profile shape and discuss the systematic effects due to
the spatial and velocity structure of the molecular gas and the effect of poor resolution of galaxy image to determine the surface brightness profile 
shape and its impact to the inference of SMBH mass. In Section \ref{sec:simulation}, we simulate observed PVDs and measure the rotation velocity to 
confirm our arguments and test the impact of systematic errors in the PVD analysis. In Section \ref{sec:discuss}, we discuss the spatial resolution and 
velocity channel width to measure the SMBH mass as a convenient guidance for ALMA proposal. In Section \ref{sec:summary} we summarise the results.  
In this work, we use the concordance LCDM cosmological model: $\Omega_m=0.3$, $\Omega_{\Lambda}=0.7$ and $H_0=100 h$ $\mbox{km s}^{-1}\mbox{Mpc}^{-1}$
where $h=0.7$.

\section{Impact of supermassive black hole to the rotation velocity}\label{sec:theory}
To measure the SMBH mass using gas kinematics, one needs to accurately measure the rotation velocity at the galactic centre and decompose into two 
components: one from the point mass due to the SMBH and the other from the extended matter distribution due to the galaxy stellar component. 
The technique is not formally valid in the case of dynamically hot, warped or inflowing/outflowing gas \citep[][]{davis_2014}. However like in \citet{davis_2014}, 
in this section, we assume that the motion of circum nuclear gas is purely circular and only governed by the SMBH and galaxy stellar mass, and the additional 
systematic effects will be discussed later.

For a flat gas disc sharing the same inclination ($i$) as the galaxy, rotation velocity of a gas parcel at radius $r$ is 
\begin{equation}\label{eq:vrot}
\vrot = \sqrt{\vgal^2 + \frac{G\mbh}{r}}~\mbox{sin}(i)
\end{equation}
where \vgal\ is a circular velocity of the galaxy due to the galaxy stellar mass distribution and 
\mbh\ is the SMBH mass in the galaxy centre. 
$\vgal=\sqrt{\frac{GM(<r)}{r}}$ where $M(<r)$ is the enclosed mass at radius $r$ and estimated by converting galaxy surface brightness distribution to 
stellar mass distribution, using a mass-to-light ratio 
($M/L$) at the observing band. Rotation velocity \vrot\ is then measured from the PVD. 

High-resolution PVD is required to accurately measure the rotation velocity and ALMA can easily achieve a sub-arcsec 
resolution which resolves the black hole sphere of influence (SOI) for most of large SMBHs with $M_{BH}>4\times10^{8}M_{\odot}$ and $i>30 \deg$ \citep[][]{davis_2014}, 
whose radius is commonly defined as 
\begin{equation}
\label{eq:rsoi}
\rsoi=\frac{G\mbh}{\sigma^2_{*}}
\end{equation}
where $\sigma_{*}$ is the stellar velocity dispersion. 
\citet{davis_2014} finds that the required angular resolution to detect the effect of black hole mass is approximately two times larger 
than \rsoi\, by calculating a maximum angular radius where the velocity difference with and without SMBH 
is detected by a factor of 5 times larger than the velocity channel width, using a typical galaxy rotation velocity (i.e., $\vgal=100$ km/s) measured by 
the $\mbox{ATLAS}^{\mbox{\tiny 3D}}$ survey at an assumed angular radius resolved by Combined Array for Research in Millimeter-wave Astronomy (CARMA).
However, as also mentioned in the introduction, estimating galaxy rotation velocity as a function of radius using more easily available photometric information is 
more useful and convenient to determine the angular resolution and velocity channel width for detecting SMBH in galaxies for a wide range of morphology
without prior knowledge of the galaxy kinematics.

In practice, observed rotation velocity \vobs\ is different from \vrot\ in equation(\ref{eq:vrot}) due to the uncertainties of position and 
velocity measurements and the complicated velocity structure in the gas. However, in this section, we assume $\vobs = \vrot$ by neglecting 
complicated velocity structure and simplifying the uncertainties in PVD by considering only interferometry beam size and velocity channel width and 
we consider other systematic effects by simulating PVDs in Section \ref{sec:simul_systematic}.

\subsection{Rotation Velocity}
We compute \vrot\ using \mbh\ and \vgal\ in equation(\ref{eq:vrot}). 
To compute \vgal, we adopt \sersic\ models for galaxy surface brightness profile to estimate 
the enclosed mass at a radius $r$. Surface brightness profile of \sersic\ model \citep{sersic1968} varies with a shape parameter $n$ and produces a profile 
continuously changing from disc to elliptical galaxies. It is also possible to extend \sersic\ model to incorporate a core-deficit of elliptical 
galaxies by using core-\sersic\ model \citep[e.g.,][]{graham_etal_2003}. Although in practice, the galaxy surface brightness profile is measured from 
the high resolution galaxy image with additional modeling of the details of the profile such as nuclear star cluster \citep[e.g.,][]{emsellem_etal_1994}, 
we use \sersic\ parametric model in this study because it is flexible to describe a wide 
range of galaxy surface brightness profiles very well and thus effective to show the impact of galaxy surface brightness profile shape on the 
SMBH mass measurement.

Surface brightness profile of \sersic\ model is
\begin{equation}
I(r) = I(0) \mbox{exp} [-b_n (r/r_{es})^{1/n}]
\label{sersic}
\end{equation}
with $I(0)$ being the central intensity, $r_{es}$ the half-light radius, and $n$ the shape parameter.

Surface brightness profile of core-\sersic\ model is
\begin{equation}
\footnotesize
I(r) = I_b [(\frac{r_b}{r})^{\gamma} u(r_b-r) + e^{b(r_b/r_e)^{1/n}-b(r/r_e)^{1/n}} u(r-r_b)]
\label{eq:core_sersic}
\end{equation} 
with being $u(r-a)$ the Heaviside step function, in case of sharp transition at $r_b$ \citep[][]{trujillo_etal_2004}.
Although this is different from the original core-\sersic\ model profile \citep[][]{graham_etal_2003}, replacing $b$ with $b_n$ and half-light 
radius $r_e$ with that of the outer \sersic\ profile $r_{es}$ gives a good approximation with much less error than 
the uncertainty in $r_e$ from the fitting process as long as $r_b \ll r_e$ \citep[][]{trujillo_etal_2004}.
So in this work, we use equation (\ref{eq:core_sersic}) for core-\sersic\ galaxy model.

To compute the enclosed mass within $r$, we integrate $I(r)$ up to $r$ being normalised by the total luminosity $\int^{\infty}_{0} I(r) dr$ and 
multiply the bolometric luminosity and mass-to-light ratio. We use a fiducial mass-to-light ratio, $M/L = 1.0$ 
and adjust an absolute magnitude parameter to obtain galaxy stellar mass that we want to control. Given galaxy stellar mass $M_{*}$ and \sersic\ index $n$, 
we assign the half-light radius of the galaxy using the relation in \citet{shen_etal_2003}: 
\begin{equation}
\footnotesize
\frac{r_{es}}{\mbox{kpc}} = \begin{cases} 0.1 \left(\frac{M_{*}}{M_{\odot}}\right)^{0.14} \left(1+\frac{M_{*}}{3.98\times10^{10}M_{\odot}}\right)^{0.25} &, \mbox{if $n<2.5$}\\ 
                            2.88\times10^{-6} \left(\frac{M_{*}}{M_{\odot}}\right)^{0.56} &, \mbox{if $n>2.5$} \end{cases}
\end{equation}
If galaxy stellar mass $M_{*}$ and \sersic\ index $n$ are given, the enclosed stellar mass $M_{*}(<r)$ is 
\begin{equation}
M_{*}(<r) = M_{*} \frac{\int^{r}_{0} I(r) dr}{\int^{\infty}_{0} I(r) dr} = M_{*}\frac{\gamma(2n,x)}{\Gamma(2n)}
\end{equation}
where $x=b_{n}(\frac{r}{r_{es}})^{1/n}$. For $n>0.36$, $b_{n}$ is approximated as follows \citep{ciotti_and_bertin_1999}.
\begin{equation}
b_n \approx 2n - \frac{1}{3} + \frac{4}{405n} + \frac{46}{25515n^2} + \frac{131}{1148175n^3} - \frac{2194697}{30690717750n^4} + O(n^{-5}).
\label{eq:b_approx}
\end{equation}

\begin{figure}
\centering
\epsfig{file=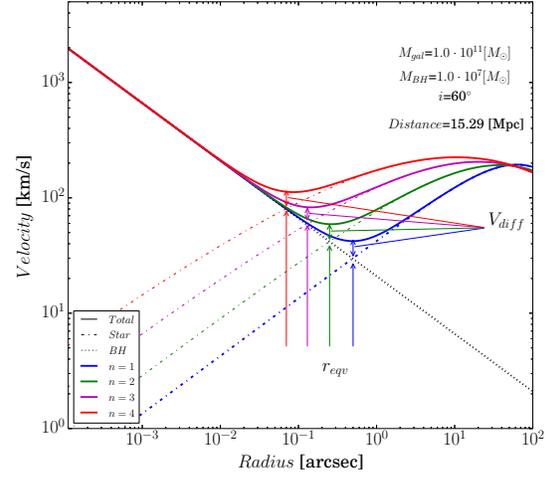,width=0.45\textwidth}
\caption{Rotation velocities of galaxies with $M_{*}=10^{11}$\msun\ containing central SMBH with $\mbh=10^7$\msun. Different colour indicates 
\sersic\ index of the galaxy and Keplerian rotation due to the SMBH is shown by the black dotted line. The radius, where the galaxy rotation and 
the SMBH rotation velocity are equal (\reqv) for each galaxy surface brightness profile, is indicated by the arrow with the same colour as the corresponding 
rotation curve. Velocity difference \vdiff\ between the total (galaxy+SMBH) and galaxy rotation velocity at \reqv\ is also shown by the arrow with the same colour scheme.
}
\label{fig:vrot}
\end{figure}

Figure \ref{fig:vrot} shows the rotation velocity of a galaxy with inclination $i=60$\deg\ and $M_{*}=10^{11}$\msun\ containing $10^7$\msun\ SMBH, 
for different surface brightness profile shapes characterised by \sersic\ index $n$. In $x$-axis, we use angular size at a given luminosity distance of the galaxy 
(roughly at a distance of Virgo cluster, $\approx 15$ Mpc) instead of physical size for easy comparison to the angular resolution 
of ALMA. Each colour in Figure \ref{fig:vrot} represents the \sersic\ index ranging from 1 to 4. Rotation velocity 
shown by the solid lines are decomposed into the rotation velocity due to the SMBH and the galaxy stellar mass distribution. Dotted lines and dot-dashed lines are 
the rotation velocity due to the SMBH mass and galaxy stellar mass respectively. We also annotate the angular radius \reqv\ where the rotation velocity due to the SMBH 
and due to the galaxy stellar mass distribution become equal (i.e., $M_{*}\frac{\gamma(2n,x)}{\Gamma(2n)}=M_{BH}$) and the velocity difference \vdiff\ at \reqv\ between 
the total rotation velocity and the rotation velocity due to the galaxy stellar mass. 
If the beam size is larger than \reqv\ or the velocity channel width is larger than \vdiff\ for the given beam size resolving \reqv, the velocity 
difference \vdiff\ is not 
resolved. So \reqv\ and \vdiff\ determines the required beam size and velocity channel width $\Delta V$ (defined as $\Delta V = \frac{1}{3}$\vdiff\ for detecting \vdiff\ at $3\sigma$) 
of the radio interferometry.

Figure \ref{fig:vrot} indicates that at fixed galaxy stellar mass, the impact of Keplerian motion due to the SMBH extends to a larger radius as galaxy surface brightness 
profile becomes less concentrated (i.e. smaller \sersic\ index). For example, 
the rotation velocity due to the SMBH at 0.5\arcs\ scale is 5 times smaller than that due to the galaxy stellar mass 
if the galaxy surface brightness profile follows the elliptical galaxy's ($n=4$ shown by red curve) while the rotation velocity due to the SMBH is same as the rotation 
velocity due to the galaxy stellar mass at the same 0.5\arcs\ angular scale if the surface brightness profile is exponential ($n=1$ shown by blue curve). 
This implies that the effect of SMBH for fixed mass \mbh\ can be detected by larger angular resolution as the galaxy surface brightness profile becomes less 
concentrated which is consistent with the finding in \citet{davis_2014}. If lowering the SMBH mass, the dotted line moves downwards and thus, for fixed 
angular resolution, the \reqv\ of larger \sersic\ index galaxy starts to be unresolved first. This implies that the effect of SMBH 
for the given galaxy stellar mass and angular resolution, can be more easily detected for the lower black hole mass if the galaxy surface brightness profile becomes less concentrated. 
This will be discussed further in the following section.

\subsection{Spatial and Velocity Resolution}\label{sec:theory_resolution}
The radius $r_{\mbox{\tiny max}}$ defined in \citet{davis_2014} where the Keplerian motion due to SMBH shows a statistically significant deviation from 
the rotation velocity due to the galaxy stellar mass distribution, is larger than the definition of the black hole sphere of influence \rsoi\ (equation(\ref{eq:rsoi})).
Using galaxy samples in the $\mbox{ATLAS}^{\mbox{\tiny 3D}}$, \citet{davis_2014} finds that
$r_{\mbox{\tiny max}} = 1.92 \rsoi$ at $5\sigma$ statistical significance level.

\begin{figure*}
\subfigure[$M_{*}=10^{10}$\msun]{\epsfig{file=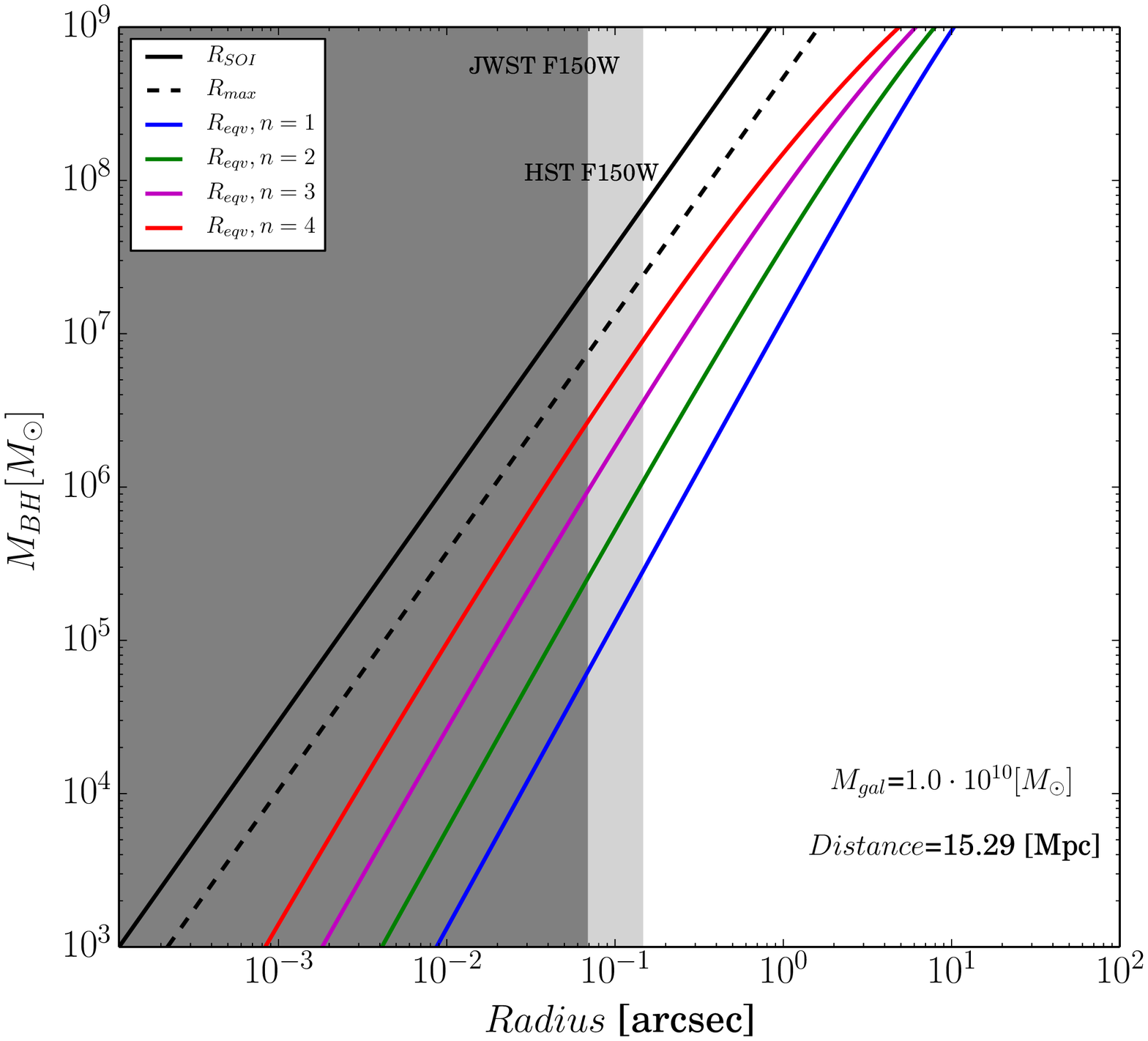,width=0.45\textwidth}}
\subfigure[$M_{*}=10^{11}$\msun]{\epsfig{file=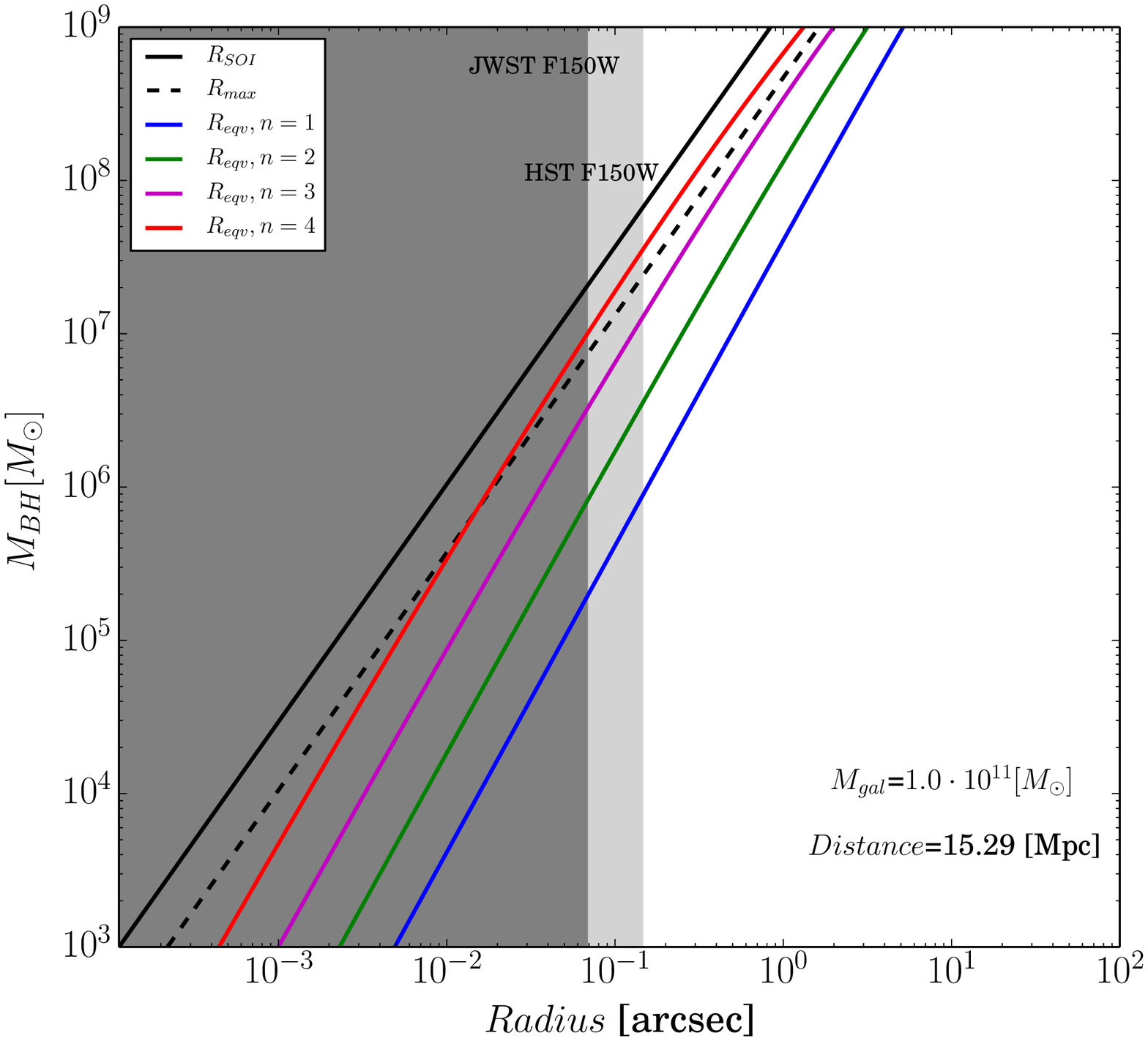,width=0.45\textwidth}}
\caption{Relation between the SMBH mass and the three different angular scales: \rsoi, $r_{\mbox{\tiny max}}$ and \reqv\ for a given galaxy stellar mass $M_{*}$. 
For \rsoi, we combine equation(\ref{eq:rsoi}) and the recent compilation of $M-\sigma$ relation \citep[][]{mcconnell_and_ma_2013}. 
For $r_{\mbox{\tiny max}}$, we use the best fit relation between \rsoi\ and $r_{\mbox{\tiny max}}$ ($r_{\mbox{\tiny max}} = 1.92 \rsoi$) 
in \citet{davis_2014}. For \reqv\, different colours indicate galaxy surface brightness profiles characterised by different \sersic\ index $n$.}
\label{fig:radius}
\end{figure*}

For given \mbh\, we compare three different radii, \rsoi, $r_{\mbox{\tiny max}}$ and \reqv\ for the same stellar mass galaxies with 
different surface brightness profiles. Black hole \rsoi\ is computed by combining equation (\ref{eq:rsoi}) and the relation between 
the SMBH mass and the stellar velocity dispersion in \citet[][]{mcconnell_and_ma_2013}. Figure \ref{fig:radius} shows these radii (i.e., angular sizes for 
given galaxy distance) for the range of SMBH masses, for a galaxy with $M_{*}=10^{10}$\msun\ (Figure \ref{fig:radius}(a)) and 
$M_{*}=10^{11}$\msun\ (Figure \ref{fig:radius}(b)). Solid black line is \rsoi, dashed black line is $r_{\mbox{\tiny max}}$ and coloured solid lines 
are \reqv\ for different \sersic\ indices ($n=1,2,3$ and $4$). The following relation provides a convenient way to determine \reqv\ for a given ratio 
between the SMBH mass and the galaxy stellar mass
\begin{equation}
\frac{M_{BH}}{M_{*}} = \frac{\gamma(2n,x)}{\Gamma(2n)}.
\end{equation}
where $x=b_{n}(\frac{r}{r_{es}})^{1/n}$ and $b_{n}$ is estimated by equation(\ref{eq:b_approx}).
The region smaller than the FWHM of HST WFC3 (0.15\arcsec\ \footnote{\small \url{http://www.stsci.edu/hst/wfc3/documents/handbooks/currentIHB/c07_ir07.html}}) 
and JWST NIRCam (0.068\arcsec\ \footnote{\small \url{http://www.stsci.edu/jwst/instruments/nircam/PSFs/}}) 
in $J$-band are shown by light and dark grey area to show the limit of angular resolution beyond which the galaxy surface brightness profile can be accurately 
measured using high resolution near-infrared image. Although the current state-of-the-art resolution image for SMBH mass measurement 
using HST/ACS $I$-band image has slightly smaller FWHM ($\approx 0.1$\arcs), the important question is whether the very inner region 
where the rotation velocity is dominated by the SMBH, is resolved both by the photometric PSF and the interferometre beam or not, and therefore the actual 
size of FWHM is not relevant for the discussion in this section.

\begin{figure*}
\subfigure[$M_{*}=10^{10}$\msun]{\epsfig{file=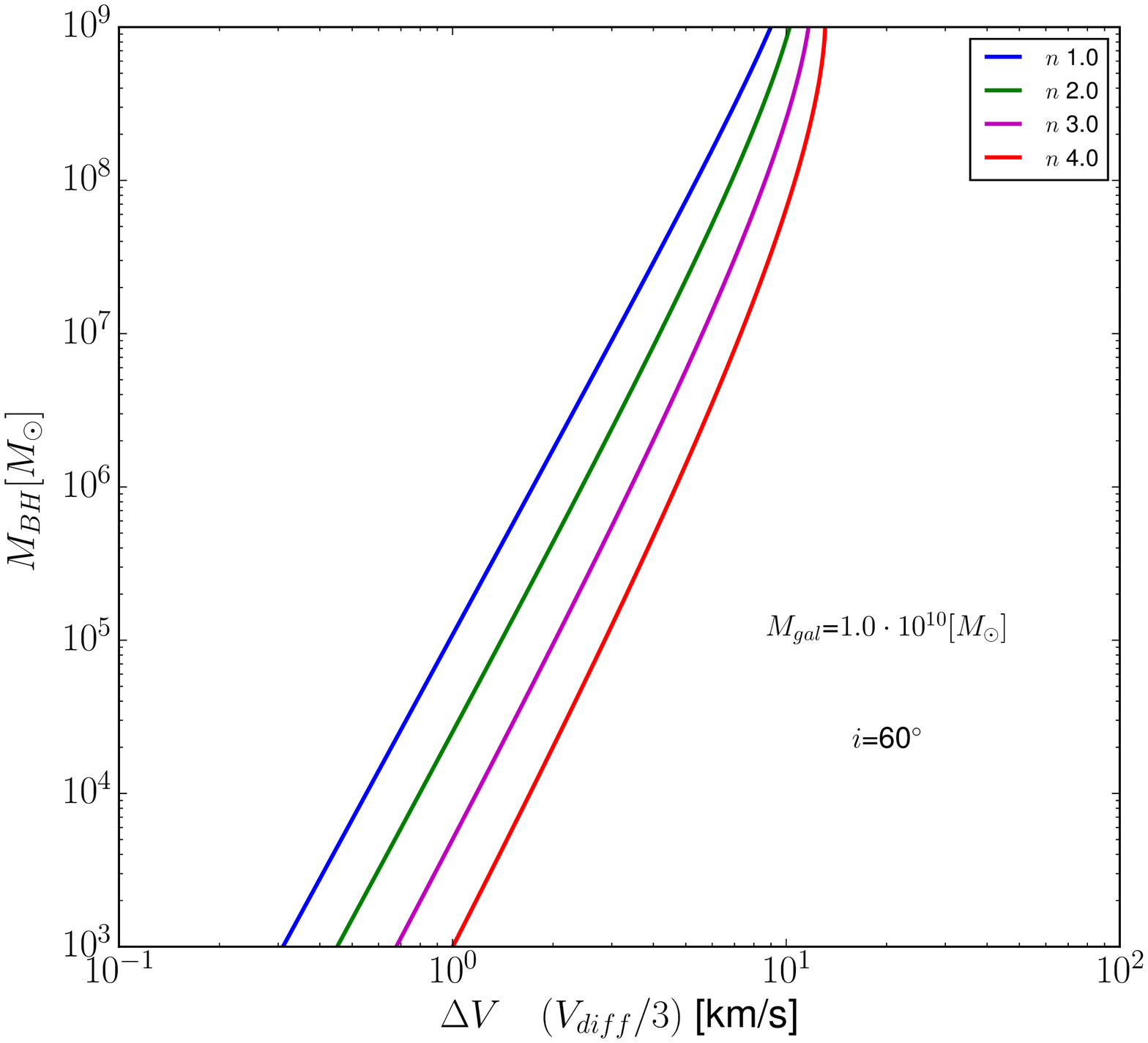,width=0.45\textwidth}}
\subfigure[$M_{*}=10^{11}$\msun]{\epsfig{file=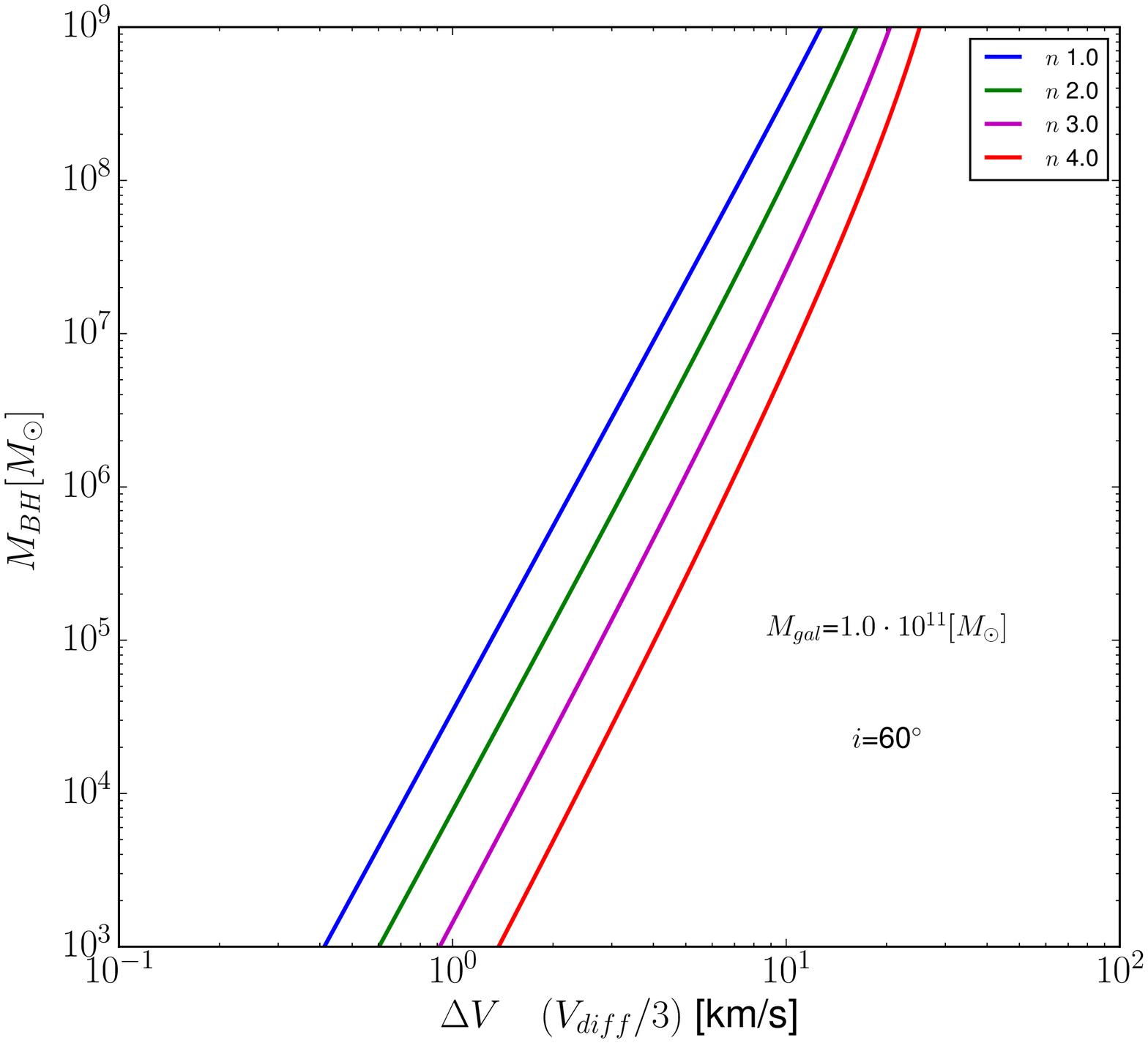,width=0.45\textwidth}}
\caption{Same as Figure \ref{fig:radius}, but for the velocity channel width $\Delta V$.} 
\label{fig:velocity}
\end{figure*}

In Figure \ref{fig:radius}, \reqv\ for all galaxy profiles is larger than \rsoi\ and all colour solid lines except for the galaxy
with $n=4$, $M_{*}=10^{11}$\msun\ and \mbh$>10^{6}$\msun\, are larger than $r_{\mbox{\tiny max}}$ by 
a factor of few for galaxy with $n=4$ and by more than an order of magnitude for galaxy with $n=1$, and this deviation becomes larger for 
the lower SMBH mass. The best agreement between $r_{\mbox{\tiny max}}$ and \reqv\
is seen for the galaxy with $M_{*}=10^{11}$\msun\ and $n=3\sim4$ for \mbh $>10^{7}$\msun\ which is similar to the galaxies
in \citet{davis_2014} used for deriving $r_{\mbox{\tiny max}}$.
However, $r_{\mbox{\tiny max}}$ was determined for $5\sigma$ statistical significance \citep[][]{davis_2014} and thus $r_{\mbox{\tiny max}}$ can be larger by 
linear factor of the inverse of significance level in case of small velocity error (i.e., channel width) compared with the rotation velocity (see equation(9) 
in \citet{davis_2014}), if lowering the significance level.
As the galaxy surface brightness profile becomes less concentrated, the required beam size for detecting SMBH becomes larger at fixed \mbh\ and the detectable SMBH mass
becomes smaller at fixed beam size. 
This is an advantage for detecting lower mass black hole residing in a small bulgeless galaxy.

In the same manner, we also show the velocity channel width $\Delta V$ ($=\frac{1}{3}$\vdiff) for different SMBH mass, for a galaxy 
with $M_{*}=10^{10}$\msun\ (Figure \ref{fig:velocity}(a)) and 
$M_{*}=10^{11}$\msun\ (Figure \ref{fig:velocity}(b)). Observed rotation velocity depends on galaxy inclination ($i$) and we assume $i=60$\deg\ 
in Figure \ref{fig:velocity}. 
The required velocity channel width for detecting SMBH becomes smaller at fixed \mbh\ and the detectable SMBH mass becomes 
larger at fixed velocity channel width, as galaxy surface brightness profile is less concentrated. This is a disadvantage for the small bulgeless galaxy. 
However the range of variation of the velocity channel width 
for different galaxy \sersic\ index is within a factor of few that is much smaller than the variation of angular resolution (i.e., order of magnitude) 
and ALMA has sufficient velocity resolution ($\le 1$ km/s) to cover the range of this channel width in Figure \ref{fig:velocity}. So the practical benefit is 
the spatial resolution. However we note that small velocity channel width does not improve the significance of the detection if velocity uncertainty including 
systematic and random motion is larger than the velocity channel width as also noted by \citet{davis_2014}.

\subsection{Effects of Spatial and Velocity Structure of Molecular Gas}\label{sec:systematic}
The above argument regarding the spatial and velocity resolution is for the rotating molecular gas
following the gravitational potential from the central SMBH and the galaxy stellar mass. In a realistic situation, we need to understand how significant 
the effects of spatial and velocity structure of molecular gas are to the observed rotation velocity. It is not trivial to quantify how these 
systematic effects influence the detection of SMBH mass using analytic arguments. However we will quantitatively discuss their effects here and 
show their significance using simulations of PVD in Section \ref{sec:simul_systematic}. 

First, we consider the systematic effect due to the spatial distribution of the molecular gas. Recent ALMA observations have revealed detail 
structure of molecular gas at a 10-100 pc scale. Most of them show disc like morphology with and without hole at the 
centre \citep[e.g.][]{izumi_etal_2013,combes_etal_2014,onishi_etal_2015,xu_etal_2015}. If the gas distribution is continuous and smooth, 
the central region of PVD is sampled well and the impact of SMBH to the rotation velocity, if significant,  can be detected. However, if the molecular gas 
distribution has a hole at the centre and the size of the hole is larger than \reqv, there will be no velocity tracers at the centre and the 
PVD analysis will not probe the region with the largest statistical significance in velocity space for detecting SMBH. 
In addition, for the same background noise and the same amount of total flux 
from the molecular gas, if the gas profile shape in central region is homogeneous across many synthesised beams, a peak flux at the 
centre where the impact of SMBH mass is observed the most becomes relatively low and the signature of SMBH will be less significant than that 
for the molecular gas with more concentrated profile, which implies that the gas disc
profile shape might affect the uncertainty of rotation velocity measurement and the resulting SMBH mass.
Since the proposed sensitivity requirement of ALMA is defined per beam, the flux variation due to the spatial geometry of gas disc within ALMA beam
may impact modeling gas kinematics given that the geometry of circum nuclear molecular gas is not known a priori.  
Also if the geometry of molecular gas disc becomes closer to the face-on, the projected line of sight velocity decreases and smaller velocity 
channel width is required to resolve the velocity difference between the pure galaxy rotation and the galaxy-SMBH rotation. In contrast, 
if the galaxy is edge on and the beam size is too large to resolve the kinematic structure along the minor axis, the velocity profile is smearing due to 
the velocity components migrated from the minor axis \citep{barth_etal_2016}.
Another potential problem might be a warp in the gas disc. Warp is prominent in HI distribution \citep[e.g.,][]{garcia_ruiz_etal_2002} but also seen in 
the nuclear molecular gas disc \citep{sofue_2001}. It introduces a bias in the inclination correction and underestimates the rotation velocity 
depending on the location of the major axis \citep{vergani_etal_2003}, which can be mitigated by modeling full 3D data cube.  

Second, we consider the systematic effect due to the velocity structure of molecular gas. In real, the velocity structure in molecular gas 
is more complicated than the pure circular motion. This velocity structure which we call non-circular motion in this work includes the gas inflow 
due to angular momentum loss \citep[e.g.][]{combes_etal_2014} and the gas outflow due to strong stellar and AGN feedback \citep[e.g.][]{garcia_burillo_etal_2014}, 
both of which have been observed by ALMA. For a set of molecular gas parcels within radio beam along the line of sight at a certain projected distance, this bulk 
motion (inflow or outflow) contributes to the line of sight velocity of the gas parcels positively or negatively depending on where the gas 
parcels are located. This will cause a velocity spread in PVD and at a given radius, make the PVD thicker along the velocity axis. As a result, 
the signature of Keplerian rotation owing to the SMBH will not be significant.
Although the driving mechanisms are different, both inflow and outflow have the same effect to the PVDs with the only difference in the velocity sign. 
In addition, there is a random motion in the circum nuclear molecular gas which also introduces a dispersion in 
the velocity axis in PVD.

\subsection{Galaxy Surface Brightness Profile Bias}\label{sec:theory_surfprofile}
To measure the SMBH mass using rotation velocity of the circum nuclear molecular gas, the surface brightness profile at 
the galactic core needs to be accurately determined \citep[e.g.,][]{ferrarese_and_ford_2005}, which requires high angular resolution to resolve \reqv. 
However, the angular resolution of galaxy image might not be sufficient for distant galaxies to accurately characterise the core 
luminosity profile while radio interferometry still has a sufficient resolution to resolve \reqv. 

Detail analysis of the high resolution HST images of early-type galaxies reveals that the galaxy core surface brightness profiles
deviate from the \sersic\ model fit and have a flat core or a power law slope \citep[e.g.,][]{trujillo_etal_2004} 
although \sersic\ model is a good fit to the overall profile in general \citep[e.g.,][]{canon_etal_1993,graham_2001}. 
If the galaxy core profile is not resolved, the stellar mass distribution at the galactic centre is systematically biased and 
as a result, the derived rotation velocity is also biased. This bias introduces a large systematic error in the 
SMBH mass measurement. In Section \ref{sec:simul_surfbias}, we illustrate how the bias in the galaxy surface brightness profile can impact the 
SMBH mass measurement.

\begin{figure*}
\subfigure[nearly flat core]{\epsfig{file=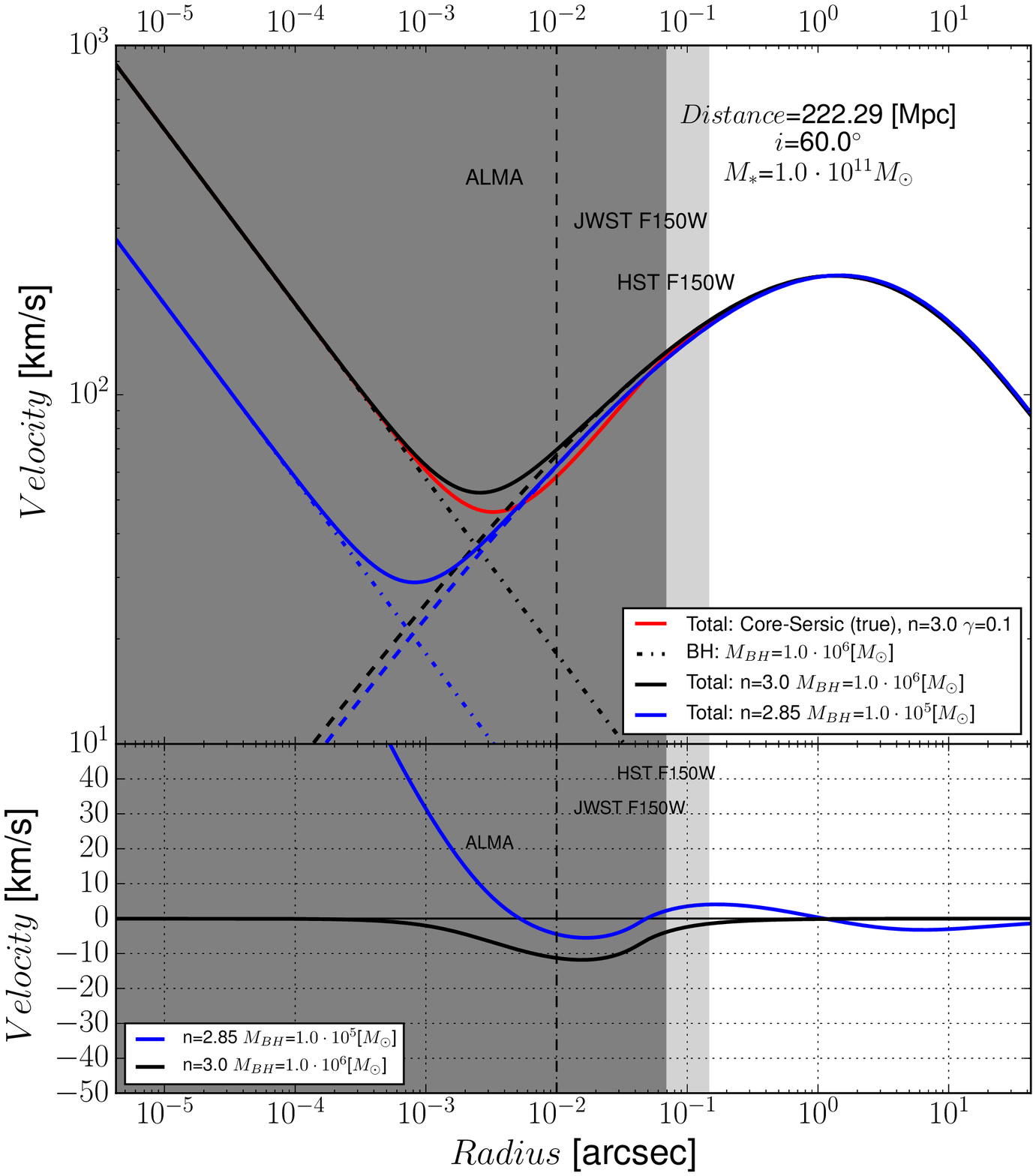,width=0.45\textwidth}}
\subfigure[power law core]{\epsfig{file=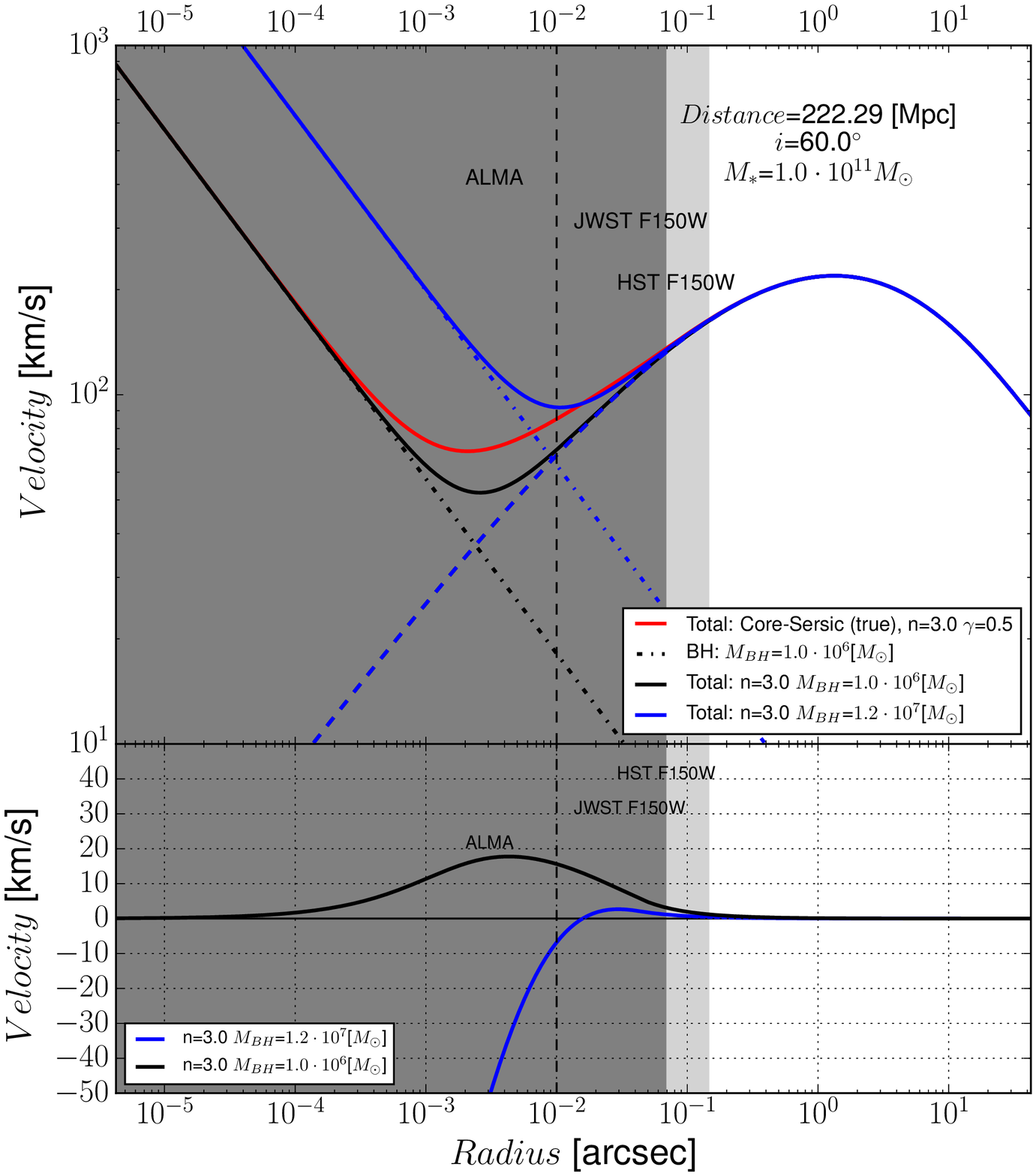,width=0.45\textwidth}}
\caption{Rotation velocity of a galaxy with $M_{*}=10^{11}$\msun\ containing $10^{6}$\msun\ SMBH. 
The surface brightness profile of the galaxy is the core-\sersic\ profile with $r_b=0.01 r_{e}$ and $n=3$ in equation (\ref{eq:core_sersic}).
Panel (a): the galaxy has a nearly flat core ($\gamma=0.1$). 
The red line shows the total rotation velocity of the true galaxy model including SMBH. The black line shows the rotation velocity of the same 
galaxy with the same \sersic\ index and the same SMBH mass but without the core. The blue line is the rotation velocity of the \sersic\ (without core) 
model galaxy with 5\% smaller \sersic\ index (i.e., $n=2.85$) and 3\% larger mass-to-light ratio (i.e., $M/L=1.03$) than the galaxy 
shown by the black line, however with a 10 times smaller SMBH mass ($10^{5}$\msun) than the true value ($10^{6}$\msun).
Panel (b): the galaxy has a core with power law profile ($\gamma=0.5$). The red and black line indicates the same case in Panel (a).
Blue line is the rotation velocity of the \sersic\ (without core) model galaxy with the same \sersic\ index (i.e., $n=3.0$) and mass-to-light
ratio (i.e., $M/L=1.0$) as the galaxy shown by the black line, however with a 12 times more massive SMBH mass ($1.2\times 10^{7}$\msun) than the true 
value ($10^{6}$\msun).}
\label{fig:galshape}
\end{figure*}

In Figure \ref{fig:galshape}, we show the rotation velocity of a galaxy with SMBH using core-\sersic\ galaxy surface brightness profile with core 
being unresolved by putting this galaxy at large distance such that at a scale of the HST and JWST seeing, the galaxy rotation velocity  
is largely dominated by the galaxy stellar mass distribution (not by the SMBH) and the difference between the rotation velocities due to 
slightly different surface brightness profiles of the galaxy can not be discriminated.
The core-\sersic\ galaxy has $M_{*}=10^{11}$\msun\ and $\mbh=10^6$\msun, and follows the surface brightness 
profile with \sersic\ index $n=3$ and core radius $r_b=0.01 r_e$ ($r_b \ll r_e$ in equation(\ref{eq:core_sersic})). 
We consider two different core profiles: nearly flat 
core using $\gamma=0.1$ and power law core using $\gamma=0.5$, (see equation (\ref{eq:core_sersic})), representing the typical values for elliptical 
galaxies seen in \citet{trujillo_etal_2004}.

Figure \ref{fig:galshape}(a) and \ref{fig:galshape}(b) shows the rotation velocities due to different galaxy surface brightness profiles and 
SMBH masses and their residuals from the true rotation velocity.
As in Figure \ref{fig:radius}, the region smaller than the PSF FWHM for HST WFC3 and JWST NIRCam is shown by light and dark grey area respectively.
The highest angular resolution of ALMA ranges from 0.006\arcs\ at 675 GHz to 0.037\arcs\ at 110 GHz. We assume a 0.01\arcs\ beam size as a typical 
ALMA resolution for the most extended array configuration and show it by vertical dashed line
in Figure \ref{fig:galshape}.

In the upper panel of Figure \ref{fig:galshape}(a) and \ref{fig:galshape}(b), we show three different rotation velocities. 
First, the red line shows the true rotation velocity of the core-\sersic\ galaxy with nearly flat core ($\gamma=0.1$, Figure \ref{fig:galshape}(a)) 
and power law core ($\gamma=0.5$, Figure \ref{fig:galshape}(b)). Second, the black line shows the rotation velocity of the galaxy 
that follows the surface brightness profile with the same $n$ as the core-\sersic\ galaxy and the same SMBH mass but without the core. 
Last, the blue line shows the galaxy rotation velocity that shows a better agreement with the true rotation velocity (red) for the scale larger than the 
ALMA angular resolution, by using very small (or no) adjustment of the \sersic\ index and mass-to-light ratio while using an order of magnitude different 
SMBH mass. In detail, the blue line in Figure \ref{fig:galshape}(a) shows the case where the galaxy has 5\% smaller \sersic\ index ($n=2.85$) and 3\% larger 
mass-to-light ratio ($M/L=1.03$) than the true values but has a 10 times smaller SMBH mass ($10^5$\msun). The blue line in Figure \ref{fig:galshape}(b) 
shows the case where the galaxy has the same \sersic\ index ($n=3.0$) and mass-to-light ratio ($M/L=1.0$) as the model galaxy shown by the black line, 
but has a 12 times more massive SMBH ($1.2\times 10^7$\msun).

In the lower panel of Figure \ref{fig:galshape}(a) and \ref{fig:galshape}(b), we show the residual velocities from the true rotation velocity (red curves in the upper panels),
for the model galaxies in the upper panels using the same colour.
For both cases of the biased galaxy surface brightness profile with largely biased SMBH mass that shows a better agreement with the true rotation velocity, 
the velocity difference at angular scale larger than the assumed ALMA resolution ($0.01$\arcs) is smaller than the typical velocity channel width adopted for measuring 
SMBH mass using molecular gas kinematics (e.g., 10 km/s in \citet{davis_etal_2013}) and their velocity residual at the spatial scale resolved by HST and ALMA
is smaller than the velocity residuals of the galaxy with the same global \sersic\ index as the core-\sersic\ galaxy (blue lines).

The rotation velocity of the true core-\sersic\ galaxy (red line) and the same \sersic\ galaxy profile without core (black line) are 
indistinguishable at the scale of HST and JWST resolution ($\approx 0.1$\arcsec) with a 1-2 km/s velocity 
difference as seen in the lower panel of Figure \ref{fig:galshape}(a) and \ref{fig:galshape}(b) as shown by black line. 
Therefore if the core of the galaxy surface brightness profile is not resolved, 
the \sersic\ galaxy profiles with the unresolved core (black lines in the upper panels of Figure \ref{fig:galshape}) may be the best determination of the galaxy stellar mass distribution. However the 
rotation velocity of this biased profile is not a good match to the true rotation velocity of the core-\sersic\ galaxy at angular scale similar to the ALMA 
beam (0.01\arcsec) and the blue line with slightly adjusted galaxy parameters ($n$ and M/L) but with order of magnitude different BH mass is a better match to the true 
rotation curve at the angular resolution of ALMA. This results in a biased SMBH mass in model fitting as demonstrated in Section \ref{sec:simul_surfbias}.

Lowering ALMA resolution less than 0.01\arcs\ in this example will remove the bias, however achieving 0.01\arcs\ resolution is 
very difficult (if not feasible) in practice for many of galaxies because of a long integration time. 
So for nearby galaxies, ALMA has sufficient spatial resolution to resolve \reqv\ and thus even though the galaxy rotation velocity is 
biased due to the unresolved galaxy core in the photometric image, ALMA can distinguish the difference between slightly different galaxy rotation velocities. 
However, for distant galaxies, the ALMA spatial resolution becomes poor and cannot differentiate the velocities due to the biased galaxy surface brightness profiles.
This implies that at the current best ALMA resolution, it is difficult to break the degeneracy between different rotation velocities due to the biased core surface 
brightness profiles not resolved by HST or JWST, if the galaxy is at a distance similar or larger than $\approx 200$ Mpc.
In principle, to avoid a bias of the SMBH mass owing to the biased galaxy surface brightness profile, the resolution of galaxy image should be similar 
to the ALMA beam size and ideally, the both needs to be comparable to \reqv\ assuming that a sufficient velocity resolution is achieved by ALMA.

\section{Detecting the impact of SMBH}\label{sec:simulation}
We have discussed the required angular and velocity resolution to measure the impact of SMBH based on the rotation velocity without considering the observational 
measurement processes. We argue that, if the galaxy stellar mass, surface brightness profile shape and inclination are known, there 
is a required spatial resolution corresponding to \reqv\ and velocity channel width corresponding to $\frac{1}{3}\vdiff$ to detect the SMBH mass that one aims 
to measure. In this section, we confirm this argument 
and show how the angular resolution and velocity channel width affect the SMBH mass detection by simulating PVDs with model rotation velocity including the observational 
measurement processes. Then we incorporate the systematic effects in the PVD simulation to test how significant
their impacts are to the measurement of rotation velocity for several representative cases.

\subsection{Simulations of Position-Velocity Diagram}\label{sec:simul_pvd}
We use \kinms\ \citep[][]{kinms_2013} to simulate PVD. \kinms\footnote{\url{https://github.com/TimothyADavis/KinMS}} is publicly 
available IDL code to simulate gas kinematics by incorporating the observational effects: beam size and velocity channel width and the properties of 
molecular gas: user defined gas density profile, random velocity dispersion, bulk motions (inflow/outflow), warp and blobs in the molecular gas. Although 
it has been originally developed to model the gas kinematics in elliptical galaxies in ATLAS$^{\mbox{\tiny 3D}}$ survey \citep[][]{kinms_2013}, it is 
also directly applicable to the analysis of molecular gas kinematics for the SMBH mass measurement \citep[][]{davis_etal_2013,onishi_etal_2015}. For detail information of 
the code, we refer to \citet[][]{kinms_2013}.

In PVD simulation, we generate the noise as follows.
We sample the spatial distribution of molecular gas using a finite number of random samples and make PVD. After sampling the distribution 
100 times using different random numbers, we estimate the standard deviation of the ensemble PVDs from the `noise-less' 
PVD generated from the density distribution sampled by a large number ($10^6$) of  random samples. Then the signal-to-noise ratio (S/N) of PVD is defined by the peak signal in the PVD 
and the standard deviation of the ensemble PVDs. We vary the number of random samples ($N_{samp}$) to obtain a range of S/N in this work. However S/N does not 
only depend on the number of random samples but also on the beam size, velocity channel width and pixel resolution; the larger the values, the larger the S/N
for the same number of random samples. We find that for most of our experiments in this work, approximately 10000 random samples gives a good S/N ranging from 60 to 100.
However, we note that the real noise in PVD is originated by the spatially correlated interferometer noise and thus depends on the spatial scale of the observation. This 
requires a more complicated simulation of the PVD using realistic array configuration, which is beyond the scope of the current work.   

\begin{figure*}
\centering
\epsfig{file=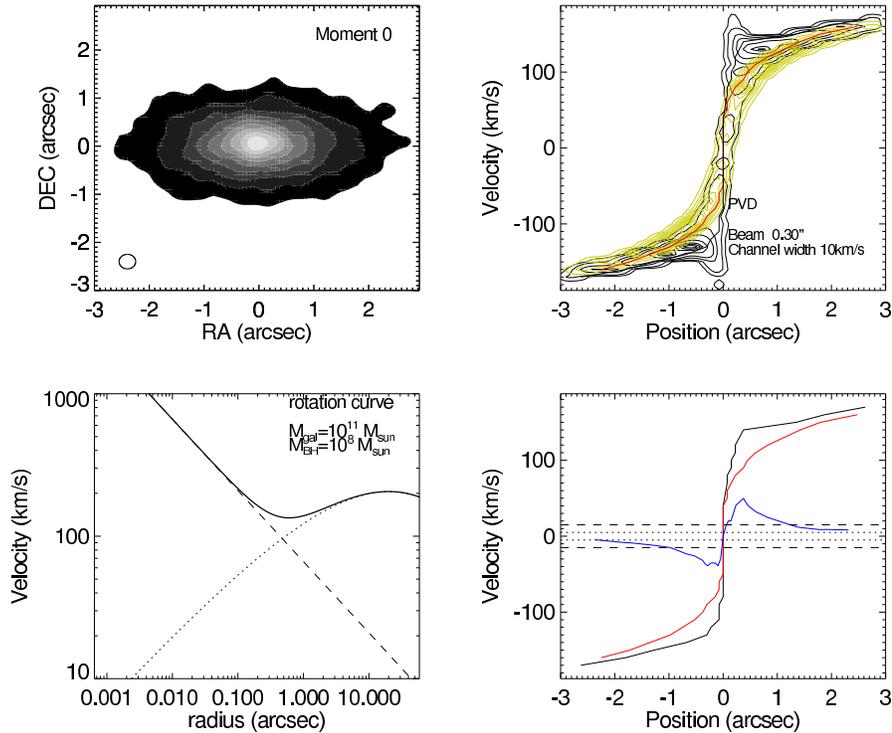,width=0.7\textwidth}
\caption{PVD simulation with S/N=100 for a galaxy with $M_{*}=10^{11}$\msun\, $i=60$\deg, $n=3$ and $\mbh=10^8$\msun.
We assume a simple exponential disc with scale radius of 1\arcs\ for the molecular gas sharing the same inclination angle with the galaxy. 0.3\arcs\ beam size and 
10km/s velocity channel width is used to generate this PVD. The top left panel shows the integrated line intensity map and the top right panel shows the simulated PVD for the 
galaxy stellar mass (yellow) on top of the PVD of the entire system (black: galaxy and SMBH). The lower left panel shows the input rotation velocity to generate this PVD and 
the lower right panel shows the rotation velocities measured from the PVD (black for the entire system and red for the galaxy without SMBH) and the residual of the two rotation 
velocities (blue line). Dashed and dotted line is $3$ and $1\sigma$ significance for the residual rotation velocity, defined by $3\times$ and $1\times$ velocity 
channel width.}
\label{fig:kinms_demo}
\end{figure*}

Figure \ref{fig:kinms_demo} shows the simulation of a $10^{11}$\msun\ galaxy with $n=3$ and $i=60$\deg\ containing $10^8$\msun\ SMBH, at $\approx 15$ Mpc distance. 
The required beam size (\reqv) and velocity channel width ($\frac{1}{3}$\vdiff) for this galaxy is 0.5\arcs\ and 12 km/s based on the discussion in Section \ref{sec:theory_resolution}. 
Molecular gas distribution around the SMBH is simply assumed to be an exponential disc with 1\arcs\ scale radius and same inclination angle as the galaxy. 
The peak S/N is $\approx 100$ using 10000 random samples. 

The top left panel shows the velocity integrated flux distribution (i.e. moment 0 map) with 
0.3\arcs\ beam. The top right panel shows PVD of the galaxy using 0.3\arcs\ beam size and 10 km/s velocity channel width. We use slightly smaller beam size and 
velocity channel width than the required values (0.5\arcs\ and 12km/s), for the purpose of clear demonstration of the impact of SMBH.
Yellow contours on top of the background black contours are the PVD due to the galaxy stellar mass without the SMBH and the background 
black contours are the PVD for the entire system (galaxy and SMBH). 
The lower left panel shows the input rotation velocity to generate the PVD showing the contribution from the SMBH and the galaxy stellar mass distribution. 
The lower right panel shows the rotation velocity of the galaxy with and without SMBH measured from the PVD and the difference of the two velocities using blue 
line enclosed by $3\sigma$ and $1\sigma$ level velocity uncertainty (defined by $3 \times$ and $1 \times$ velocity channel width) as shown by the dashed and dotted line.

Rotation velocities are derived from the yellow and black contours by using `envelop-tracing' methods \citep[][]{sofue_2001}, which makes use of the terminal velocity 
in PVD along the major axis. The terminal velocity is defined by a velocity at which the intensity becomes equal to $I_{t} = \sqrt{(\eta I_{max})^2 + I_{lc}^2}$ on 
PVD, where $I_{max}$ and $I_{lc}$ are the maximum intensity and intensity corresponding to the lowest contour level. Using $\eta=0.5$, this defines a velocity at 
50\% level of the intensity profile at a fixed position for sufficiently large signal-to-noise ratio or a velocity along the lowest contour level if the signal-to-noise
ratio is small \citep[][]{sofue_2001}. This will capture the circular velocity that lies at the outer envelope of PVD. 

We caution that the velocity determined by 
tracing the outer envelope of PVD is biased toward the material along the line of sight and does not very well trace the Keplerian rise in the PVD seen in Figure \ref{fig:kinms_demo}.
Some of the issues discussed in this section can be removed by fitting the 2D PVD intensity map itself or the entire 3D data cube.
The detailed comparison of the performance of different methods is beyond the scope of this work.
However, we note that our approach tracing the outer envelope of PVD is conservative and therefore ensures that more sophisticated fitting 
methods provide a better constraint to the SMBH mass.

For illustration,
we select a typical example of SMBH and host galaxy. The assumed SMBH mass ($10^8$\msun) and galaxy stellar mass ($10^{11}$\msun) are 
not in the extreme range \citep[e.g., dynamically measured $10^{10}$\msun\ SMBH in $10^{11}$\msun\ galaxy reported by][]{vdbosch_etal_2012}. 
In the upper right panel of Figure \ref{fig:kinms_demo}, we are able to detect a signature of the SMBH as revealed by the high velocity tip of black contour at 
the centre indicative of the Keplerian rotation. If looking at the residual rotation velocity shown by the blue line 
in the lower right panel of Figure \ref{fig:kinms_demo}, the maximum deviation at the centre is much larger than $3\sigma$ velocity uncertainty and the deviation is 
larger than $1\sigma$ range for the entire range of the spatial scale. In this case, it is relatively easy to claim a detection of the SMBH mass. However for 
given \mbh\ and galaxy stellar mass, the significance of this deviation will be smaller with increasing beam size and velocity channel width of the radio 
interferometry. Since it is already obvious that the SMBH with large enough mass will be easily detected as shown in this example, we will test
the case of the lowest detectable SMBH masses for the given galaxy parameters inferred from Figure \ref{fig:radius} and \ref{fig:velocity}.

\subsection{Effect of Noise}\label{sec:simul_snr}
Before we investigate the impact of systematic effect to the PVD analysis, we discuss the effect of noise first.
We take a fiducial galaxy with $M_{*}=10^{11}$\msun\ including the SMBH with $\mbh=10^7$\msun\, of which surface brightness follows a \sersic\ profile with $n=3$ and $i=60$\deg.
Then we assume that the gas surface density distribution is an exponential disc with 1\arcs\ scale radius and sample the density distribution using a finite number 
of random samples as discussed above. For this galaxy, the minimum required beam size and velocity channel width is 0.15\arcs\ and 8 km/s respectively inferred from 
Figure \ref{fig:radius} and \ref{fig:velocity}. For this beam size and channel width, we make PVDs with 4 different signal-to-noise ratios; S/N=10, 30, 60 and 120 determined 
by 150, 1400, 6000 and 24000 random samples respectively. Note that S/N is roughly proportional to $\sqrt{N_{samp}}$ as expected from statistics. 

\begin{figure*}
\centering
\subfigure[S/N=10]{\epsfig{file=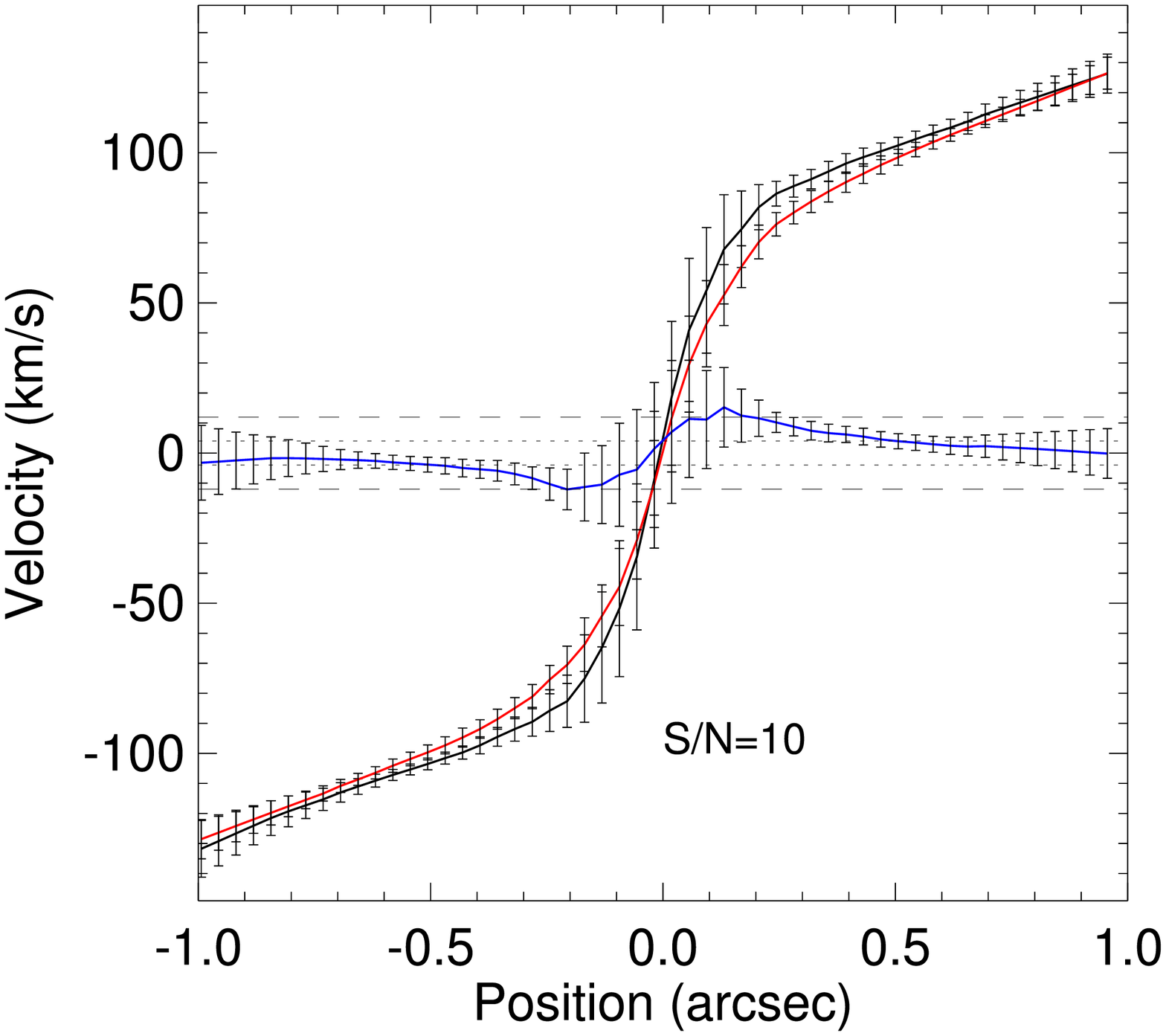,width=0.4\textwidth}}
\subfigure[S/N=30]{\epsfig{file=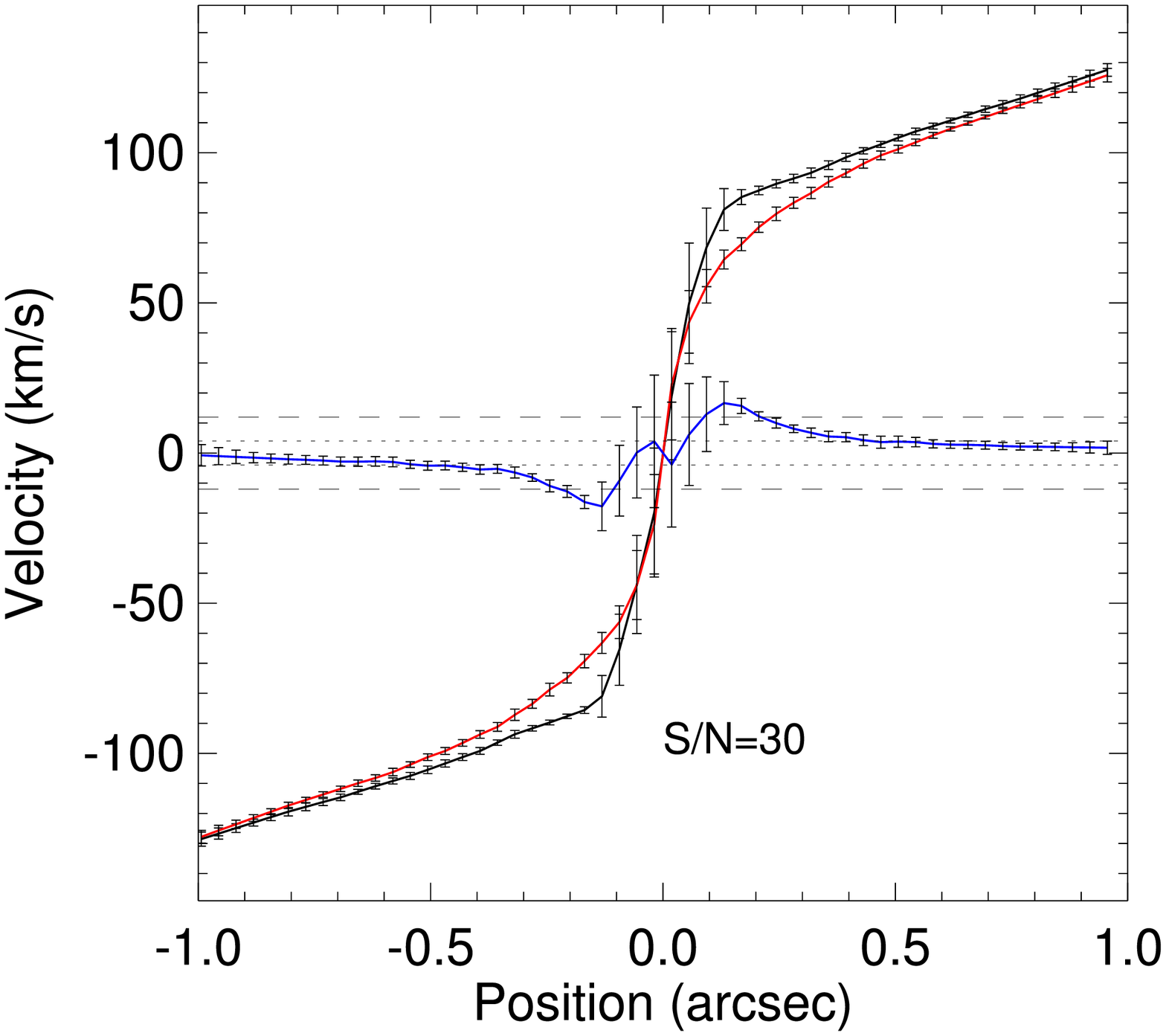,width=0.4\textwidth}}
\subfigure[S/N=60]{\epsfig{file=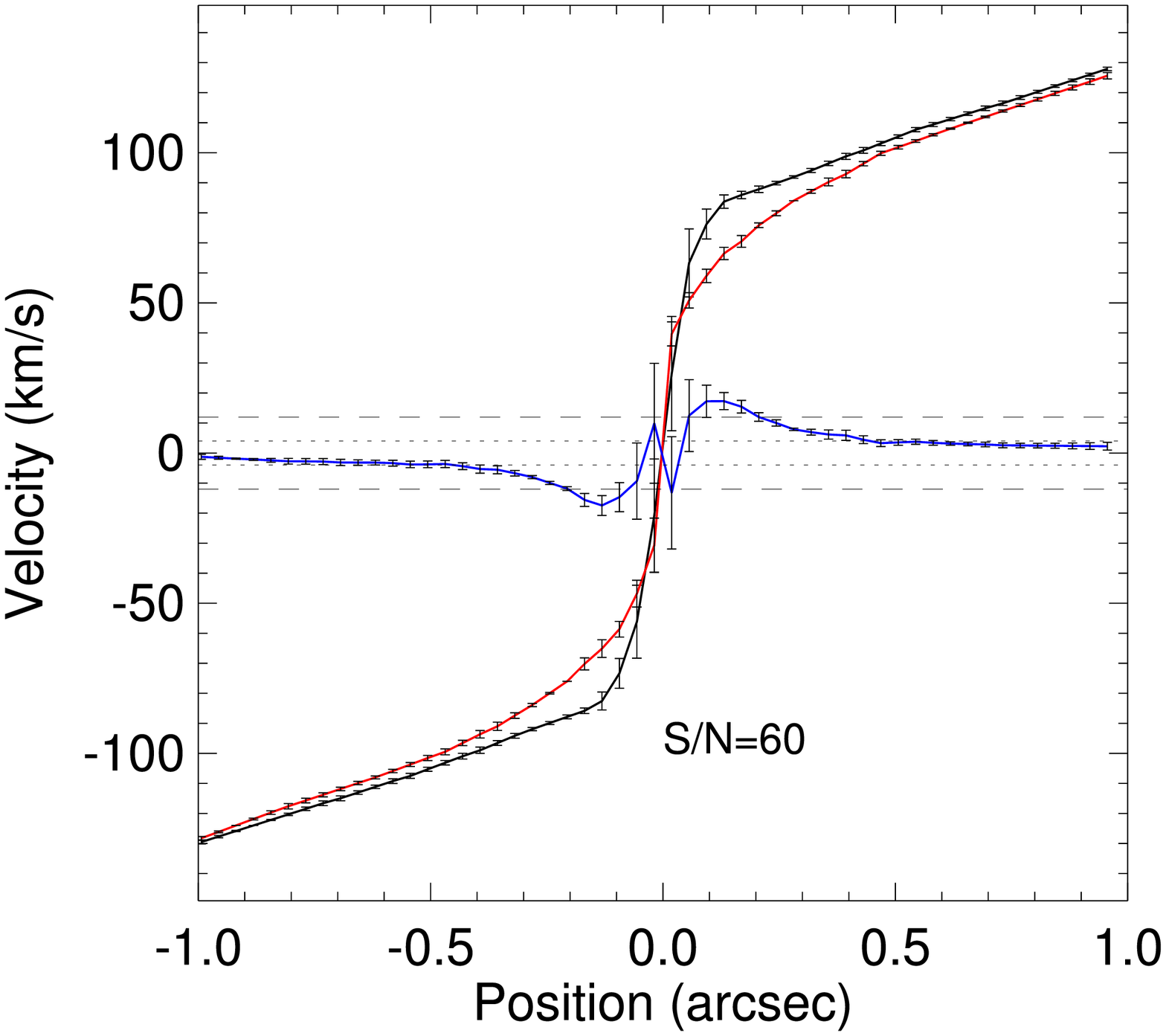,width=0.4\textwidth}}
\subfigure[S/N=120]{\epsfig{file=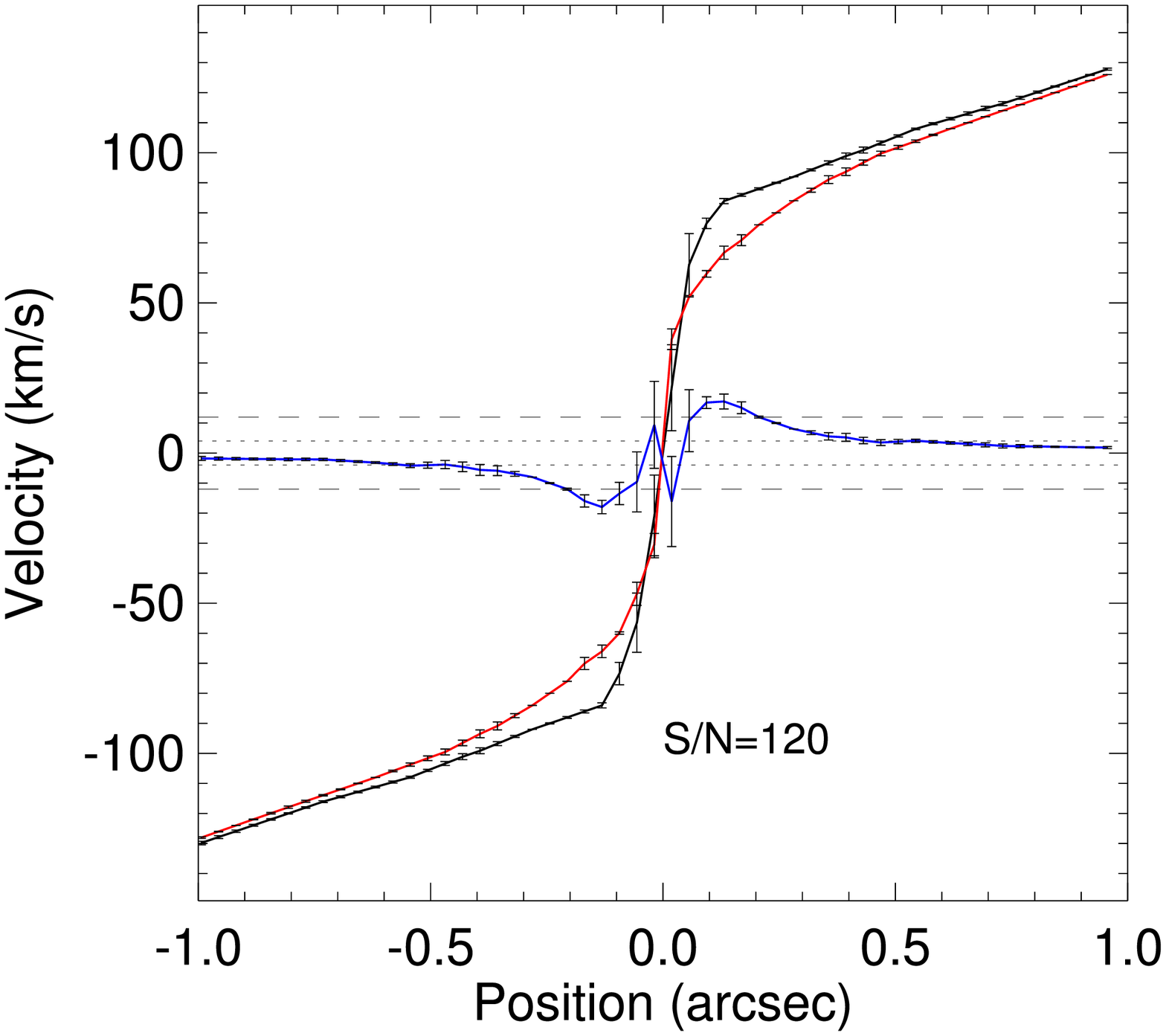,width=0.4\textwidth}}
\caption{Rotation velocities and their residuals of the simulated galaxy with $M_{*}=10^{11}$\msun\ and $\mbh=10^{7}$\msun\ measured using 0.15\arcs\ beam and 8 km/s velocity 
channel width. Each panel shows the same galaxy with different signal-to-noise ratio, S/N=10, 30, 60 and 120. The black and red solid line show the rotation 
velocity of the galaxy with and without SMBH and the blue line is the difference of the two rotation velocities.
Rotation velocity (solid line) and associated error bar has been determined by mean and standard deviation of the realisation of 100 ensemble PVDs. 
The velocity uncertainties due to the velocity channel width corresponding to 1 and 3 $\sigma$ significance, 
$\frac{1}{3} \vdiff$ and \vdiff\ are shown by the dotted and dashed line.
}
\label{fig:kinms_snr}
\end{figure*}

Figure \ref{fig:kinms_snr} shows the rotation velocity of the galaxy with SMBH (black), without SMBH (red) and the residual velocity between 
the two (blue). For each S/N, we generate 100 ensemble PVDs and each solid line with error bars is the mean of this ensemble measurements with 
the standard deviations of the ensemble PVDs. When S/N is low (S/N=10), both the black and red line has large errors and thus the residual velocity is largely 
uncertain; the standard deviation in the central region is larger than $3\Delta V$. As S/N increases, the uncertainty 
of the rotation velocity measurement becomes smaller. When S/N=60, it becomes much smaller than $\Delta V$ in the outer region and comparable to the $\Delta V$ in the centre. 
This noise simulation confirms our intuition that the rotation 
velocity difference with and without SMBH becomes less significant as S/N decreases and it will increase the uncertainty of the SMBH mass measurement.

In the following experiments to investigate the required spatial and velocity resolution (Section \ref{sec:simul_resol}) and to demonstrate the impact of systematic 
effects to the PVD analysis (Section \ref{sec:simul_systematic}), we ensure that PVD has a reasonably good 
S/N (S/N $\approx 60$) by choosing the appropriate number of random samples for given beam size and velocity channel width for each experiment.

\subsection{Spatial and Velocity Resolution}\label{sec:simul_resol}
We simulate PVDs of the same galaxy (i.e., a galaxy with $M_{*}=10^{11}$\msun, \mbh$=10^{7}$\msun\ and $i=60$\deg) in Figure \ref{fig:kinms_snr} 
for four different \sersic\ indices ($n=1,2,3$ and 4) and investigate a significance of the 
difference between the rotation velocity of the galaxy with and without SMBH, using different beam sizes and velocity channel widths.

\begin{figure*}
\centering
\subfigure[$n=1$]{\epsfig{file=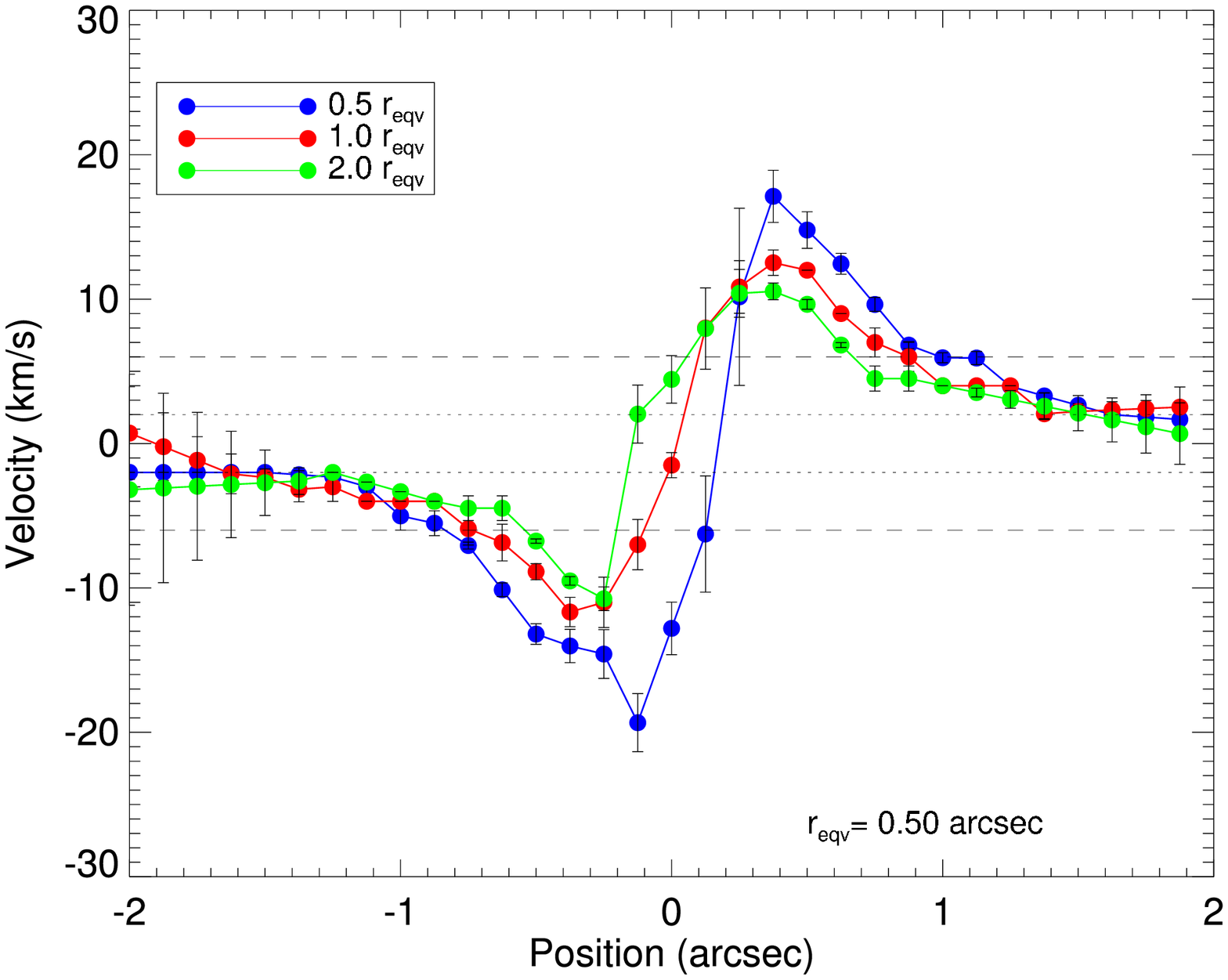,width=0.4\textwidth}}
\subfigure[$n=2$]{\epsfig{file=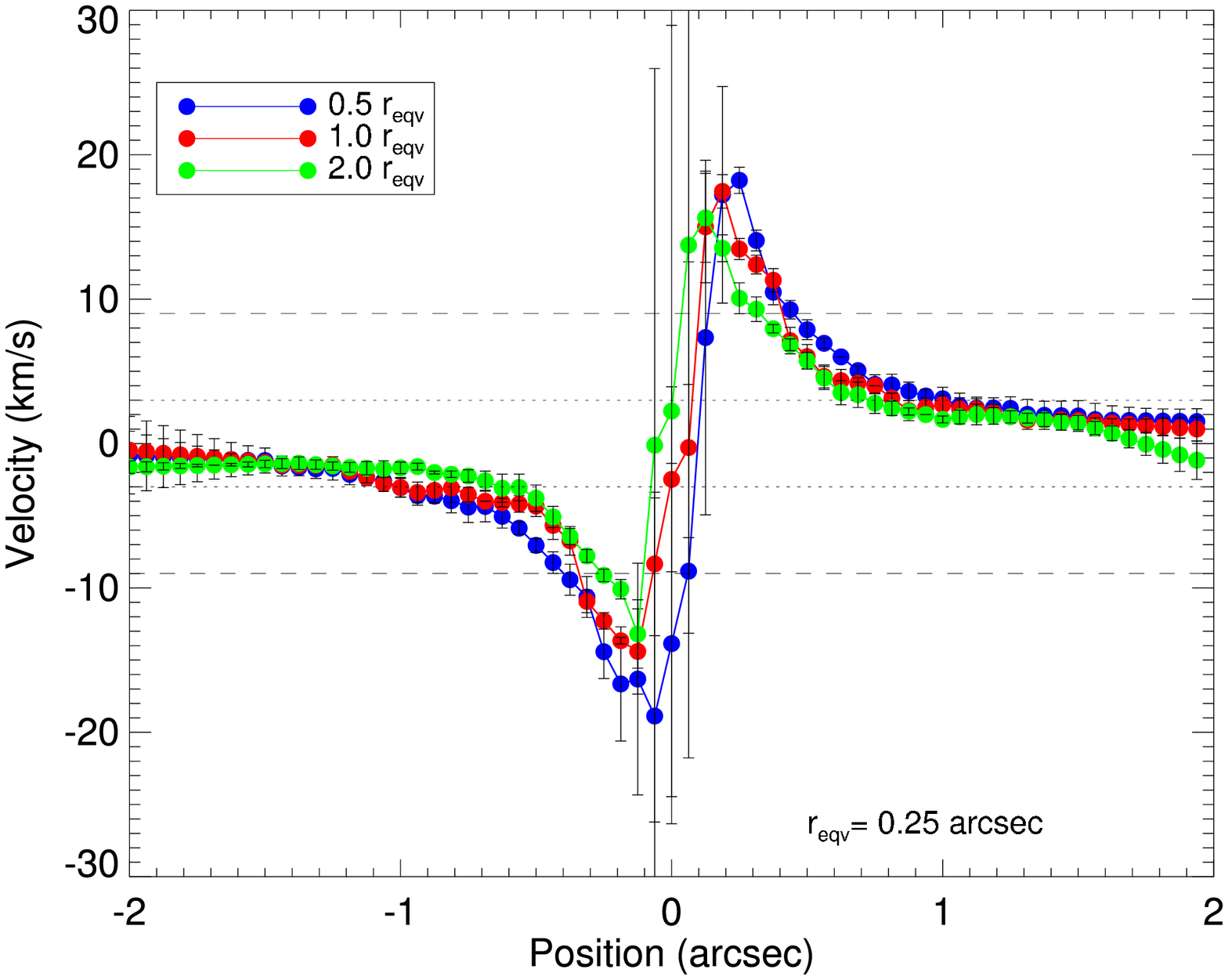,width=0.4\textwidth}}
\subfigure[$n=3$]{\epsfig{file=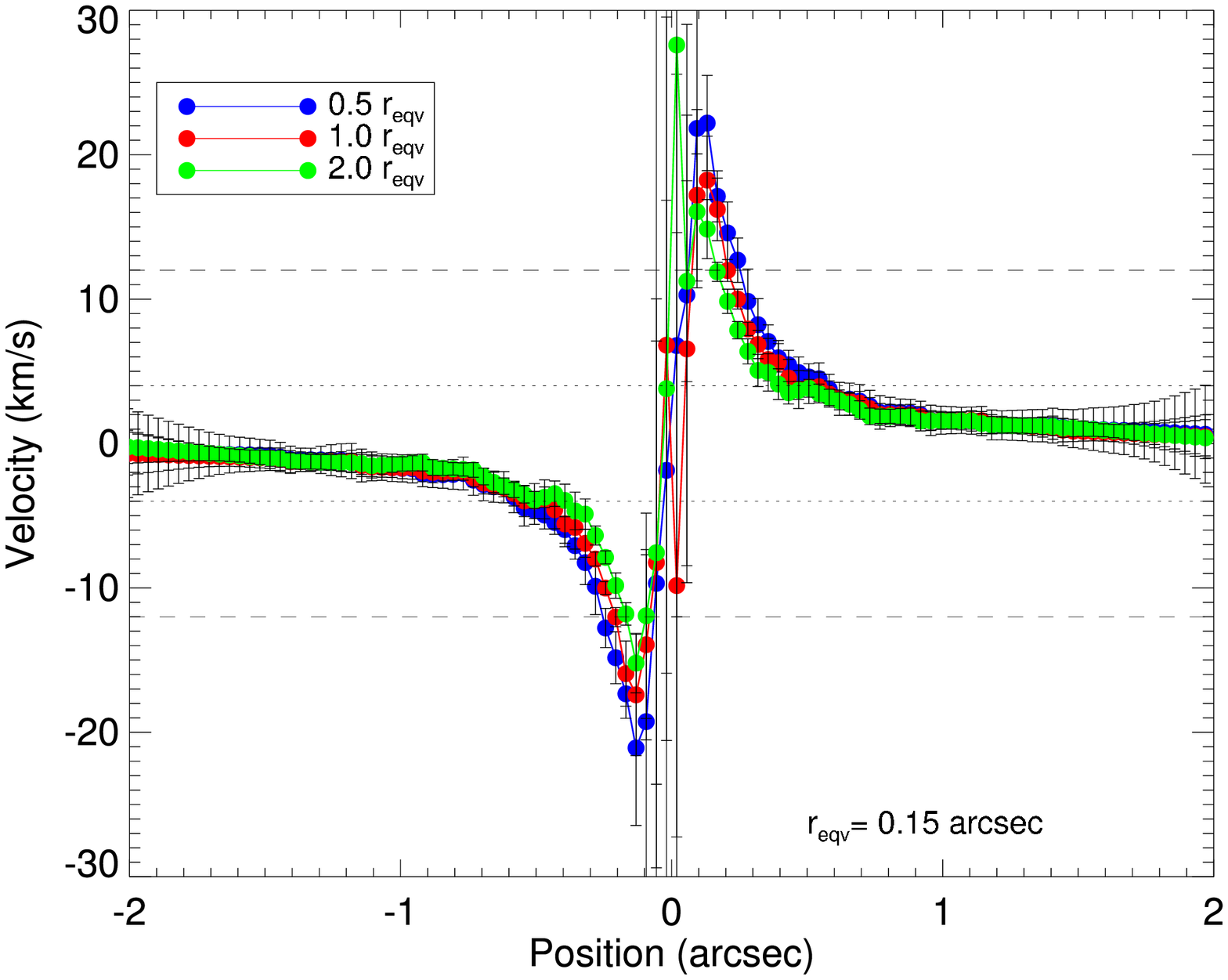,width=0.4\textwidth}}
\subfigure[$n=4$]{\epsfig{file=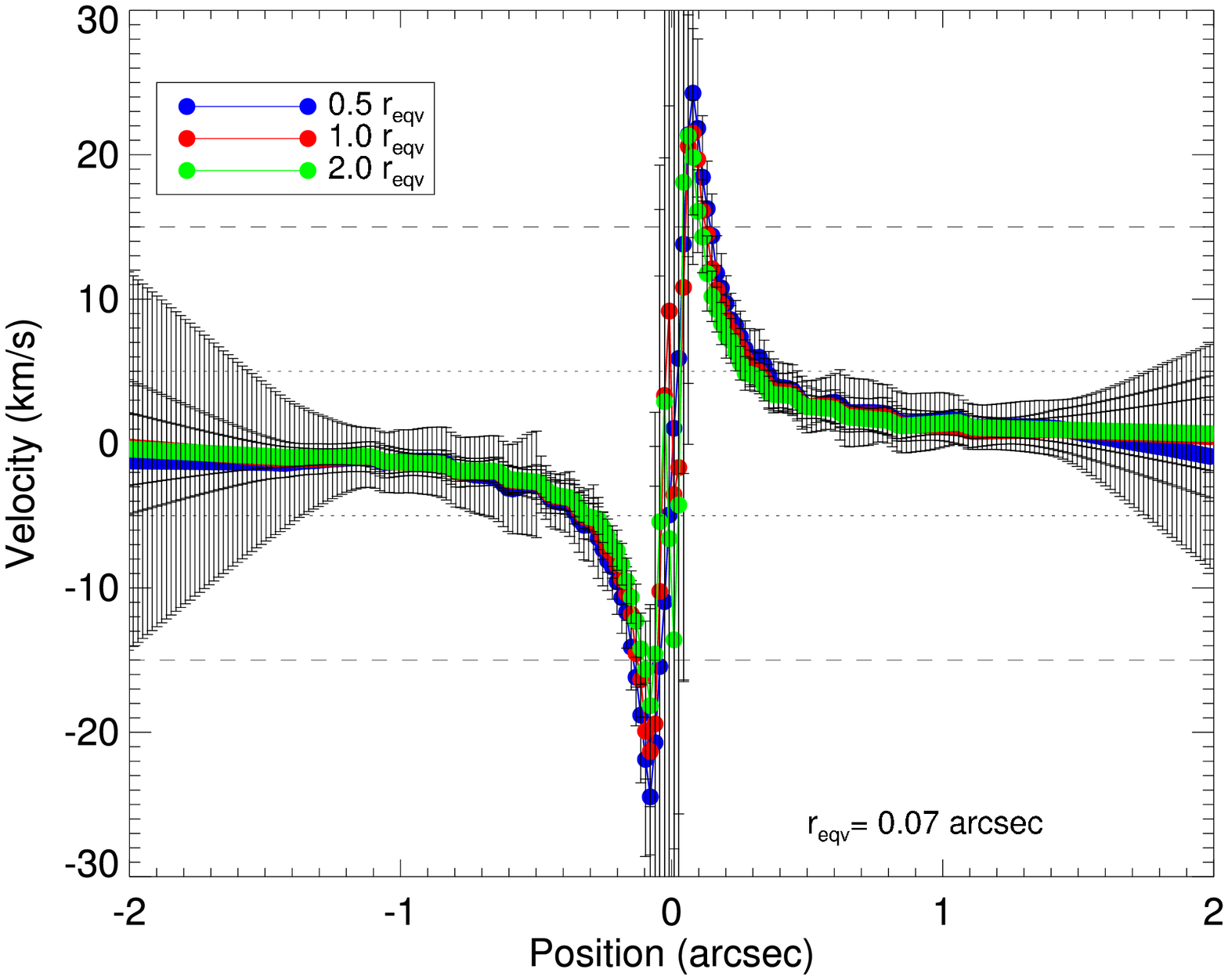,width=0.4\textwidth}}
\caption{Difference of the rotation velocity with and without SMBH, for the simulated galaxy with $M_{*}=10^{11}$\msun\ and $\mbh=10^{7}$\msun\ measured using the same velocity channel 
width $\frac{1}{3} \vdiff$ but using different beam sizes shown by different colours. Each panel shows the same galaxy with different \sersic\ index from 1 to 4. 
Error bar has been determined by the standard deviation of the realisation of 100 ensemble PVDs. 
The velocity uncertainties associated with 1 and 3 $\sigma$ significance, $\frac{1}{3} \vdiff$ and \vdiff\ are shown by the dotted and dashed line.}
\label{fig:kinms_delr}
\end{figure*}

\begin{figure*}
\centering
\subfigure[$n=1$]{\epsfig{file=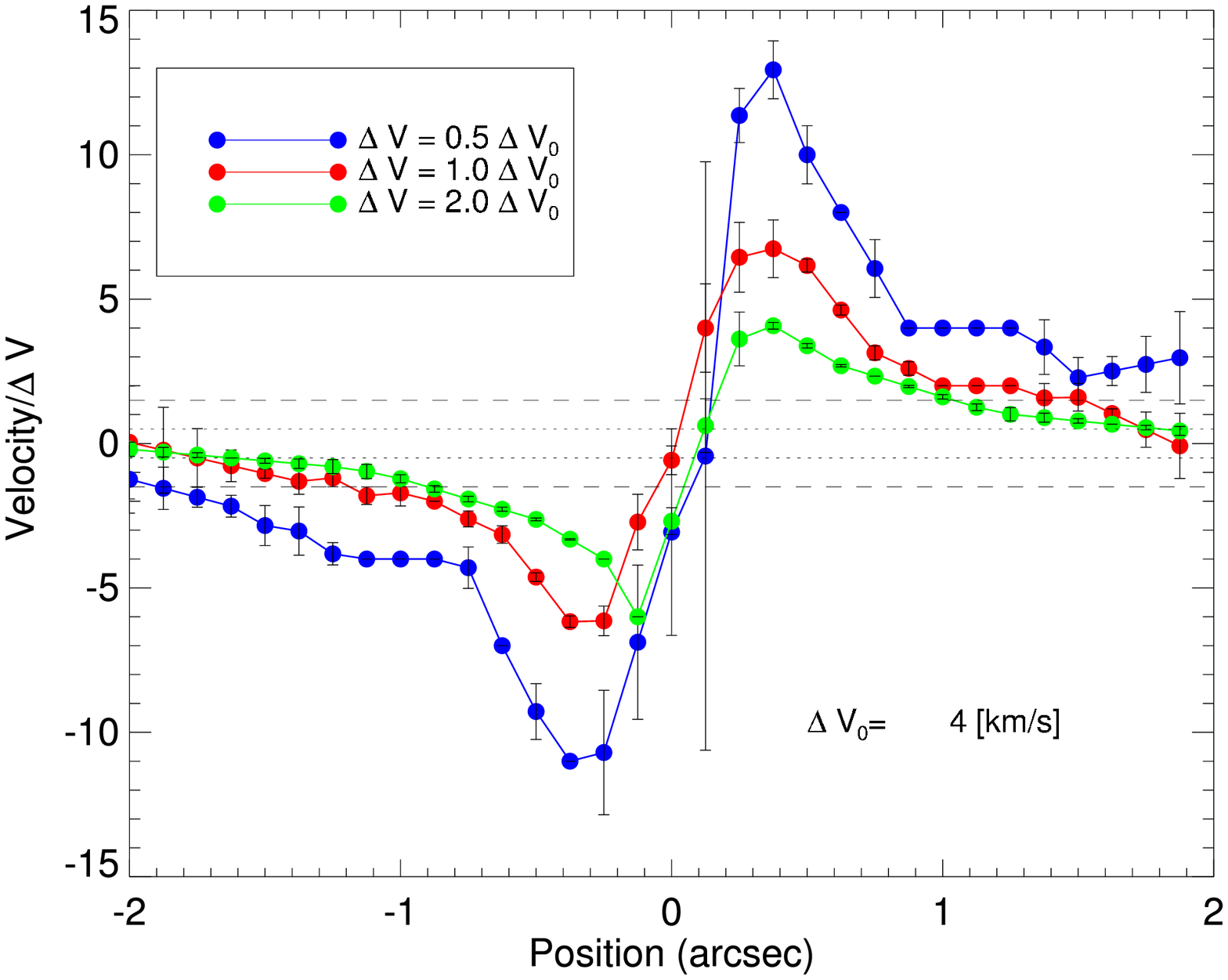,width=0.4\textwidth}}
\subfigure[$n=2$]{\epsfig{file=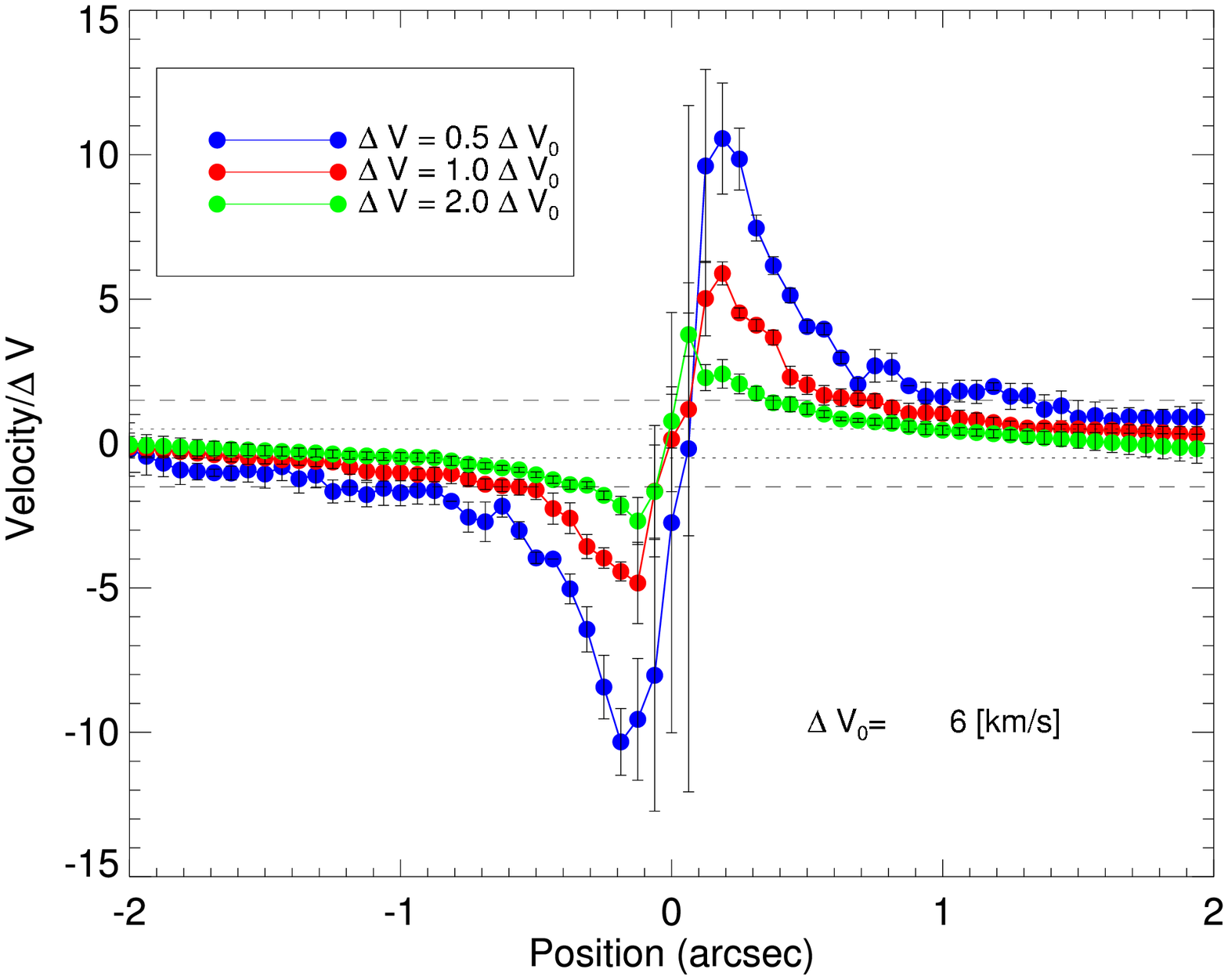,width=0.4\textwidth}}
\subfigure[$n=3$]{\epsfig{file=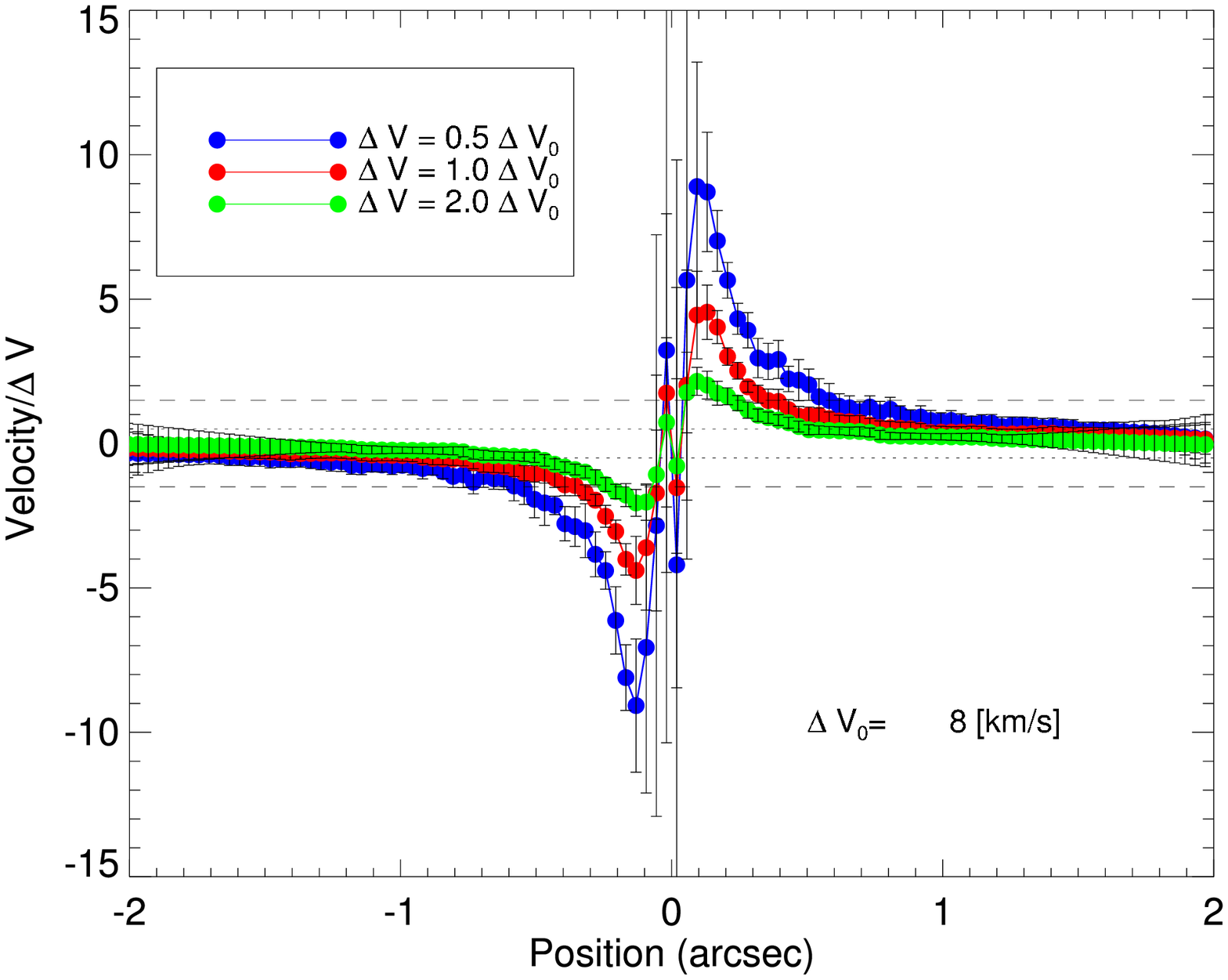,width=0.4\textwidth}}
\subfigure[$n=4$]{\epsfig{file=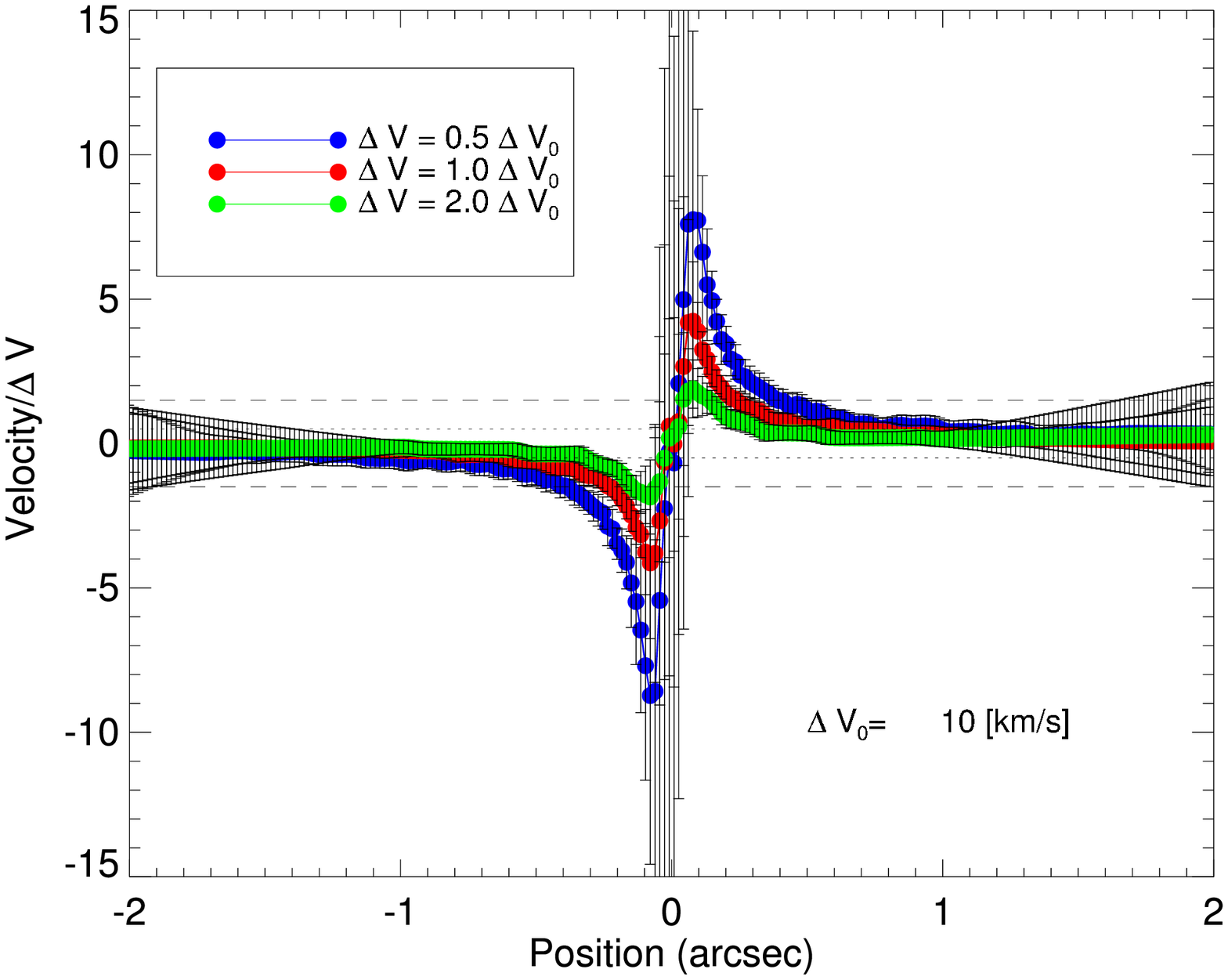,width=0.4\textwidth}}
\caption{Difference of the rotation velocity with and without SMBH, for the simulated galaxy in Figure \ref{fig:kinms_delr} measured using the same beam size \reqv, but using 
different velocity channel width shown by different colours. Each panel shows the same galaxy with different \sersic\ index from 1 to 4. 
Error bar has been determined by the standard deviation of the realisation of 100 ensemble PVDs. The residual 
velocities for each panel are scaled by the velocity channel width being used to generate the PVD for each panel. 
The uncertainties associated with 1 and 3 times the velocity channel width are shown by the dotted and dashed line.}
\label{fig:kinms_delv}
\end{figure*}

Each panel in Figure \ref{fig:kinms_delr} shows the residual rotation velocity of the galaxy with the 
corresponding \sersic\ index, for different beam sizes. The velocity channel width $\Delta V$ is fixed to $\frac{1}{3}$\vdiff. 
Beam size varies from 0.5 to 2.0 \reqv\ (as shown by different colours) for the fixed velocity channel width used in each panel.
In each panel, the width of $\Delta V$ and $3\Delta V$ are shown by the dotted and dashed line respectively to show the $1\sigma$ and $3\sigma$ significance of 
the residual rotation velocity. Error bar has been determined by the standard deviation of the realisation of 100 ensemble PVDs.

The residual velocity with 0.5\reqv\ beam size shown by the blue symbol always shows a significant difference at the central region in all panels of Figure \ref{fig:kinms_delr}; 
the maximum residual is significantly larger than $3\Delta V$. Although the 1.0\reqv\ beam size (red line) shows smaller maximum residual than that for the 0.5\reqv\ beam size, 
the residual is still significant; larger than or similar to $3\Delta V$. However as the beam size increases (i.e., $>2\reqv$), the maximum residual decreases 
and the significance of the residual rotation velocity does not strongly indicate a rotation velocity due to the SMBH.

Each panel in Figure \ref{fig:kinms_delv} shows the case where different velocity channel widths have been used to generate the PVD of the same galaxy for the fixed 
1.0\reqv\ beam size. Like in Figure \ref{fig:kinms_delr}, the error bar has been determined by the standard deviation of the realisation of 100 ensemble PVDs. 
The residual velocities for each panel are scaled by the velocity channel width being used.
The velocity channel width $\Delta V$ has been adjusted with respect to the required velocity channel width for each panel ($\Delta V_0 = \frac{1}{3}$\vdiff).
Similarly to Figure \ref{fig:kinms_delr}, as the velocity channel width increases from 0.5 to 2.0$\Delta V_0$, the residual velocity
becomes less significant. We find that the choice of $\Delta V \approx 1.0\Delta V_0$ shows a significant ($>3\sigma$) velocity difference for most 
types of galaxies that we simulated in this work.

In summary, 
regarding the effect of spatial resolution, if the radio interferometry beam size is smaller than or similar to \reqv, the PVD is 
clearly resolved and the excess of the rotation velocity due to the SMBH is detected above $3\sigma$ level when the velocity channel width resolves $\frac{1}{3}$\vdiff. 
Then regarding the effect of velocity resolution, if the velocity channel width is smaller than $\frac{1}{3}\vdiff$, the effect of SMBH is detected above 
$3\sigma$ level if the scale of \reqv\ is resolved by the interferometry beam.

\subsection{Systematic Effects}\label{sec:simul_systematic}
In the previous section, we discussed the spatial and velocity resolution for the PVD analysis using rotation velocity of pure circular motion without systematic effect
and confirmed our argument that \reqv\ and \vdiff\ set the required beam size and velocity channel width.
In this section, we demonstrate the impact of possible systematic effects to the PVD analysis: spatial structure (i.e., gas density distribution) and 
velocity structure (i.e., inflow/outflow, random motion and warp) as discussed in Section \ref{sec:systematic}. We do not perform an extensive search in the parameter 
space for all these systematic effects and only show the PVD simulations with S/N$\approx$60 of a galaxy with $M_{*}=10^{11}$\msun\ that follows a \sersic\ profile with $n=3$
and hosts $10^7$\msun\ SMBH, using the required angular resolution (0.15\arcs) and velocity channel width (8 km/s). We show the results of simulations using selective
parameters for each effect. However our findings and discussions regarding these systematic effects are valid for the given angular resolution and velocity 
channel width of simulated PVD and they can be applied to other galaxies with different $n$ and SMBH masses. 

For investigating the systematic effects, we note that one of the disadvantages of using PVD is that disc inclination, kinematic position angle and centroid parameters 
are needed to be well constrained \citep{barth_etal_2016}. Using more sophisticated methods including 
fitting 3D data cube might be more reliable and can resolve some of the issues related with the gas disc geometry. However, we also note that fitting PVD, on the other hand, has 
benefited from the better sensitivity to the central velocity upturn and may give better constraints on $M_{BH}$ than fitting 3D data cube \citep{barth_etal_2016}. 

\subsubsection{Geometry of molecular gas distribution}
We consider three different types of density profiles of molecular gas: the same exponential disc profile as the previous simulations but with different scale radius, 
the same 1\arcs\ scale radius exponential disc but with different inclination angles and the density profile with an inner scale radius for density truncation which mimics 
the density distribution of circum nuclear ring.

\begin{figure*}
\centering
\subfigure[$R_s=0.15$\arcs]{\epsfig{file=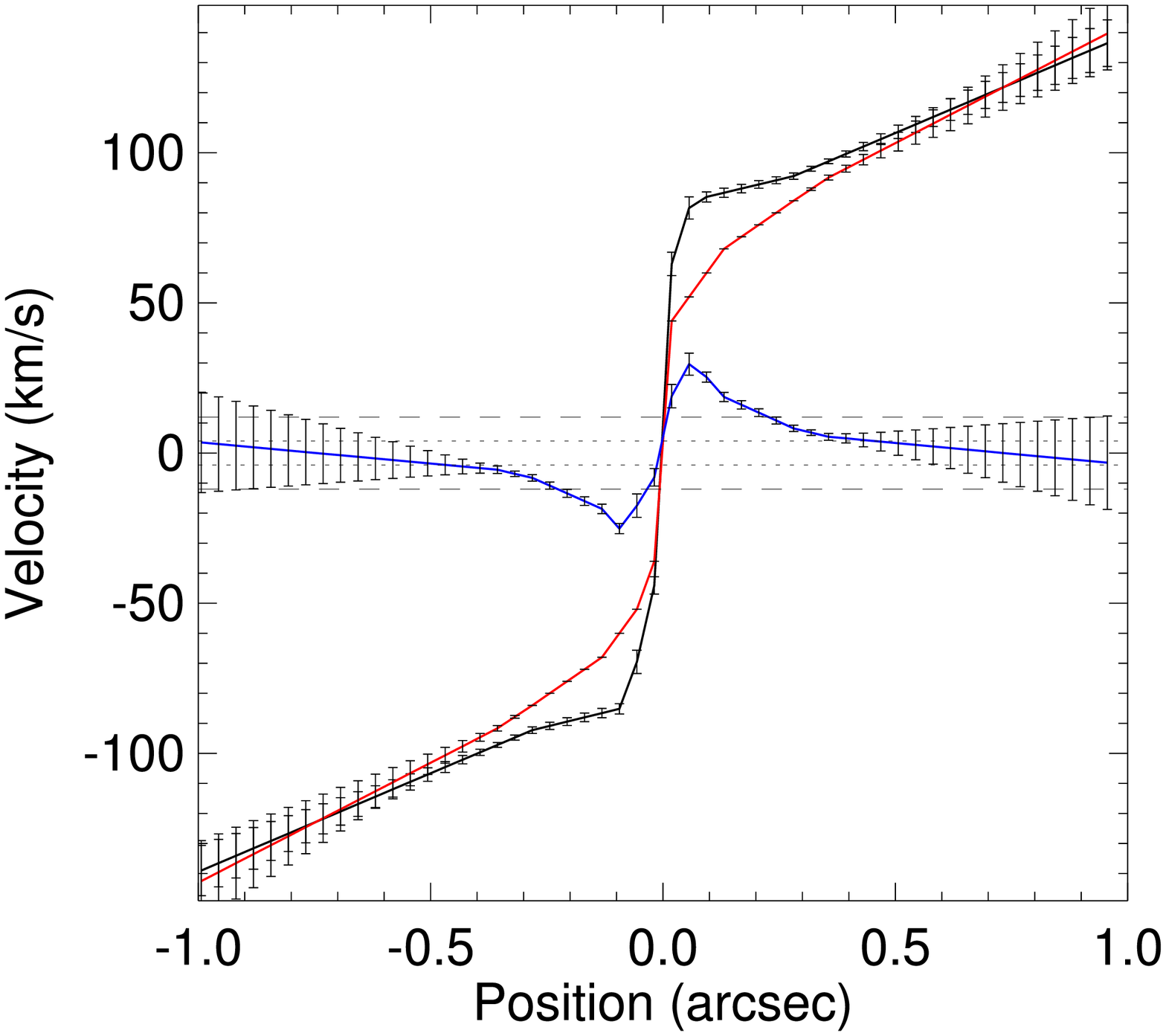,width=0.4\textwidth}}
\subfigure[$R_s=0.30$\arcs]{\epsfig{file=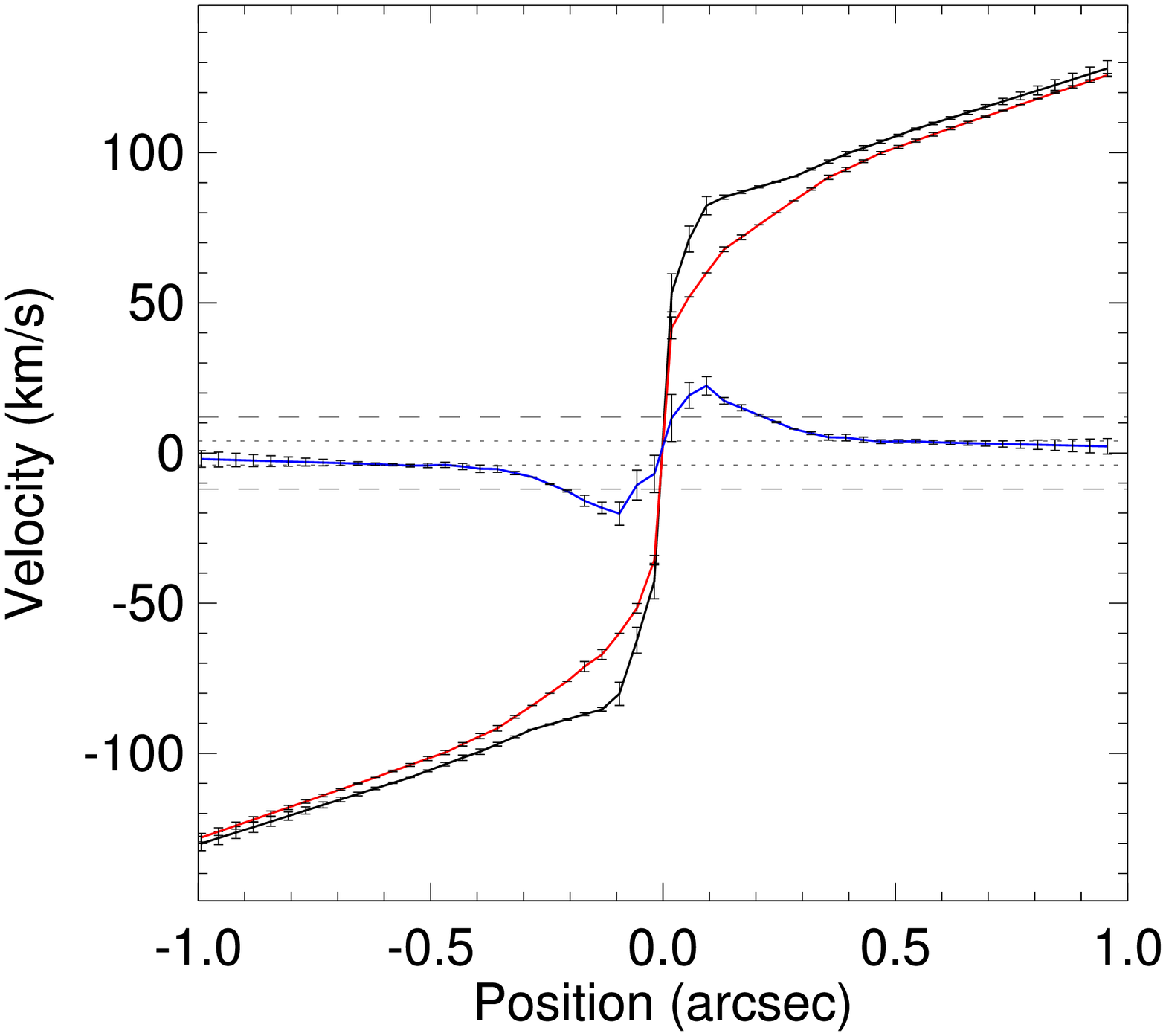,width=0.4\textwidth}}
\subfigure[$i=30$\deg]{\epsfig{file=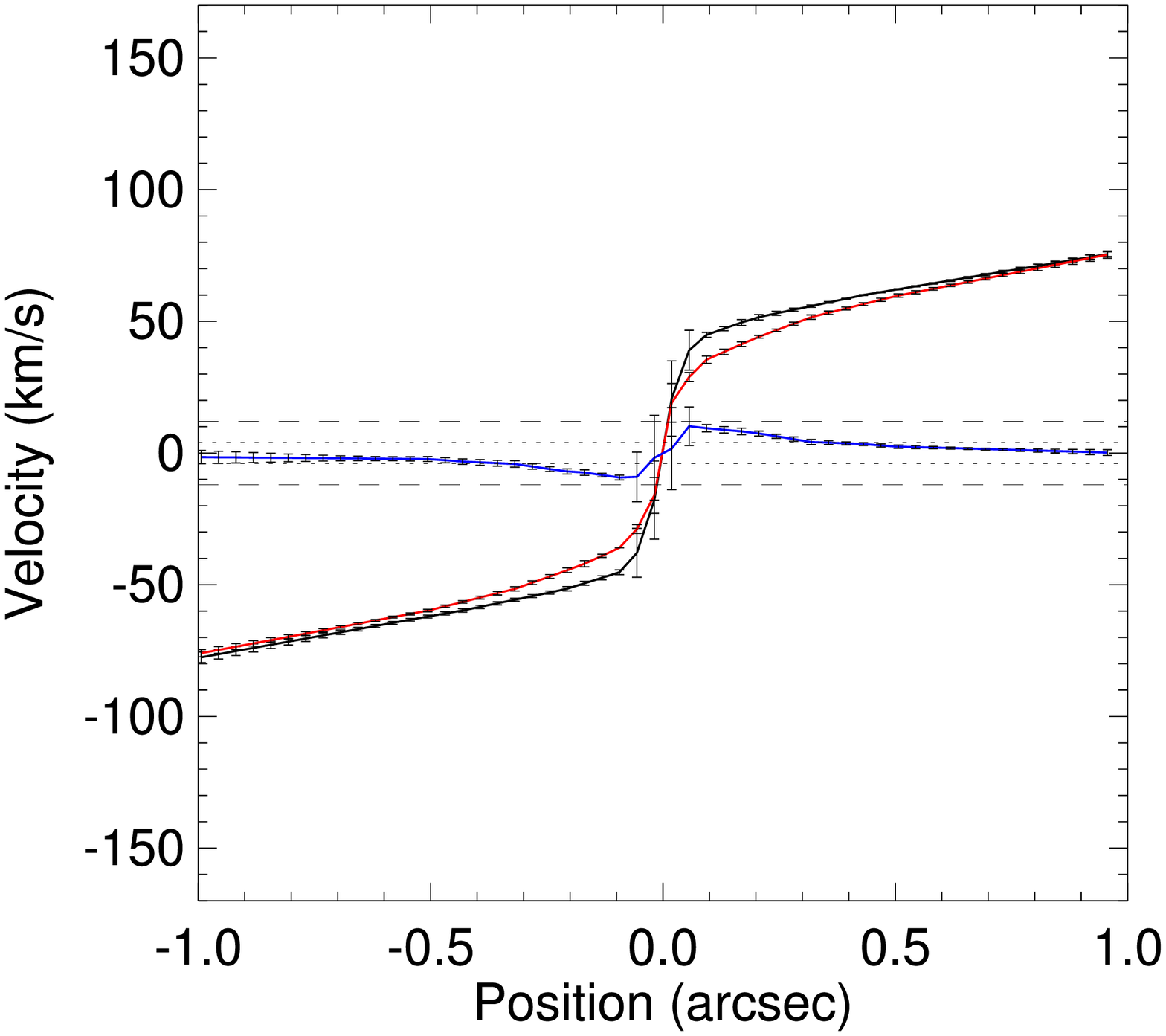,width=0.4\textwidth}}
\subfigure[$i=80$\deg]{\epsfig{file=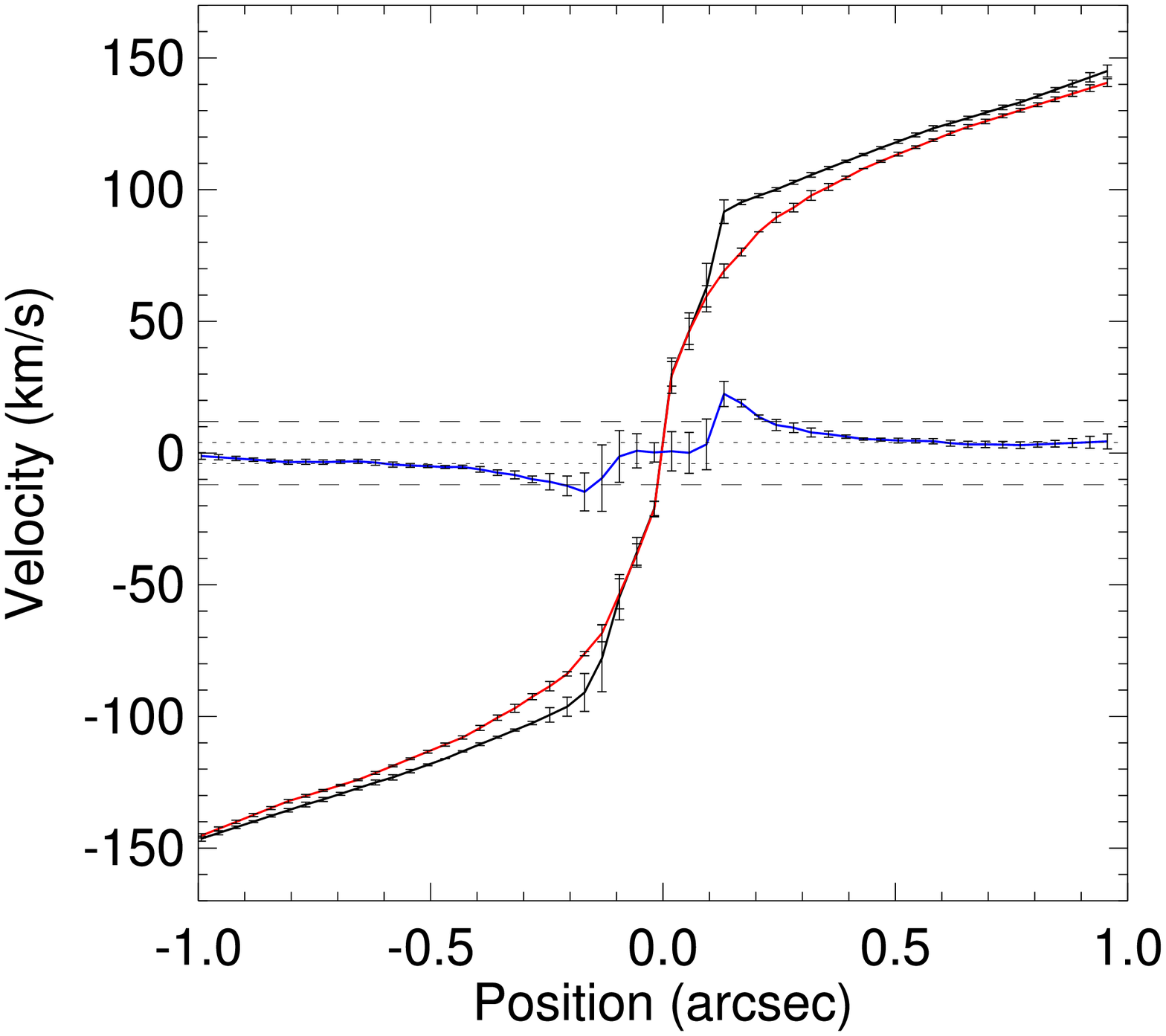,width=0.4\textwidth}}
\caption{Rotation velocities and their residuals of the simulated galaxy in Figure \ref{fig:kinms_snr}(c) but using different scale radius and inclination for the gas disc.
The same beam size \reqv\ (0.15\arcs) and velocity channel width (8km/s) are used for all simulations. 
Error bar has been determined by the standard deviation of the 
realisation of 100 ensemble PVDs. The residual velocity for each panel are scaled by the velocity channel width being used to generate the PVD for each panel.
The uncertainties associated with 1 and 3 times the velocity channel width are shown by the dotted and dashed line.}
\label{fig:kinms_rs}
\end{figure*}

\begin{figure*}
\centering
{\epsfig{file=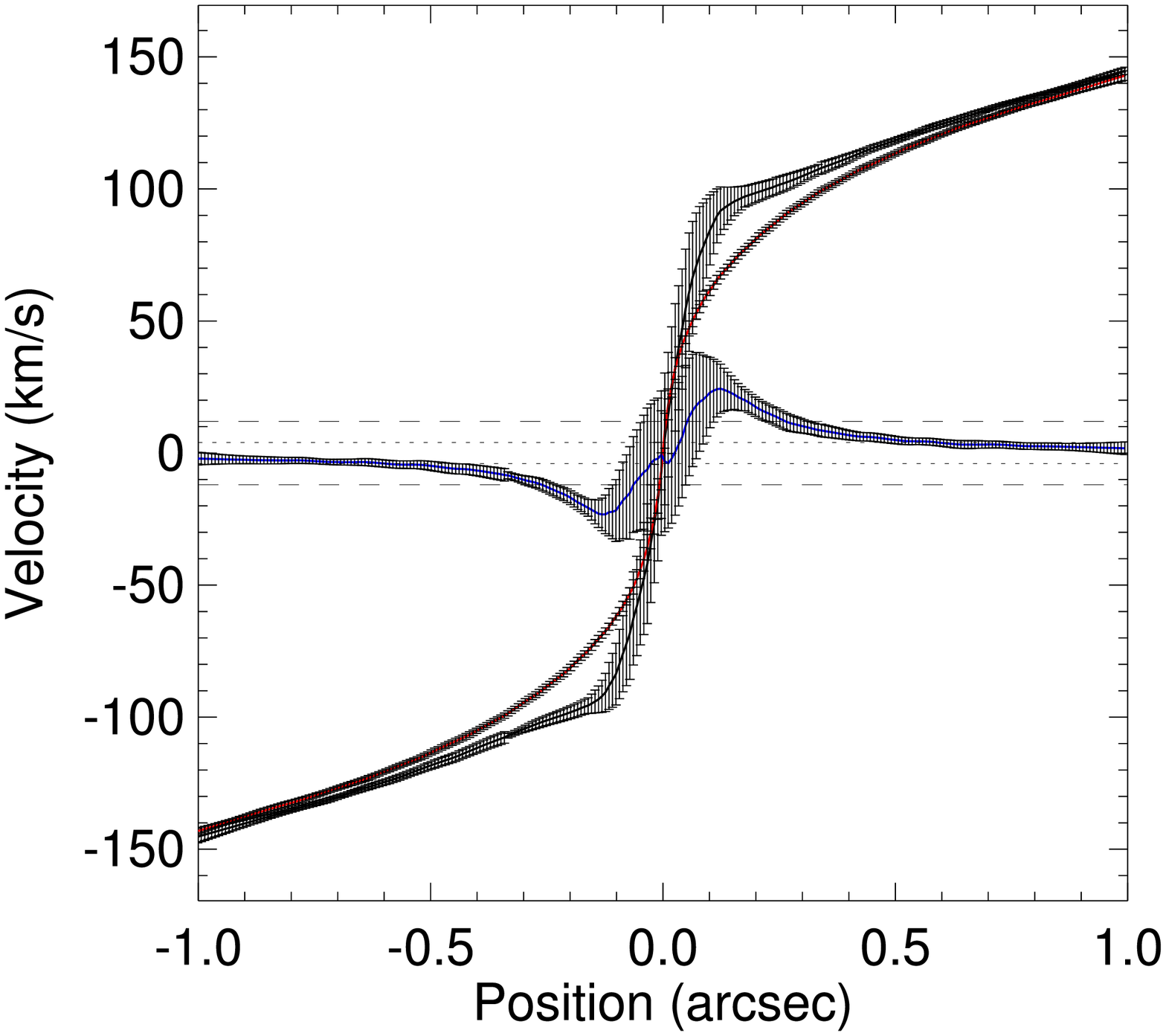,width=0.4\textwidth}}
\caption{Same as Figure \ref{fig:kinms_rs}(d) but using higher angular resolution resolving projected \reqv\ along 
the minor axis, \reqv\ $\mbox{cos}(i)$. The beam size is 0.03\arcs.}
\label{fig:kinms_inc}
\end{figure*}

First, we perform the test for different gas density profiles by changing the scale radius to adjust a compactness of the profile. 
Figure \ref{fig:kinms_rs}(a) and (b) show the rotation velocity of the gas disc following exponential density profile with 0.15 and 0.3 \arcs\ scale radius 
and the residual velocity between the rotation velocity of galaxy with and without central SMBH.
The inclination angle is fixed to $i=60$ \deg. In this test, the density distribution of the molecular gas is normalised 
to the integrated density and therefore the surface density profile of the molecular gas with smaller scale radius has higher peak flux value than that with 
larger scale radius. Since more signal is concentrated in the central region of the molecular gas disc with 0.15\arcs\ scale radius compared with the disc
with 0.3\arcs\ radius, the measured velocity for density profile with 
smaller scale radius has smaller error at the central region where the impact of SMBH is the strongest while it has larger error at the outer region due to low S/N 
as seen in Figure \ref{fig:kinms_rs}(a) and (b) respectively. Comparing the three figures with the similar S/N (S/N$\approx 60$): 
Figure \ref{fig:kinms_rs}(a) with 0.15\arcs\, Figure \ref{fig:kinms_rs}(b) 
with 0.30\arcs\ and Figure \ref{fig:kinms_snr}(c) with 1.0\arcs\ scale radius, we note that the error of residual velocity at the centre becomes smaller 
with decreasing scale radius of the molecular gas disc. 
Also the maximum of the residual velocity is significantly larger than $3 \Delta V$ for the gas disc with the smallest scale 
radius (Figure \ref{fig:kinms_rs}(a)). We find that for the same S/N, angular resolution, 
velocity channel width and inclination angle, more compact molecular gas density distribution gives a larger significance of the velocity difference for 
the detection of SMBH as long as the angular resolution is sufficiently small to resolve \reqv.

Second, we show the difference of inclination angle for the same galaxy shown in Figure \ref{fig:kinms_snr}(c). Figure \ref{fig:kinms_rs}(c) and (d) show 
the rotation velocity for the two different inclination angles: 30\deg\ and 80\deg, instead of 60\deg\ as shown in Figure \ref{fig:kinms_snr}(c). 
If the galaxy has small inclination ($i=30$\deg), the significance 
of the velocity difference at the centre between the rotation velocity with and without SMBH is less than the case of the large inclination angle 
because of the velocity projection. Maximum of the residual velocity in Figure \ref{fig:kinms_rs}(c) is less than $3\sigma$ and lower than that of the higher inclination
molecular gas disc shown in Figure \ref{fig:kinms_rs}(d) and Figure \ref{fig:kinms_snr}(c).
On the other hand, if the inclination angle becomes even larger and makes the galaxy close to edge on ($i=80$\deg) as seen in Figure \ref{fig:kinms_rs}(d), 
the synthesised beam includes more molecular gas components with lower velocities migrated from the minor axis and thus overall velocity measurement at the centre is weighted 
more by the lower velocity components. 
This issue regarding the high inclination angle is recently discussed by \citet[][]{barth_etal_2016} which suggest that the projected \rsoi\ along the minor axis has 
to be resolved to measure the accurate SMBH mass. We remake Figure \ref{fig:kinms_rs}(d) using smaller beam $\reqv\ \mbox{cos}(i)$ to resolve 
the projected \reqv\ along the minor axis, as shown in Figure \ref{fig:kinms_inc} which shows that the residual velocity at the central region that was 
diluted by the lower velocity component as shown in Figure \ref{fig:kinms_rs}(d) reveals the velocity difference more clearly although it has large error due to the small beam. 
We find that for the same S/N, angular resolution, velocity channel width and scale radius of the molecular gas density profile, the significance of the 
velocity difference increases as galaxy becomes more inclined, however, for highly inclined galaxy, the beam size should be small enough to resolve a spatial 
scale with a velocity gradient not much larger than the velocity channel width to avoid the velocity smearing within the beam. 
Resolving the scale of projected \reqv\ along the minor axis is possible, however the resulting ALMA beam size ($\approx 0.03$\arcs) will
be a practical limitation for observing galaxy with high inclination. This issue of high inclination angle can be significantly alleviated 
by modeling PVD in position and velocity space together (i.e., 2D pixel distribution in PVD) or even completely removed by modeling 3D data cube, if the 
signal-to-noise ratio of the data is sufficient.

Finally, to test for the circum nuclear molecular ring, we use the following model density profile, which has often
used to model proto-planetary disc.
\begin{equation}
\Sigma = \Sigma_0 \left(\frac{r}{R_s}\right)^{-\gamma} \mbox{exp}{ \left(\frac{r}{R_s}\right)^{2-\gamma} } \sqrt{1-\frac{R_{in}}{r}}
\label{eq:gasprof}
\end{equation}
This profile exponentially decreases at $r>R_s$ and has a power law profile at the centre as same as the commonly used profile \citep[e.g.,][]{andrews_etal_2009} however 
it has an additional cut-off radius $R_{in}$ below which the profile truncates very sharply.

\begin{figure*}
\centering
\subfigure[$R_{in}=0.15$\arcs]{\epsfig{file=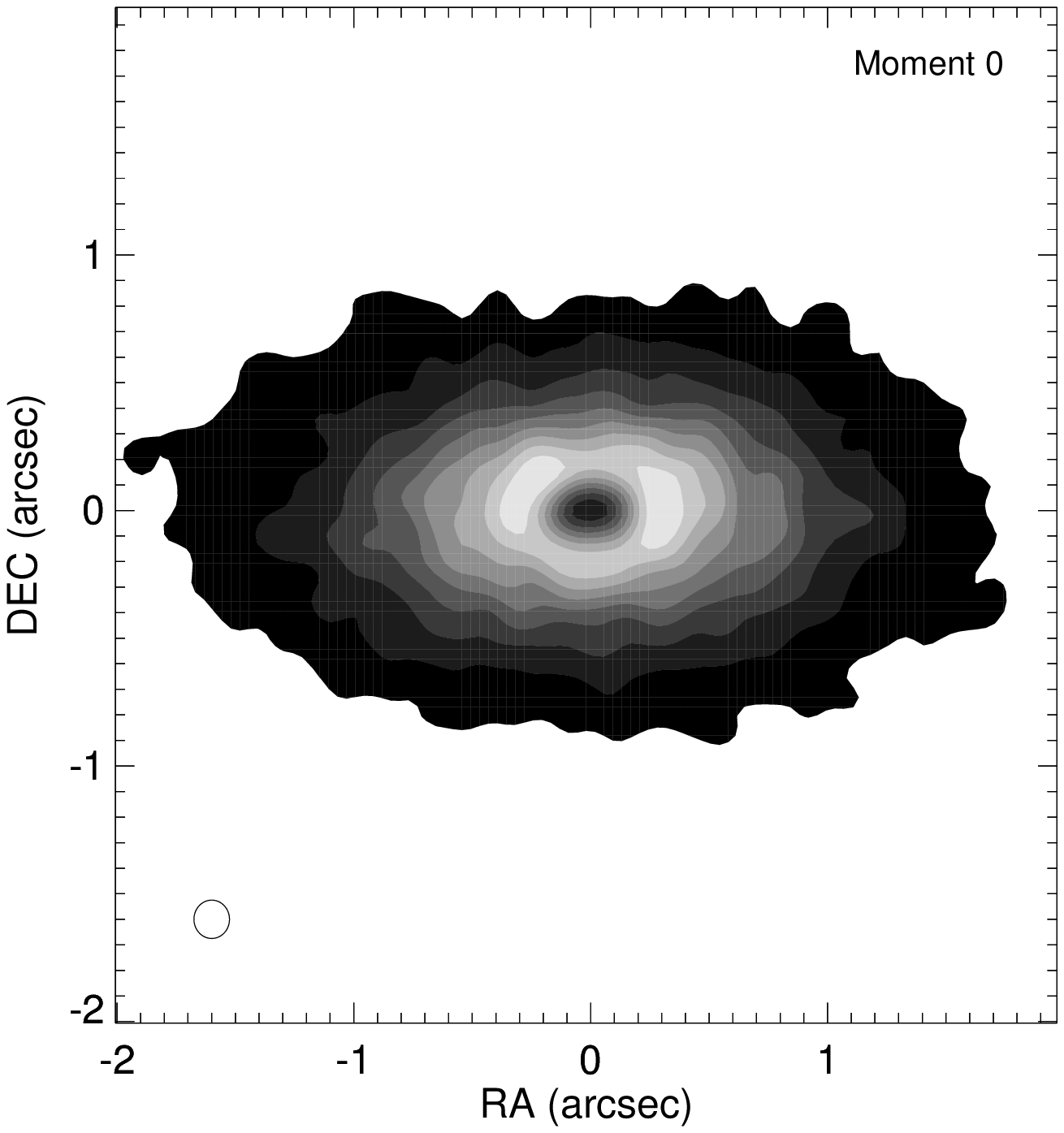,width=0.4\textwidth}}
\subfigure[$R_{in}=0.15$\arcs]{\epsfig{file=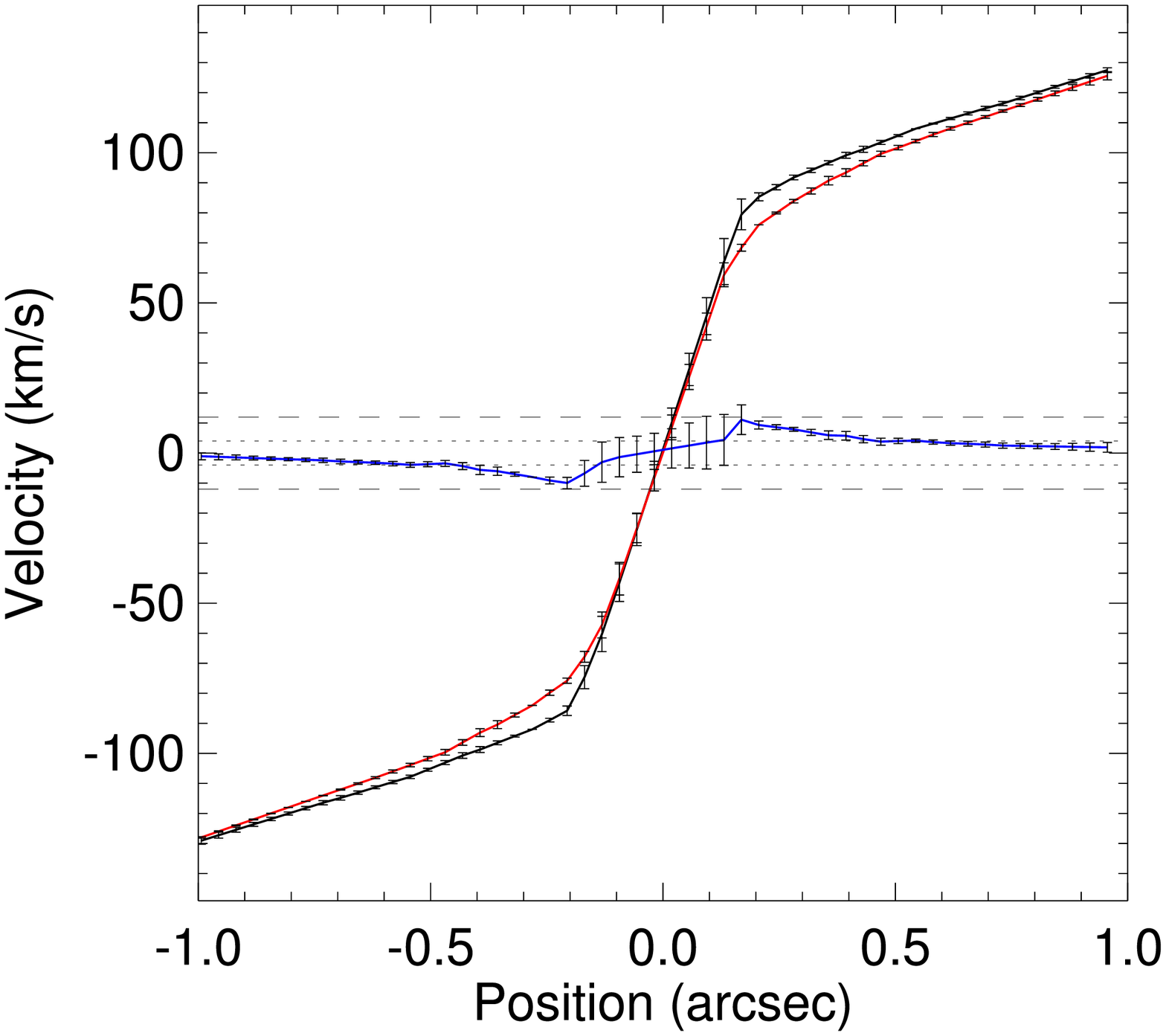,width=0.4\textwidth}}
\subfigure[$R_{in}=0.30$\arcs]{\epsfig{file=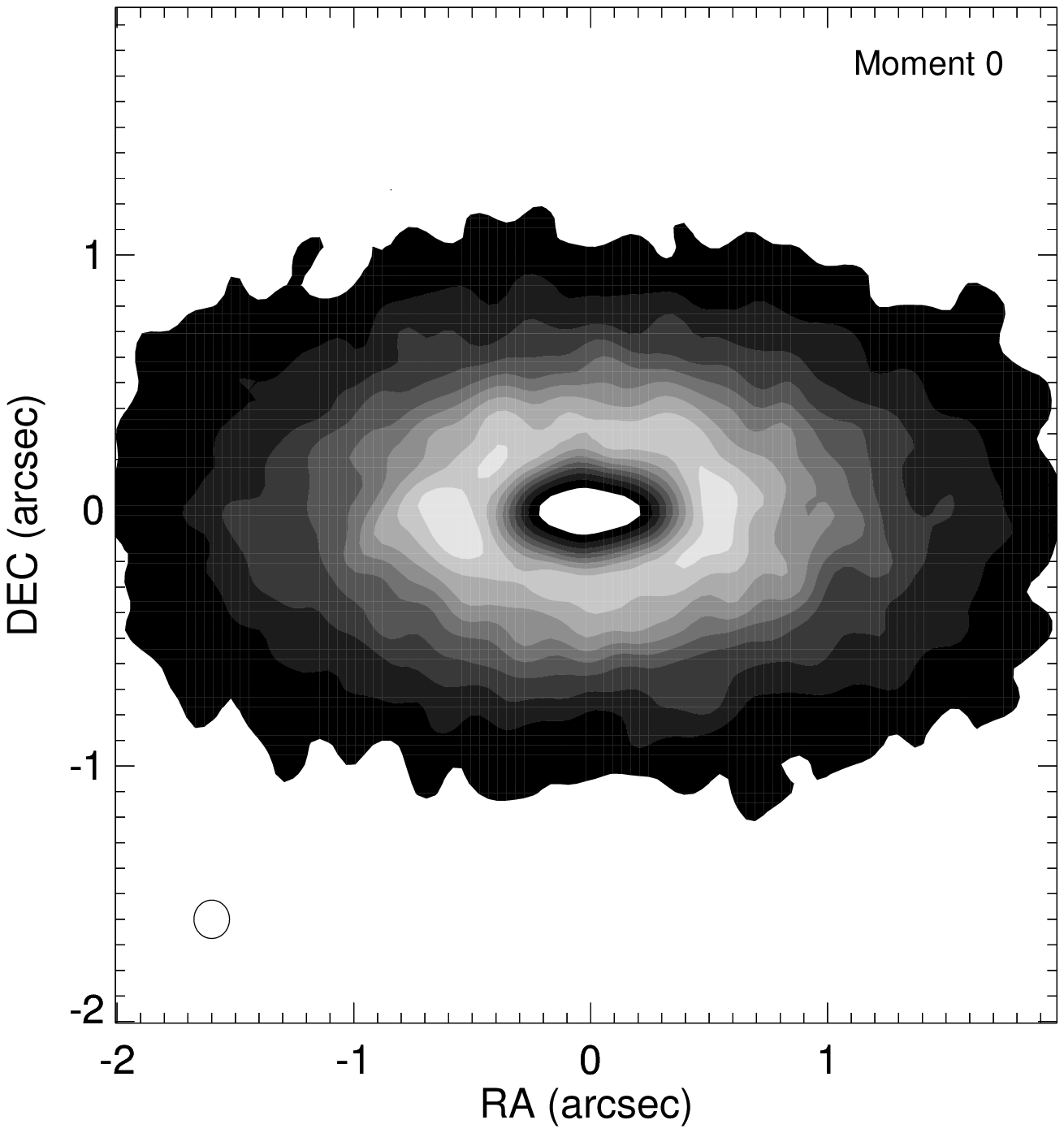,width=0.4\textwidth}}
\subfigure[$R_{in}=0.30$\arcs]{\epsfig{file=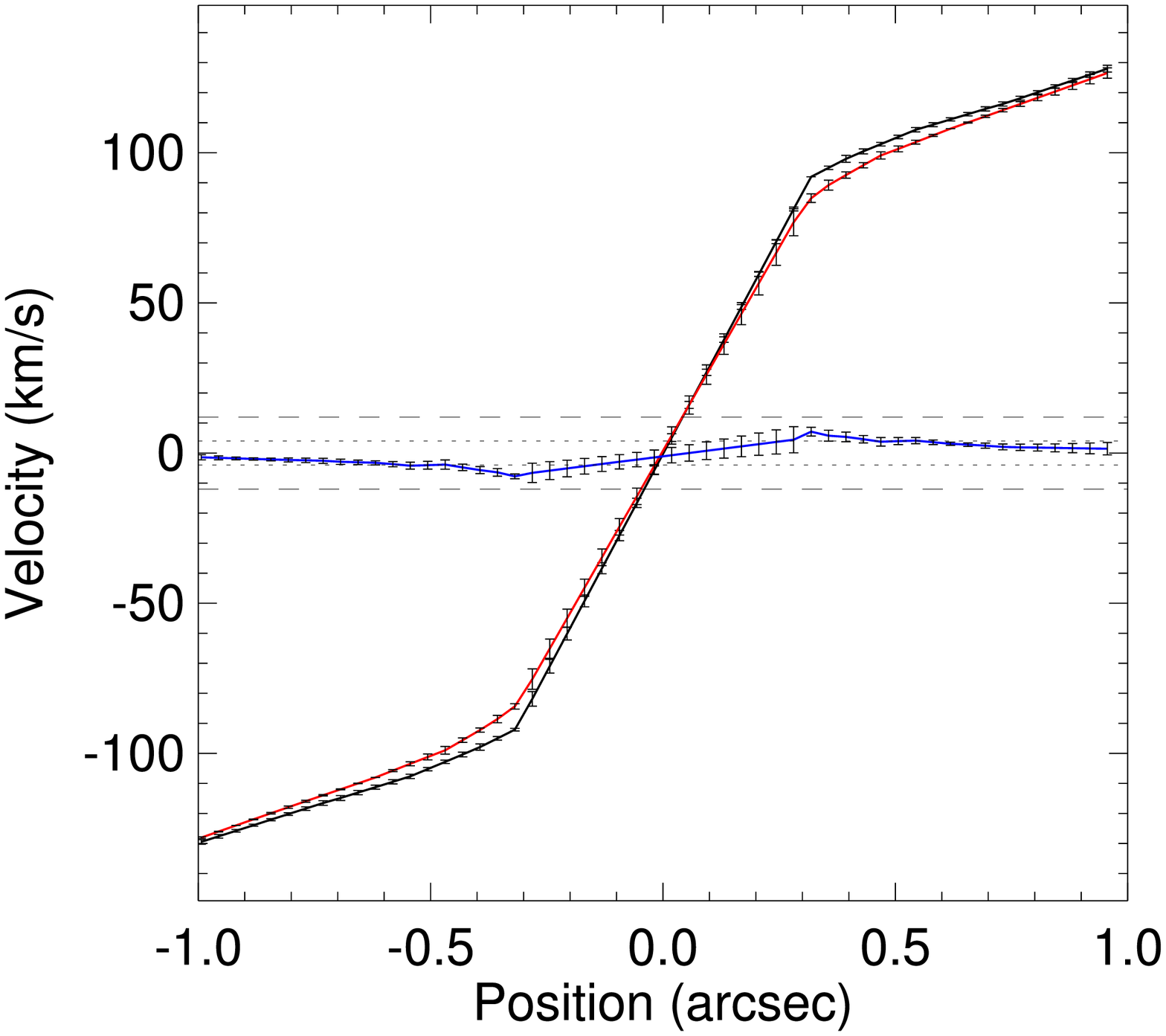,width=0.4\textwidth}}
\caption{Moment 0 map (right panel) and rotation and residual velocity (left panel) of the simulated galaxy in Figure \ref{fig:kinms_snr}(c) measured using the required beam size \reqv\ (0.15\arcs) and
velocity channel width (8km/s), but using gas density distribution with inner hole. For fixed $R_s=1$\arcs\ in equation(\ref{eq:gasprof}), $R_{in}=0.15$ and 0.30 \arcs\ 
was used for upper and lower panel respectively. $R_{in}=0.15$\arcs\ is same as the required beam size for this galaxy to detect the SMBH. Error bar has been determined by the 
standard deviation of the realisation of 100 ensemble PVDs. The residual velocities are scaled by the velocity channel width being used to generate the 
PVD for each panel. The uncertainties associated with 1 and 3 times the velocity channel width are shown by the dotted and dashed line. Note that if there is no velocity 
measurement in the centre, the velocity measurement is linearly interpolated to the centre.}
\label{fig:kinms_rin}
\end{figure*}

Figure \ref{fig:kinms_rin}(a) and (b) show the moment 0 map and PVD for the same galaxy used in this section but for the molecular gas with density profile following equation 
(\ref{eq:gasprof}) with $R_s=1.0$\arcs and $R_{in}=0.15$\arcs. Similarly Figure \ref{fig:kinms_rin}(c) and (d) shows the case with $R_s=1.0$\arcs and $R_{in}=0.30$\arcs. 
As $R_{in}$ becomes larger, the velocity tracer in the centre disappears and the central region with the strongest impact of SMBH cannot be probed if the hole 
size is similar to or larger than \reqv. Figure \ref{fig:kinms_rin}(b) and (d) demonstrate this by showing that for the circum nuclear gas ring, significance of the 
residual velocity at the centre is much smaller than that for the circum nuclear gas disc if the hole size is similar to or larger than \reqv. 
For the case of large $R_{in}=0.3$\arcs, the velocity at the centre in Figure \ref{fig:kinms_rin}(d) was linearly interpolated using the velocity measurements in the outer region.

\subsubsection{Non-circular velocity structure}
We consider three different systematic velocity structures: outflow, random motion and disc warp and demonstrate how they distort the PVD from pure 
circular motion. 

\begin{figure*}
\centering
\subfigure[Small outflow velocity]{\epsfig{file=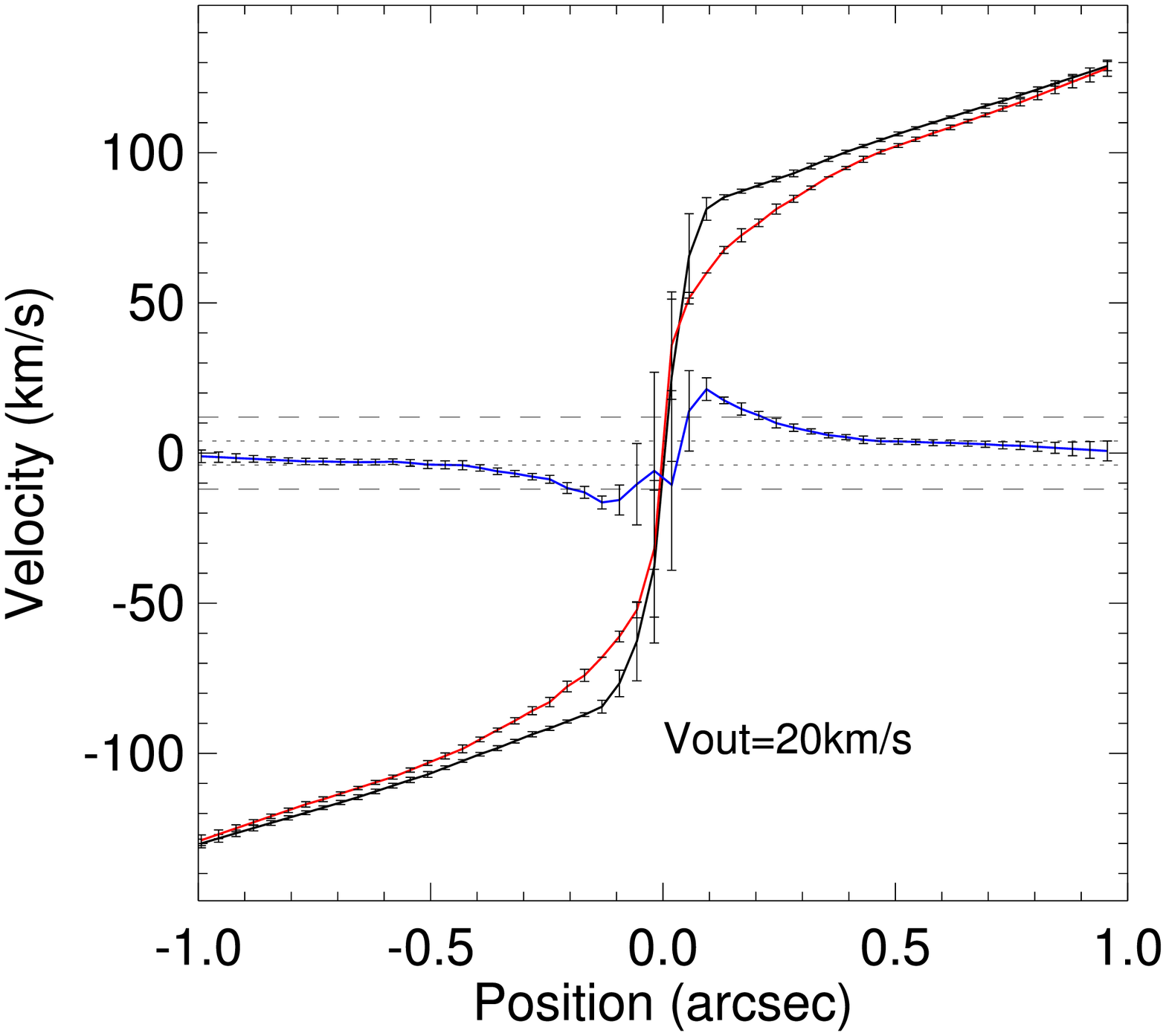,width=0.4\textwidth}}
\subfigure[Large outflow velocity]{\epsfig{file=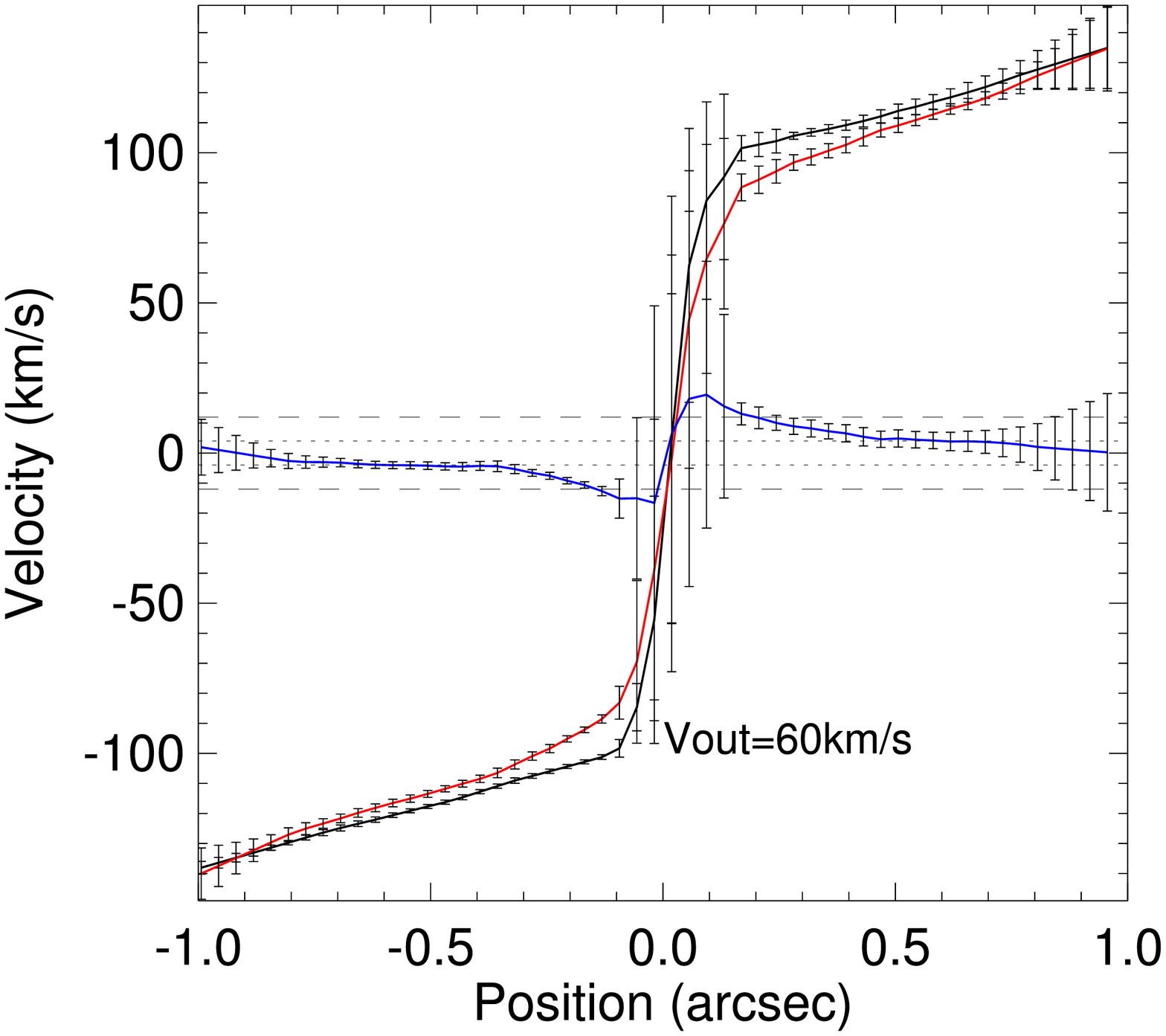,width=0.4\textwidth}}
\subfigure[Small random velocity]{\epsfig{file=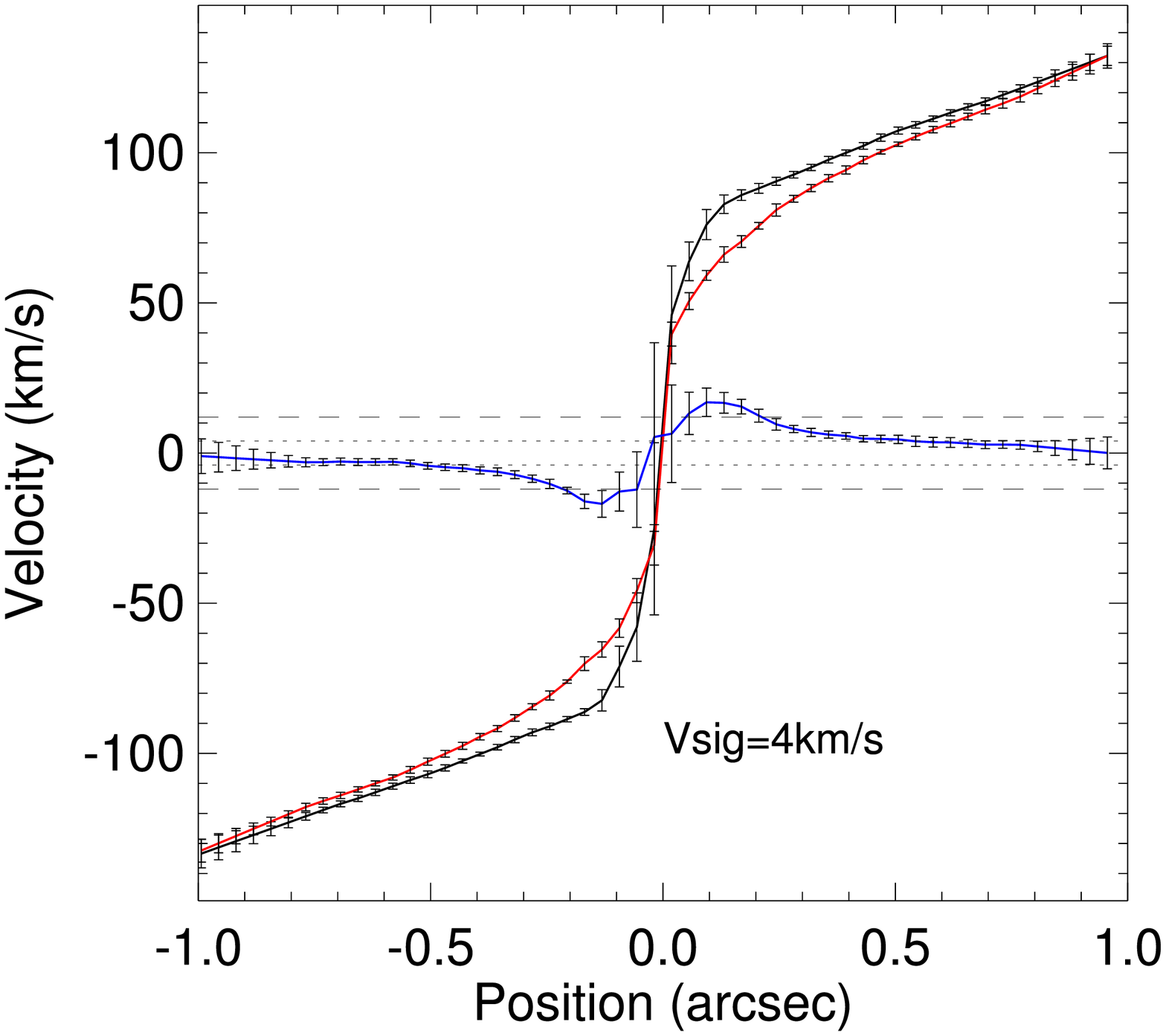,width=0.4\textwidth}}
\subfigure[Large random velocity]{\epsfig{file=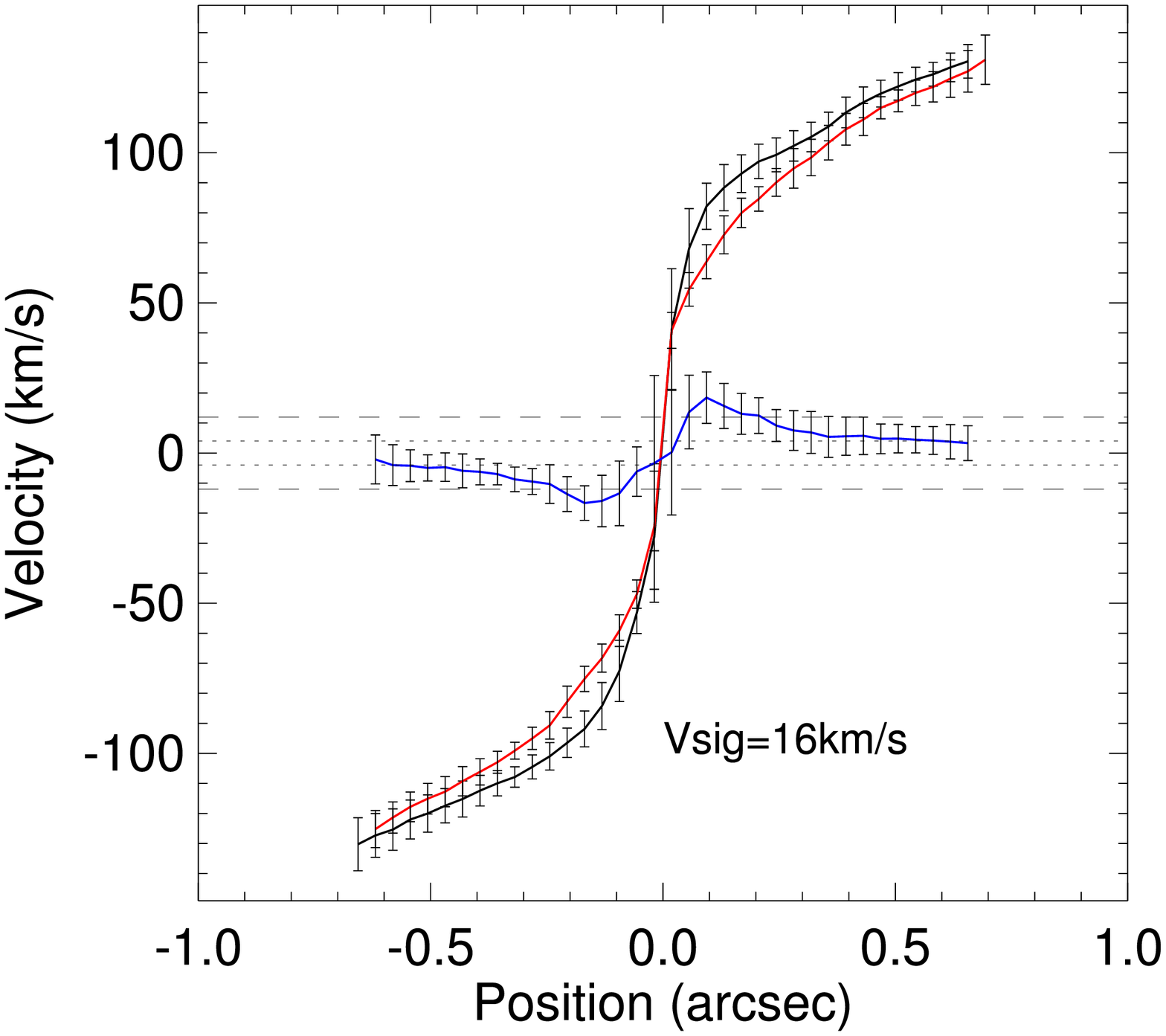,width=0.4\textwidth}}
\caption{Rotation velocities and their residuals of the simulated galaxy in Figure \ref{fig:kinms_snr}(c) measured using the required beam size \reqv\ (0.15\arcs) and
velocity channel width (8km/s), but using different outflow and random velocities for the gas disc. Error bar has been determined by the standard deviation of the
realisation of 100 ensemble PVDs. The residual velocity velocities for each panel are scaled by the velocity channel width being used to generate the PVD for each panel.
The uncertainties associated with 1 and 3 times the velocity channel width are shown by the dotted and dashed line.}
\label{fig:kinms_velsys}
\end{figure*}

First, we consider the impact of outflow velocity in the circum nuclear molecular gas. Inflow and outflow are included in the velocity field in \kinms\ 
using the formalism of KINEMETRY\citep[][]{krajnovic_etal_2006}. We note that KINEMETRY only models the smooth inflow and outflow in the plane of molecular gas disc 
and more violent molecular gas outflows that we do not consider in this work may have distinct position angles from the main body of the gas disc.  
Figure \ref{fig:kinms_velsys}(a) and (b) show the rotation velocity of galaxy with and without SMBH and the difference of the two velocities for the same galaxy 
used in Figure \ref{fig:kinms_snr}(c), which however has small (20 km/s) and large (60 km/s) constant radial 
outflow velocity in the input velocity profile. \vdiff\ for this galaxy is 24 km/s. If outflow velocity is small compared with \vdiff\ as seen in 
Figure \ref{fig:kinms_velsys}(a), velocity structure is still dominated by the circular motion and the residual velocity shows a $>3 \sigma$ deviation 
at angular scale \reqv. However, if the outflow velocity 
is significantly larger than \vdiff\ as seen in Figure \ref{fig:kinms_velsys}(b), error bar of the residual velocity in the central region is larger than 
3$\sigma$ deviation and signature of the maximum deviation seen in Figure \ref{fig:kinms_velsys}(a) is not significant anymore.

Second, we consider the impact of random velocity dispersion in the molecular gas. Figure \ref{fig:kinms_velsys}(c) and (d) show the rotation velocity of galaxy with and
without SMBH and the difference of the two velocities for the same galaxy used in Figure \ref{fig:kinms_snr}(c) which, however has a small (4 km/s) and large (16 km/s) random 
velocity dispersion in the input velocity profile, compared with the 8km/s velocity channel width used in Figure \ref{fig:kinms_snr}(c). 
If the random velocity dispersion is smaller than the velocity channel width (Figure \ref{fig:kinms_velsys}(c)), 
the velocity broadening in PVD is mild and significance of the residual velocity is not much contaminated by errors.
However if the random velocity is larger than the channel width (Figure \ref{fig:kinms_velsys}(d)), 
the residual velocity shows larger error and decreasing significance of the velocity difference due to SMBH. 
Therefore, to detect the SMBH for the given spatial resolution and galaxy mass profile, a care should be given to the choice of velocity channel 
width, which should be significantly smaller than the random velocity dispersion in the molecular gas, which is, however, difficult to know a priori and needs to be 
included as a part of modeling if necessary.

\begin{figure*}
\centering
\subfigure[Warp disc]{\epsfig{file=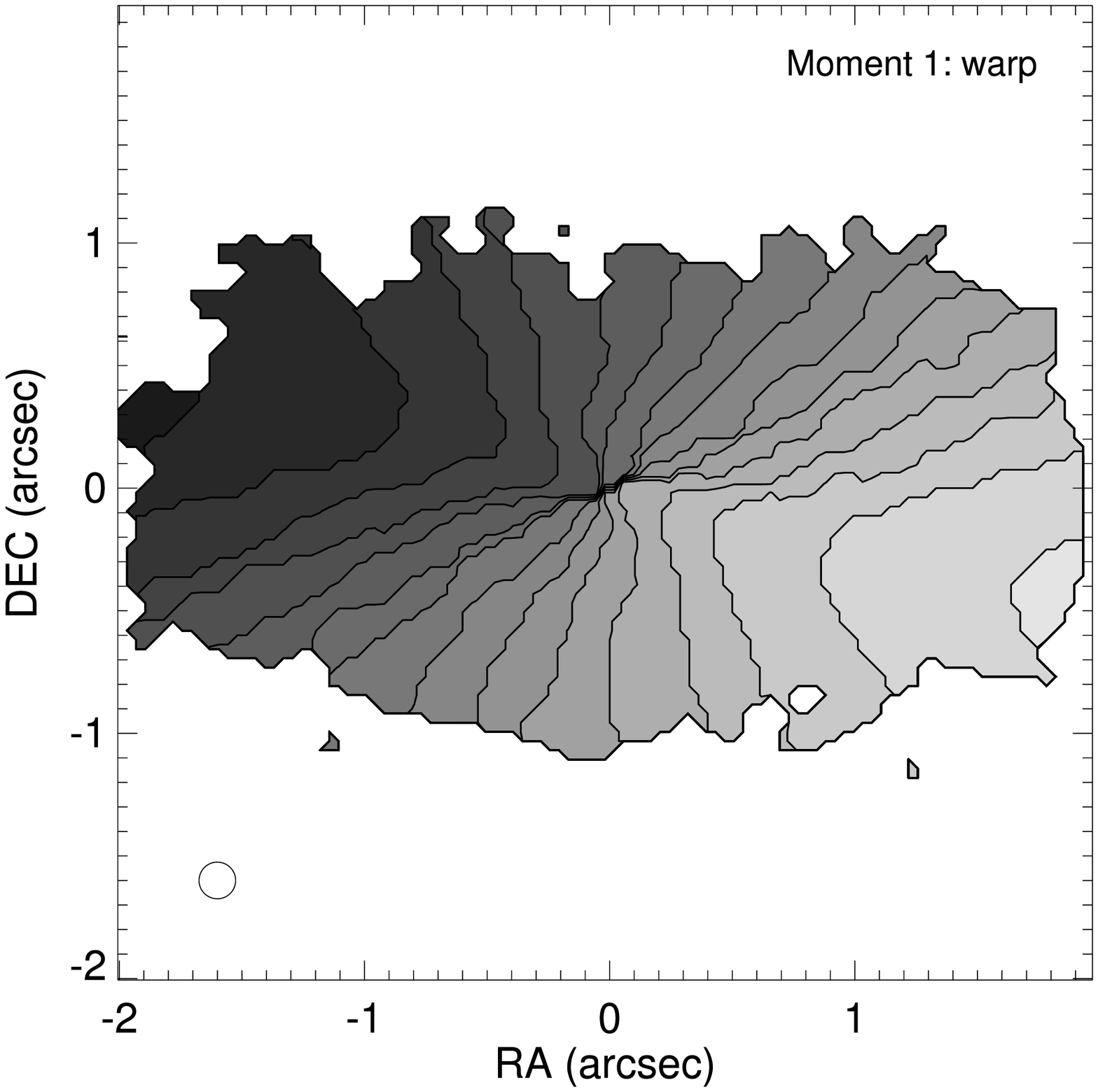,width=0.39\textwidth}}
\subfigure[Warp disc]{\epsfig{file=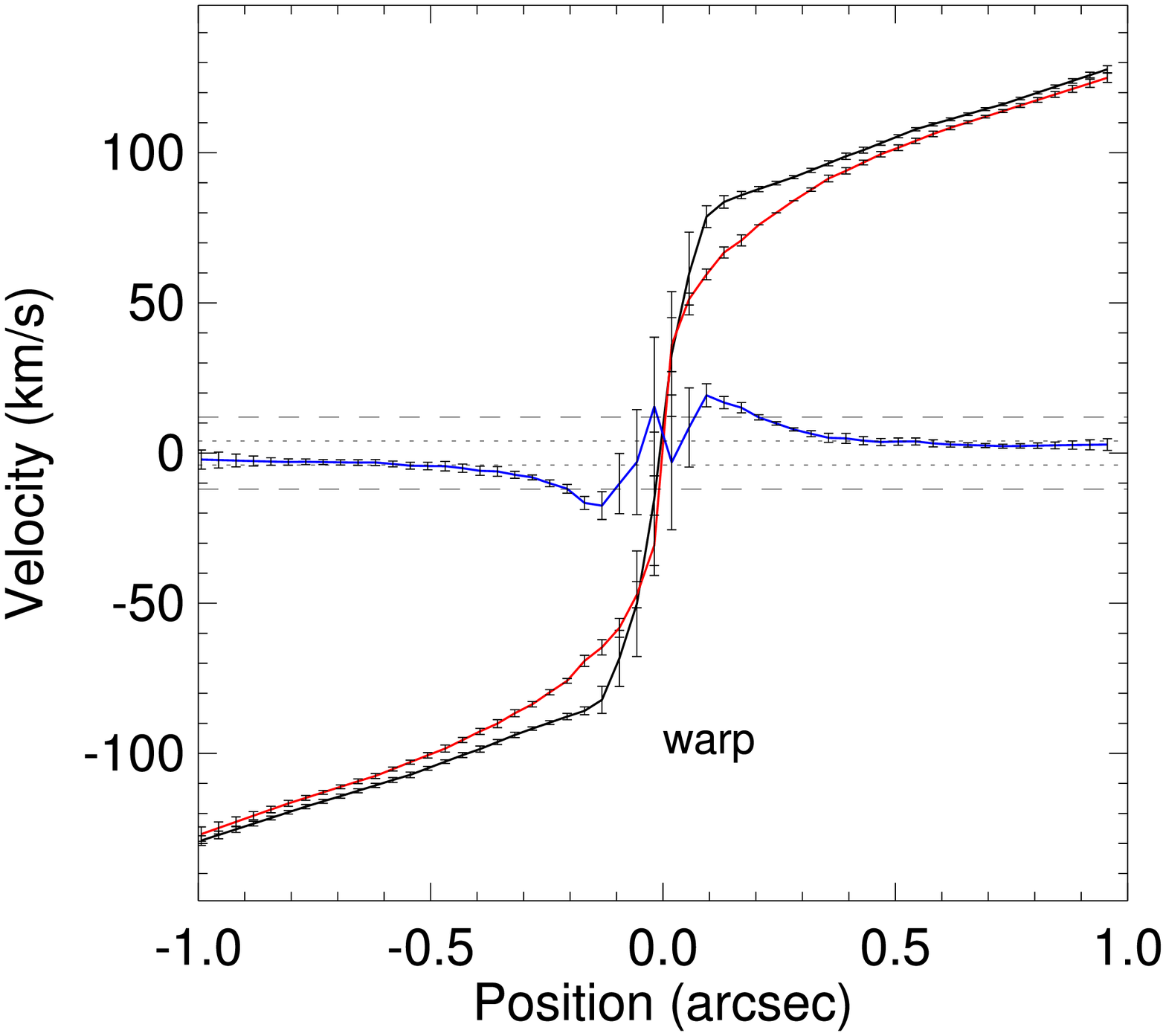,width=0.41\textwidth}}
\caption{
Velocity field (left panel) and rotation and residual velocity (right panel) of the simulated galaxy in Figure \ref{fig:kinms_snr}(c) measured using the required beam size \reqv\ (0.15\arcs) and
velocity channel width (8km/s), but using velocity structure for the warped gas disc. If projected, the warped disc appears with varying position angle. The difference of 
the position angle at the inner most and the outer most region is 30\deg. Error bar has been determined by the
standard deviation of the realisation of 100 ensemble PVDs. The residual velocities for each panel are scaled by the velocity channel width being used to generate the
PVD for each panel. The uncertainties associated with 1 and 3 times the velocity channel width are shown by the dotted and dashed line.
}
\label{fig:kinms_warp}
\end{figure*}

Last we consider a warp disc of molecular gas.
If the warp disc is projected 
onto the sky, the position angle of the disc gradually varies from the centre to the outer region. As a result, warp disc introduces an uncertainty to the determination of 
the kinematic major axis in PVD. Figure \ref{fig:kinms_warp}(a) shows the 
velocity field (or moment 1 map) of the warp disc simulated by changing position angle from 240\deg\ to 270\deg\, as a function of radius. The major axis is determined by the position angle at the 
outer region (i.e. $i=270\deg$) and the inner most region has a 30 degree offset. Like other tests, every other parameters are the same as the ones in 
Figure \ref{fig:kinms_snr}(c). Unlike the regular symmetric `spider diagram' due to pure rotation, the velocity field of the warp disc shows a distortion at the centre. 
Because of the incorrect major axis in PVD analysis, the rotation velocity in the 
region where the warping is occurring can be underestimated depending on the velocity measurement methods \citep{sofue_2001}.
Figure \ref{fig:kinms_warp}(b) shows the rotation velocity of galaxy with and without SMBH and the difference of the two velocities.
Although `envelope tracing' method from this test traces the terminal velocity that was not much affected by warping as shown by comparison with Figure \ref{fig:kinms_snr}(c),
more sophisticate methods \citep[e.g.,][]{jozsa_etal_2007} are required to perform detail modeling of the warped disc to infer the SMBH mass. 

\subsection{Surface Brightness Profile Bias}\label{sec:simul_surfbias}
\begin{figure*}
\centering
\subfigure[nearly flat core]{\epsfig{file=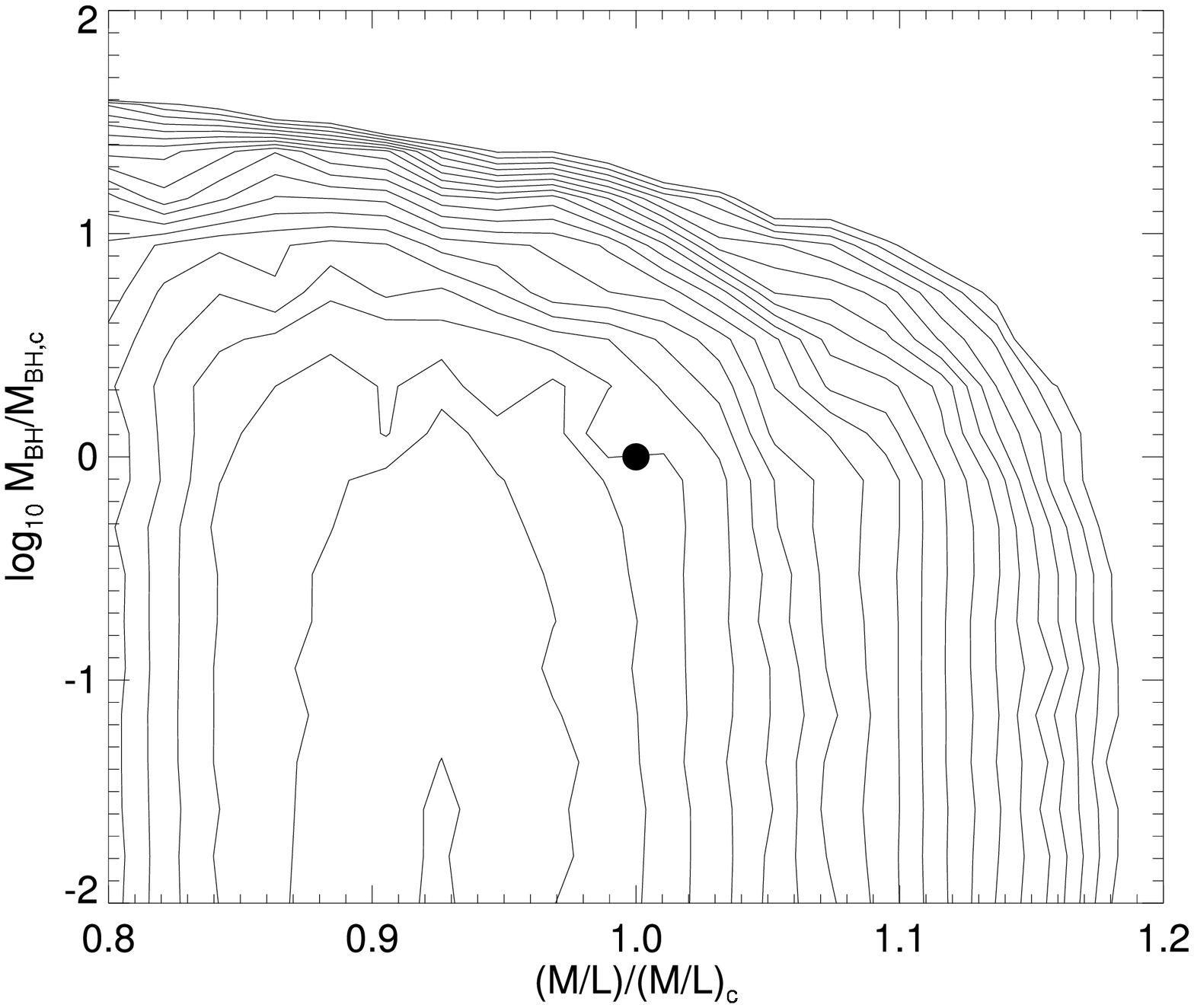,width=0.45\textwidth}}
\subfigure[power law core]{\epsfig{file=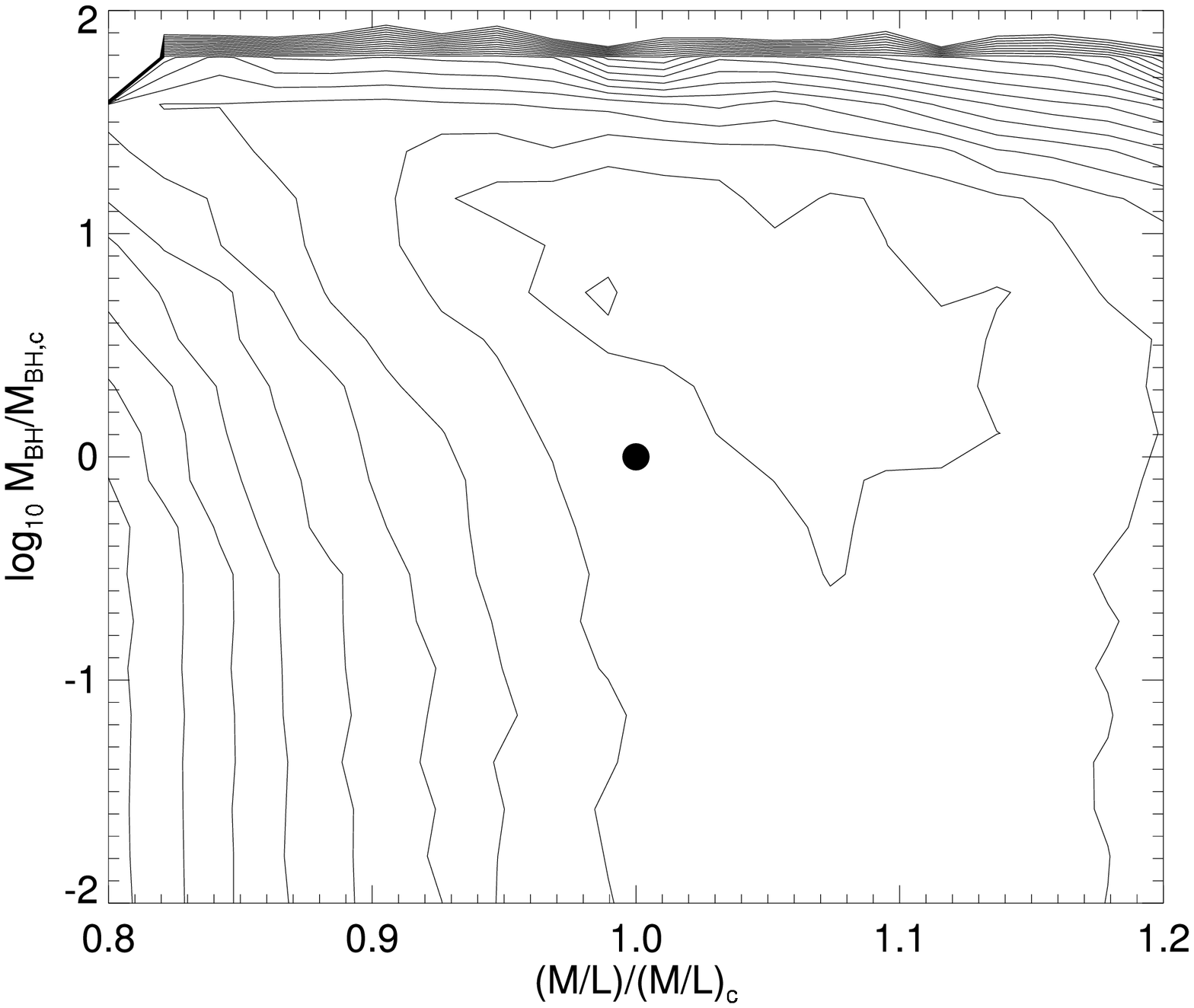,width=0.45\textwidth}}
\caption{Contour of $\chi^2$ per degree of freedom of the model rotation velocity for the core-\sersic\ galaxy with nearly flat core seen 
in Figure \ref{fig:galshape}(a) and power law core seen in Figure \ref{fig:galshape}(b). Galaxy surface brightness is assumed to be accurately determined without 
bias for radius larger than the seeing of galaxy image. Two free parameters: SMBH mass and mass-to-light ratio are normalised by the true value used for simulating 
each galaxy. Black point in the middle of each panel indicates the location 
of the true parameter. The largest probability in the $\chi^2$ is offset from the location of the true parameter, biasing the SMBH mass by an order of magnitude with a small
change of the mass-to-light ratio. 
}
\label{fig:kinms_galshape}
\end{figure*}

In Section \ref{sec:theory_surfprofile}, we show that the systematic error of galaxy surface brightness profile due to the poor resolution of galaxy image may 
bias the inference of the SMBH mass as illustrated in Figure \ref{fig:galshape}. 
We simulate the two galaxy PVDs using the same rotation velocities of the core-\sersic\ profiles in Figure \ref{fig:galshape}. 
However, in this demonstration, we shift the rotation velocities in Figure \ref{fig:galshape} along the $x$-axis by moving the galaxy closer and simulate PVD using 0.15\arcs\ beam size 
instead of 0.01\arcs\ beam size shown in Figure \ref{fig:galshape}, by assuming that the 
galaxy photometric image does not resolve the galaxy core profile and ALMA beam also cannot resolve the difference of the true rotation velocities
and the biased velocity due to the galaxy surface brightness profile bias. We simulate the PVD using the same 0.15 \arcs\ beam 
size and 8 km/s velocity channel width for the two core-\sersic\ galaxies one with $\gamma=0.1$ (Figure \ref{fig:kinms_galshape}(a)) and the other 
with $\gamma=0.5$ (Figure \ref{fig:kinms_galshape}(b)). 

Then we measure the SMBH masses in these galaxies by modeling their rotation velocities to demonstrate the effect of bias in the measurement of galaxy surface brightness profile. 
For this purpose, we assume that galaxy surface brightness profile is accurately determined beyond the radius larger than the seeing of optical or NIR image and fix the \sersic\
index parameter to the true value since the cores of these galaxies are assumed to be not resolved. This is the biased galaxy surface brightness profile used for demonstration. 
Then two free parameters are the SMBH mass and the mass-to-light ratio. On two-dimensional grid space of the SMBH mass
and mass-to-light ratio, we compute a $\chi^2$ per degree of freedom by comparing the true rotation velocity and model rotation velocity generated from the biased galaxy 
surface brightness profile with varying mass-to-light ratio and SMBH mass. We assume a constant error of the data point related only to the velocity channel 
width, $\sigma_{v}=\sqrt{0.5} \Delta V$ \citep[][]{davis_2014}. In addition to the channel width, the real error also includes the uncertainty of velocity measurement using 
galaxy surface brightness profile however, we neglect this uncertainty since we assume that the galaxy surface brightness profile is accurately determined down to the scale
of image seeing and only mass-to-light ratio changes the galaxy rotation velocity.

Figure \ref{fig:kinms_galshape}(a) and (b) respectively shows the contour of $\chi^2$ per degree freedom of the model rotation velocity for the simulated core-\sersic\ 
galaxy in Figure \ref{fig:galshape}(a) and (b). Two free parameters: SMBH mass and mass-to-light ratio are normalised by the true SMBH and mass-to-light
ratio. Black point in the middle of each figure indicates the location of the true parameter. For galaxy with nearly flat core shown in 
Figure \ref{fig:kinms_galshape}(a), the best $\chi^2$ value region is systematically offset from the location of the true parameter in such that the SMBH mass is $>10$ times
smaller and the mass-to-light ratio is $\approx 5$\% smaller than the true value. For galaxy with power law core shown in Figure \ref{fig:kinms_galshape}(b), the best $\chi^2$ 
region is also offset from the true parameter in such that the SMBH mass is $\approx 10$ times larger and the mass-to-light ratio is $\approx 3$\% smaller than the true value.

Figure \ref{fig:kinms_galshape}(a) and \ref{fig:kinms_galshape}(b) demonstrate that if the galaxy core profile in the photometric image is not resolved and the determination of 
stellar mass distribution in the centre is slightly biased, 
the SMBH mass can be biased by more than an order of magnitude compensated by small change in the mass-to-light ratio. 
This systematic bias in the SMBH mass is much larger than the fitting error \citep[20-80 percent e.g.,][]{davis_etal_2013,onishi_etal_2015} and can 
be even larger depending on the galaxy profile shape and the angular resolution of the galaxy image. This confirms our argument in Section \ref{sec:theory_surfprofile} 
and implies that the spatial resolution of galaxy image and the beam size of radio interferometry should be similar and the both have to be comparable to the \reqv\ of 
the target galaxy to obtain accurate SMBH mass without large bias.

\section{Discussion}\label{sec:discuss}
Since only few SMBH mass measurements using molecular gas kinematics with high resolution radio interferometry have been 
reported \citep[e.g.][]{davis_etal_2013,onishi_etal_2015,barth_etal_2016} at the time of writing, it is difficult to validate our arguments using real observation. 
Nevertheless, we find that \citet{davis_2014} analyses molecular gas kinematics of 
NGC 4526 using different beam sizes to demonstrate that reliable inference of the SMBH mass is possible with the beam size larger than \rsoi. The beam size (i.e., 0.25\arcs) 
used in the original paper reporting the mass of SMBH in NGC 4526 \citep{davis_etal_2013} is close to the SMBH \rsoi\ inferred by $M-\sigma$ relation \citep{gultekin_etal_2009} 
and the consistent measurement of the SMBH mass has been obtained using up to 1\arcs\ beam \citep[][]{davis_2014}, which is 4 times larger than \rsoi. Interestingly, we note that 
\reqv\ for NGC 4526 also corresponds to 1\arcs\ based on the assumed SMBH mass ($2\times10^{8}$\msun) and the luminosity distance (16.4 Mpc) used in \citet{davis_etal_2013}, if 
adopting the known \sersic\ index \citep[$n=2.7$,][]{krajnovic_etal_2013} and the galaxy stellar mass \citep[$M_{*}=10^{11}$\msun][]{capetti_etal_2009} from 
the literature. It suggests that \reqv\ as the minimum spatial resolution required for the SMBH mass measurement also makes sense to the real observational data.

Based on the simple discussions of the galaxy rotation velocity and the analysis of simulated PVDs, we argue that for given galaxy stellar mass ($M_{*}$), surface brightness 
profile shape ($n$), inclination ($i$) and luminosity distance ($D_L$), there is a minimum requirement of the spatial and velocity resolution to resolve PVD to detect the SMBH mass 
that one aims to detect, which will be an upper limit if not detected. This allows us to investigate the capability of ALMA in the parameter space of the beam size and velocity 
channel width, for the given galaxy observational parameters ($M_{*}$, $n$, $i$ and $D_L$).

\begin{figure*}
\centering
\subfigure[]{\epsfig{file=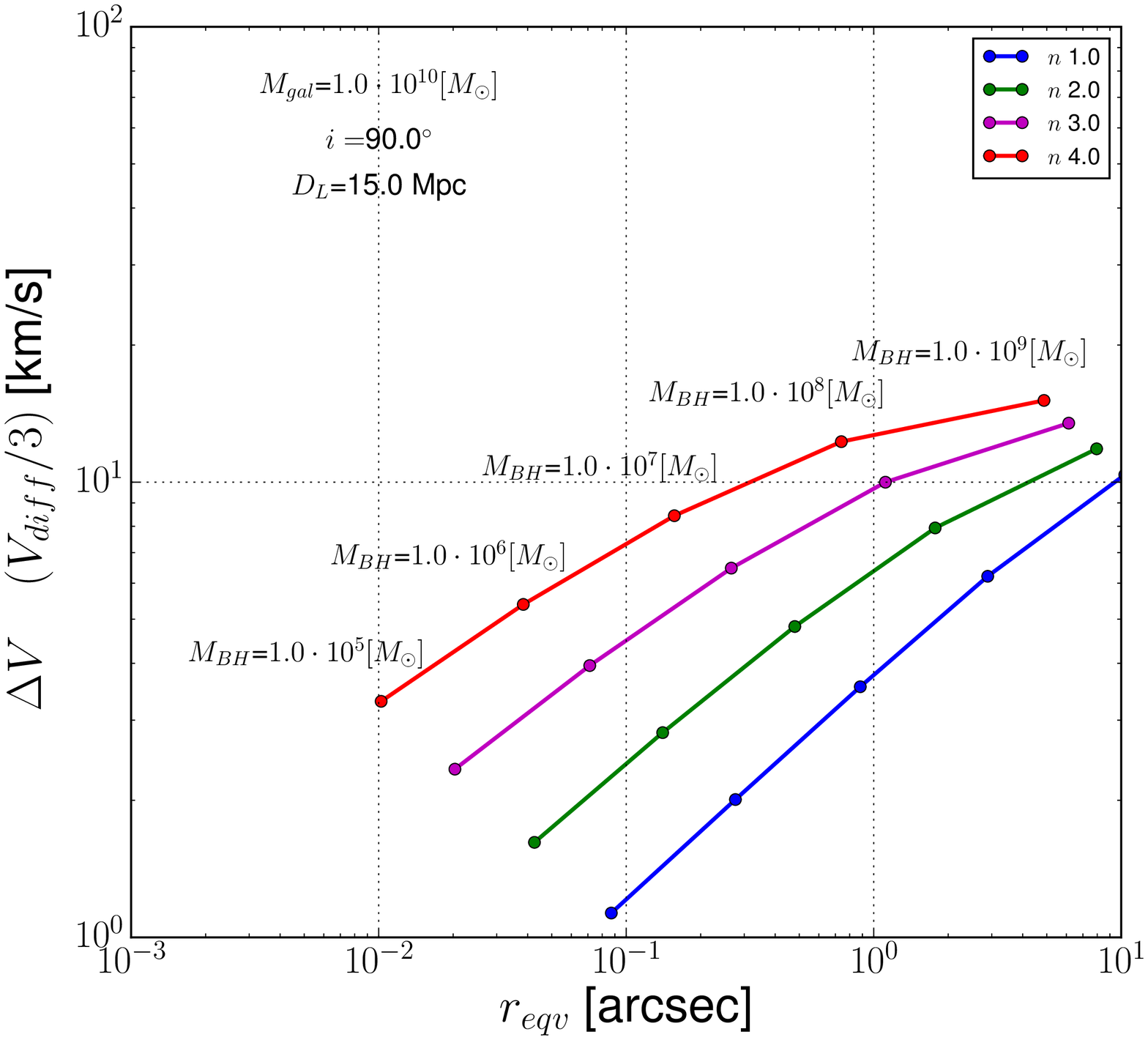,width=0.45\textwidth}}
\subfigure[]{\epsfig{file=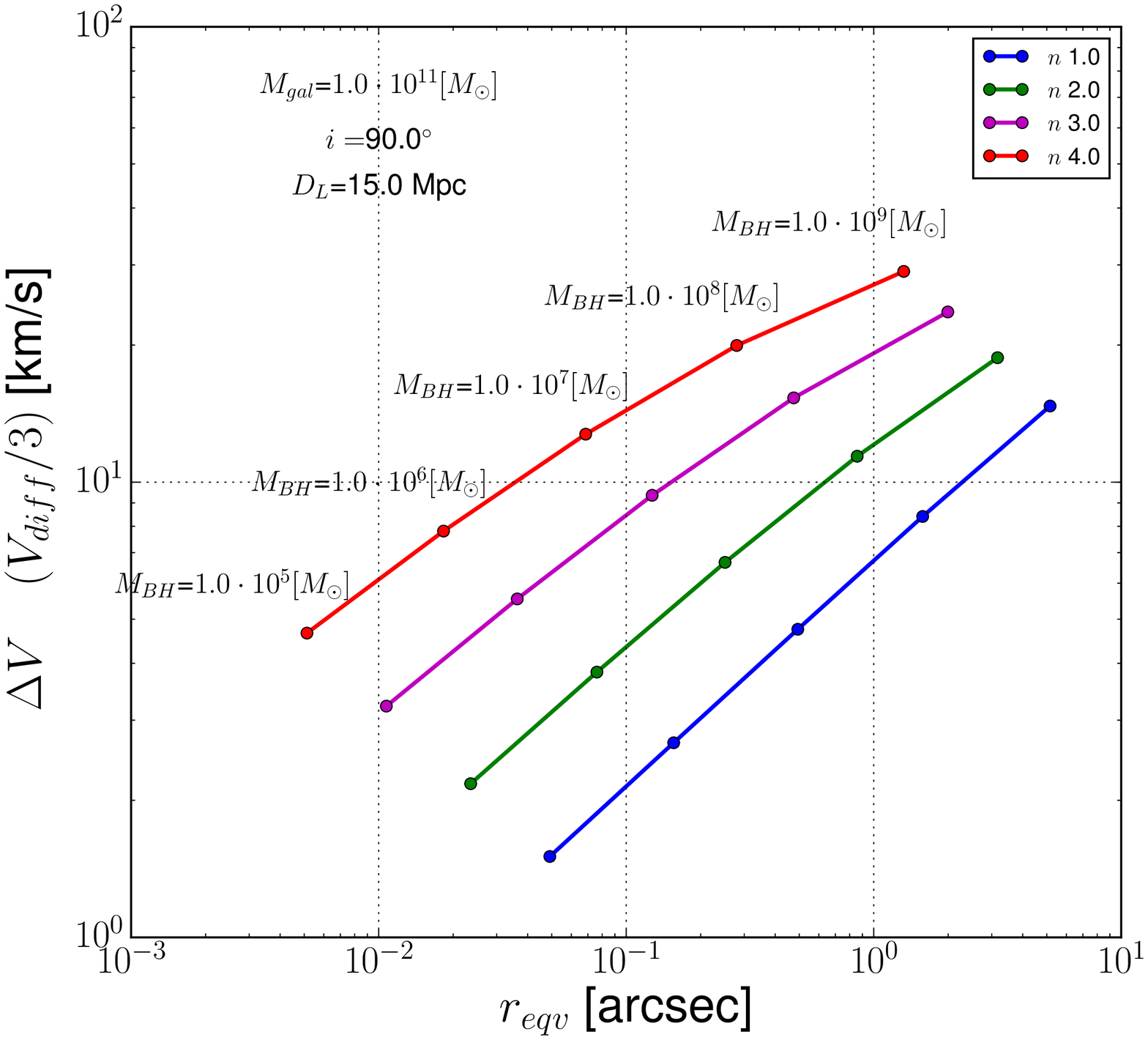,width=0.45\textwidth}}
\caption{Parameter space of the required beam size and velocity channel width for detecting the SMBH mass of which location is shown by the circle with 
appropriate annotation. A fiducial galaxy with $M_{*}=10^{10}$\msun\ (left) and $M_{*}=10^{11}$\msun\ (right) with inclination $i=90$ is assumed to be  
at 15 Mpc distance. Colours indicate galaxy surface brightness profiles determined by \sersic\ index ($n=1,2,3,4$).
Using relevant galaxy inclination and distance, one can shift the points and estimate the required beam size and velocity channel width to detect
the SMBH mass that one aim to detect.}
\label{fig:reqv_vel}
\end{figure*}

In Figure \ref{fig:reqv_vel}, we show the locations of \mbh\ in the parameter space of the required beam size and velocity channel width to resolve the PVD to 
detect the SMBH. Figure \ref{fig:reqv_vel}(a) and \ref{fig:reqv_vel}(b) shows the case of a galaxy with $M_{*}=10^{10}$\msun\ and $M_{*}=10^{11}$\msun\ respectively.
Different colours indicate galaxy \sersic\ indices and each round symbol indicates the location of the SMBH mass as being annotated.
We set luminosity distance $D_L=15$Mpc ($z=0.0035$ for adopted cosmology in this study) and $i=90$\deg\ in Figure \ref{fig:reqv_vel}, and if needed, the 
value of $\Delta V$ can be adjusted by multiplying $\mbox{sin}(i)$ for given galaxy inclination and the value of \reqv\ can be adjusted by multiplying the 
ratio between angular diameter distance at $D_L=15$Mpc and at $D_L$ where the galaxy is actually located, using available redshift. 
However we emphasise that $\Delta V$ and \reqv\ are in practice limited by the systematic velocity structure and angular resolution of the photometric image as we discussed in 
previous sections.

Keeping that in mind, as the SMBH mass that one aims to detect decreases, both $\Delta V$ and \reqv\ also decreases for all types of galaxies as expected. 
In more detail, for the galaxy with lower \sersic\ index, the required angular and velocity resolution can be routinely achievable by ALMA. 
For example, Figure \ref{fig:reqv_vel}(b) implies that small SMBH with mass as low as $10^5$\msun\ within a galaxy with $M_{*}=10^{11}$\msun\ and $i=60$\deg\ 
whose surface brightness profile follows \sersic\ model with $n=3$, can be detected at a distance up to the Virgo cluster (15 Mpc) using 0.01\arcs\ beam size 
and 2.5 km/s velocity channel width based on the assumption that the galaxy surface brightness profile is determined without large bias and the velocity uncertainty 
in the circum nuclear molecular gas is properly modeled with small residual error ($<2.5$ km/s).
For fixed SMBH mass, \reqv\ becomes larger and $\Delta V$ becomes smaller, as the galaxy surface brightness profile becomes less concentrated.
The effect is more prominent for the spatial resolution, and therefore the disadvantage of decreasing $\Delta V$ (increasing velocity resolution) 
is better compensated by the advantage of increasing beam size. For late-type galaxies, one can take this advantage of large \reqv\ to reach the lower SMBH mass by 
increasing velocity resolution. High-velocity resolution increases the data volume but is easily achievable by ALMA unlike the spatial resolution which is limited by 
the physical array configuration. 

Based on the results in this work, we argue that for a typical 1D velocity dispersion \citep[e.g., 7 km/s in][]{mogotsi_etal_2016} of 
molecular gas in nearby galaxies, ALMA can detect the SMBH mass larger than $10^{7}$\msun\ for nearly all types of galaxies with $M_{*}=10^{11}$\msun\ for $n>2$ 
within 15 Mpc distance, using high angular resolution (0.05\arcs) and velocity channel width ($\Delta V=7$km/s). If using larger $\Delta V \approx 10$ km/s, minimum 
detectable BH mass is $\approx 10^{6}$\msun\ for $n=4$ \sersic\ profile galaxies. This is consistent with the finding from Figure 7 in \citet{davis_2014}.  

In principle, the same exercise can be done when proposing ALMA observation to measure the gas kinematics to detect the SMBH. Although real galaxy 
surface brightness profile does not follow \sersic\ model, this exercise provides a rough estimate of the angular resolution and velocity channel width for ALMA to 
detect the SMBH with expected mass, for a wide range of galaxy morphology types. This study will serve as a simple but useful 
technical justification for ALMA proposal with the science goal of measuring the SMBH mass using molecular gas kinematics.

\section{Summary}\label{sec:summary}
By generalizing and extending the work of \citet{davis_2014}, we studied a potential merits of the technique using molecular gas kinematics to measure the SMBH mass by combining 
analytic argument and realistic PVD simulations by considering the relevant spatial and velocity resolutions, the
systematic effect in the spatial and velocity structure of the circum nuclear molecular gas and the impact of biased galaxy surface brightness profile shape. 
The simple analytic argument suggests that the effect of SMBH can be detected at a spatial scale where the 
rotation velocity due to the SMBH and the galaxy stellar mass distribution becomes equal, \reqv\ which is larger than \rsoi\ by a factor of few for the early-type galaxies
and by an order of magnitude for the late-type galaxies. We find that the increased \reqv\ for less concentrated galaxies 
is an advantage for measuring the SMBH mass in the late-type galaxies. The velocity 
channel width also has to be $\frac{1}{3}$\vdiff\ to resolve the velocity difference 
with $3\sigma$ significance level.

However systematic effects due to the spatial and velocity structure in the circum nuclear molecular gas affect the rotation velocity measurement.
We find that the signature of the SMBH is more clearly detected: (1) if the molecular gas surface density profile is more compact, (2) if the gas disc
is inclined as much as possible but the projected \reqv\ along the minor axis is still being resolved by the observing beam size, and (3) if 
there are enough gas clouds in the centre to trace the kinematics.
We also find that the systematic motions of molecular gas affect the galaxy PVD. Therefore, if existing, gas outflow should be smaller than \vdiff\ and random velocity 
dispersion should be smaller than the velocity channel width. Disc warp, if existing, introduces an uncertainty in the kinematic major axis and may distorts the kinematics
along the major axis, which needs to be considered in the rotation velocity measurement. In addition, we illustrate the impact of systematic error introduced by the incorrect measurement of 
galaxy surface brightness profile due to the insufficient galaxy image resolution to resolve the core surface brightness profile. Depending on the shape of galaxy 
surface brightness profile and the resolution of galaxy photometric image, the SMBH mass can be largely biased and this systematic error can be larger than the fitting error. 
Therefore both the resolution of photometric image and the radio interferometry beam have to be small enough to resolve the \reqv\ for given galaxy in order to minimise the bias.

We use the IDL program, \texttt{KinMS} \citep[][]{kinms_2013} to simulate the observed PVDs of galaxies with SMBH including observational measurement 
processes with noise and systematic effects. Analysis of the measured rotation velocity from the simulated PVD demonstrates the validity of our arguments and 
confirms our intuitions of the impact of SMBH to the rotation velocity of the circum nuclear molecular gas disc. 
This work provides useful guidance to the analysis of galaxy PVD for the SMBH mass measurement and technical justification 
for ALMA proposal to observe the kinematics of molecular gas in the galactic centre to measure its SMBH mass.

\section*{acknowledgment}
The author thanks the anonymous referee for constructive comments that greatly improved the paper.
This study was motivated by the comments from the TAC on the author's ALMA observing proposal.
The author also thanks Tim Davis for kindly making his code \kinms\ publicly available and acknowledges the support 
from the ALMA Local Expertise Group (Allegro) at Leiden. 

\bsp

\bibliographystyle{mn2e}
\bibliography{reference}

\label{lastpage}

\end{document}